\documentclass[11pt]{article}
\pdfoutput=1 

\usepackage{jheppub}
\usepackage{amssymb,amsmath,amsthm,graphicx}
\usepackage[english]{babel}
\usepackage{graphics, color}
\usepackage{subfigure}
\usepackage{ragged2e}
\usepackage{latexsym}
\usepackage{tensor}
\usepackage{adjustbox}
\usepackage{placeins}

\hypersetup{
 pdfauthor={Nicholas M. K. Ross, Toby Wiseman},
 citecolor=black,linkcolor=black,urlcolor=black}
\usepackage[utf8]{inputenc}
\allowdisplaybreaks[1]

\def \dd {\mathrm{d}}

\newcommand{\vect}[1]{\boldsymbol{#1}}
\newcommand{\mat}[1]{\boldsymbol{#1}}

\newcommand{\be}{
\begin{eqnarray}
}

\newcommand{\nl}{
\nonumber \\
}

\newcommand{\ee}{
\end{eqnarray}
}

\newcommand{\cavpanel}[1]{%
  \adjincludegraphics[width=0.47\textwidth, trim={{0.04\width} {0.0\height} 0 {0.08\height}}, clip]{#1}}

\newcommand{\dSpanel}[1]{%
  \adjincludegraphics[width=0.47\textwidth, trim={{0.04\width} {0.0\height} 0 {0.08\height}}, clip]{#1}}
  
\begin{document}
\title{`Forward-backward' flows and integration cycles for semiclassical Euclidean gravity}
\author{Nicholas~M.~K.~Ross} \emailAdd{n.ross25@imperial.ac.uk}
\author{and\;Toby~Wiseman}
\emailAdd{t.wiseman@imperial.ac.uk }

\affiliation{Abdus Salam Centre for Theoretical Physics, Blackett Laboratory, Imperial College, London SW7 2AZ, United Kingdom}

\abstract{ 

Integrating Euclidean semiclassical gravity over real Riemannian metrics is subtle due to the conformal factor problem. 
A perturbative prescription was given by Gibbons--Hawking--Perry (GHP) where the integration contour for the conformal perturbations is rotated to be imaginary.
A potential non-perturbative remedy is that the integration cycles associated to semiclassical saddle points should be deformed into the space of complex metrics in the sense of Picard--Lefschetz theory.  Here we propose a non-perturbative construction of such complex integration cycles.
This employs a `forward-backward' gradient flow  of the Einstein--Hilbert action, starting at a saddle point, along which the action has monotonically increasing real part and constant imaginary part. 
The flow is not parabolic. In a neighbourhood of a saddle point, the real part of the conformal factor flows forward and that of the conformal class backwards, with the opposite true of their imaginary parts. 
Starting the flow at a saddle point, then roughly speaking, one specifies the real part of the conformal class and the imaginary part of the conformal factor at the end of the flow. Solving the flow one finds the complex metrics along a one parameter path within the integration cycle of that saddle, and varying this final data generates the integration cycle, the Lefschetz thimble of the saddle point.
We find that near to a real saddle point this reproduces the GHP prescription.
We characterise the neighbourhood of a saddle where such forward-backward flows may exist, finding a weaker restriction than that of Kontsevich--Segal--Witten.
We numerically demonstrate flows explicitly for 4-dimensional gravity, constructing sections through the integration contours for the saddles given by Euclidean de Sitter, and for Schwarzschild in a cavity. 

}

\maketitle

\section{Introduction}\label{sec:intro}

The Einstein--Hilbert action for semiclassical Euclidean gravity famously suffers from the `conformal factor problem', namely that conformal deformations of the metric lower the action without bound~\cite{Gibbons:1978ac,Mazur:1989by}. 
Nonetheless the Euclidean gravitational saddle point approximation has been tremendously useful in understanding semiclassical physics, such as black hole thermodynamics~\cite{Gibbons:1976ue,Gibbons:1994cg} and entanglement entropy~\cite{Ryu:2006bv,Lewkowycz:2013nqa}. 
There remain mysteries associated to precisely what the Euclidean path integral computes, for example as discussed recently in~\cite{Maldacena:2024spf}.

One perspective is that the Euclidean path integral does not exist as a well defined object, even in a semiclassical limit. Instead only the saddle points themselves, and their fluctuations can be made sense of, and this emerges from a Lorentzian path integral. Thus perhaps one should not consider off-shell Euclidean geometries, even in a semiclassical regime. Indeed Marolf has studied the semiclassical Lorentzian path integral, arguing results usually derived in a Euclidean setting can be deduced directly in the Lorentzian setting~\cite{Marolf:2022ybi}.

However the alternate viewpoint we explore here is that discussed by Witten in~\cite{Witten:2010cx,witten2010newlookpathintegral,witten2022notecomplexspacetimemetrics}, namely that the semiclassical Euclidean path integral may exist in its own right, and that the conformal factor problem is resolved by choosing an appropriate complex deformation of the integration cycle over real Euclidean metrics so that the path integral converges and can be computed from saddle points of the action. The underlying manifold remains real, but the metric -- the integration variable in the path integral -- is continued to a complex tensor field over that real manifold.\footnote{When the metric is generalized from being Riemannian, this underlying manifold plays a role in determining whether the origin of the theory is Euclidean or Lorentzian.} This is precisely what the perturbative prescription put forward by Gibbons, Hawking and Perry (GHP) proposed~\cite{Gibbons:1978ac}. For fluctuations about a real saddle point they state that the correct integration cycle is such that the conformal modes of the metric perturbation should be Wick rotated to be integrated along the imaginary axis, while transverse-traceless modes are integrated on their usual real contour.

Beyond perturbation theory the suggestion is that in the semiclassical approximation the original integration over real metrics can be written as a sum over complex integration cycles, where each is associated to a saddle point in the sense of Picard--Lefschetz theory. These complex integration cycles  form the Lefschetz thimbles for the saddle points, with each being the collection of paths of gradient ascent of the real part of the action, $S$, in the space of complex metrics that originate at the saddle, and along which the action has constant imaginary part. 
Then integration over such a complex integration cycle is a steepest descent of the integrand $e^{- \frac{1}{\hbar} S}$ and may be convergent, when it would be dominated by the saddle point contribution in the semiclassical regime, recovering our naive expectations. 

It is worth emphasising that while Einstein gravity is not UV complete, and hence its path integral is not well defined without a completion (see for example~\cite{Sen:2026bzy}), the semiclassical approximation we are considering here is that of the low energy effective theory. Whatever the true physics of quantum gravity, once the UV degrees of freedom are integrated out, we will arrive at a path integral involving the  metric of the low energy  Einstein effective description, and it is the integration contour of this effective theory that we are concerned with.

A key element of this picture that has been stubbornly missing is a non-perturbative construction of the complex integration cycle for a given saddle point -- i.e. its Lefschetz thimble. The GHP prescription is restricted only to linear perturbations. Indeed even in that setting Marolf and Santos have highlighted the need to improve GHP as in many situations the conformal mode and transverse-traceless modes may be coupled by boundary conditions~\cite{Marolf_2022, Santos_2022, liu2023stabilitysaddleschoicescontour}. In that work they have made a proposal for how to more generally perform the Wick rotation.

The purpose of this paper is to provide the first non-perturbative construction of the integration cycle for a (possibly complex) saddle point of the Euclidean Einstein--Hilbert action. Indeed we will find that the essence of the GHP perturbative prescription emerges very naturally from our non-perturbative construction. 
Our task is to construct a complex gradient ascent flow in the sense of Picard--Lefschetz theory for complex metrics and the Einstein--Hilbert action.
Starting at a saddle point of the action, this flow then constructs a path in the space of complex metrics. 
Choosing different end point data for the flow gives a set of flow paths, and the collection of all possible paths starting at the saddle generates its Lefschetz thimble, at least in the allowed complex neighbourhood where the flows exist.

We begin by considering a related question, namely whether we can construct a gradient ascent flow of the action for Euclidean gravity with real metrics. A canonical flow of the action is Ricci flow, but whilst this is a gradient flow, it is neither a gradient descent nor a gradient ascent flow, as the metric on superspace that governs the flow has indefinite signature~\cite{Headrick_2006,Tseytlin:2006ak}. Indeed the fact that it need not  be  monotonic is precisely related  to the conformal factor problem.
It is straightforward to write an analogous flow that is a gradient ascent of the action; one simply flows by the Einstein tensor rather than (minus) the Ricci tensor. 
We term this the Einstein flow, and in $n$-dimensions it is given by,
\be
\label{eq:EinsteinFlow}
\frac{\partial}{\partial\lambda} g_{\mu\nu} = 2  G_{\mu\nu} + (n-2) \Lambda g_{\mu\nu}  + \nabla_{(\mu} \xi_{\nu)} 
\ee
where $\lambda$ is the flow time, and we include a cosmological constant $\Lambda$, so that fixed points of the flow are saddle points of the Einstein--Hilbert action with $R_{\mu\nu} = \Lambda g_{\mu\nu}$. Here we have also included a flow dependent diffeomorphism term generated by $\xi_\mu$. 
However the price one pays for flowing by the Einstein tensor is that the flow is no longer parabolic, and hence such flows have not been seriously considered previously. 

In this work we emphasize that we may still make sense of this as a `forward-backward' flow, where some degrees of freedom propagate forwards in the flow and others propagate backwards. For the Einstein flow the conformal factor propagates in the usual direction (relative to Ricci flow) but the conformal class flows backwards. Thus the data for the flow is the conformal factor at the initial flow time, and the conformal class at the final flow time. 
Such a flow cannot  be thought of as an integration of initial data at the initial flow time as for a parabolic flow. However we expect solutions nonetheless still exist, and instead must be found by solving the entire flow all at once.
We will show with explicit numerical examples how this works in practice, and in particular how to deal with gauge fixing which is necessary to yield a  solution. We will numerically construct Einstein flows for deformations of Euclidean de Sitter, and also the flow from the large to the small Euclidean Schwarzschild solution in a cavity.

Having explored forward-backward flows for real metrics, we then employ a similar strategy for complex metrics. As we will discuss, even for a real scalar field theory on a fixed Riemannian manifold, then after analytically continuing this to complex scalar fields the canonical complex gradient ascent  that constructs Lefschetz thimbles -- we term this the `Lefschetz flow' -- already gives a `forward-backward' flow, rather than a parabolic flow. A natural generalization of the Einstein flow yields a complex gradient  ascent  flow for the gravitational action for complex metrics on a real manifold. We term this the `Einstein--Lefschetz flow' and  for an $n$-dimensional complex symmetric metric $g_{\mu\nu}$ it is given by,
\be
\label{eq:ELflow}
\frac{\partial}{\partial \lambda} g_{\mu\nu} =  2  \sqrt{\frac{\det(\bar{g}_{\alpha\beta})}{\det(\mathfrak{g}_{\rho\sigma})}} \, \mathfrak{g}_{\mu\rho}\,\mathfrak{g}_{\nu\sigma}    \left( \overline{G}^{\rho\sigma} + \frac{(n-2)\Lambda}{2}  \overline{g}^{\rho\sigma} \right) + \nabla_{(\mu} \xi_{\nu)} 
\ee
where $\mathfrak{g}_{\mu\nu}$ is a real Riemannian metric and $\xi_\mu$ is a complex covector field that generates a complex diffeomorphism along the flow. Here $\Lambda$ is the cosmological constant, so that saddle points of the complex Einstein--Hilbert action, where $R_{\mu\nu} = \Lambda g_{\mu\nu}$, are fixed points of this flow. 
This produces flows of complex geometries where the action $S$ has monotonically increasing real part, and constant imaginary part. Hence $e^{-\frac{1}{\hbar}S}$ decreases in modulus and has constant phase along the flow, as required to construct Lefschetz thimbles for the path integral. 
We require that $\mathfrak{g}_{\mu\nu} \to g_{\mu\nu}$ when the metric is real, in which case this reduces back to the real Einstein flow. 

Unsurprisingly this again is a forward-backward flow provided that the metric is not `too' complex; as for the real Einstein flow, the real conformal class flows backwards, and the real conformal factor flows forwards. Then their imaginary counterparts flow in the opposite sense.
Starting the flow at a saddle point of the action, which is a fixed point of the flow, we argue that consistent data for the flow is roughly given by  fixing the imaginary part of the conformal factor and the real part of the conformal class at the final flow time.
We characterise how `complex' the metric may become so that the flow retains its forward-backward character and is defined by these initial and final boundary conditions.
We will restrict our complex metrics so that,
\be
\label{eq:ELconditions}
\Re\left( \left( \sqrt{\det(g_{\mu\nu})} g^{\mu\nu} q_\mu q_\nu \right)^{1/2} \right) > 0
\ee
for any non-zero real covector $q_\mu$. We note this implies  $\det(g_{\mu\nu}) \ne 0$ so that the metric is invertible.
Given a real frame $\hat{\theta}^a_\mu$ we may write the metric as $g_{\mu\nu} = \vect{\hat{\theta}}_\mu \mat{g} \vect{\hat{\theta}}_\nu$ using matrix notation for the frame indices, so $\mat{g}$ is a complex symmetric matrix valued function. Then we may decompose this as,
\be
\label{eq:ELmetric}
\mat{g} = e^\phi e^{\frac{1}{2} \mat{X}'} e^{i \mat{Y}' } e^{\frac{1}{2} \mat{X}'}
\ee
for real symmetric traceless matrices $\mat{X}'$ and $\mat{Y}'$, with the decomposition giving a unique $\Re(\phi)$ and $\mat{X}'$. We choose the real reference metric $\mathfrak{\mat{g}}$ to be,
\be
\label{eq:ELreference}
\mathfrak{\mat{g}} = e^{\Re(\phi)} e^{\mat{X}'}
\ee
which is indeed Riemannian.
Here $\phi$ parameterizes the conformal factor, and we may say that $\mat{X}'$ is associated to the real part of the complex conformal class, and $\mat{Y}'$ to the imaginary part.  From a characteristic analysis we will find that the appropriate data in $3$ and $4$-dimensions for the flow from a fixed point to a finite flow time is to specify  $\Im(\phi)$ and $\mat{X}'$ at the final flow time -- which roughly can be thought of as fixing the imaginary conformal factor and real class. This analysis relies on choosing a suitable gauge for the metric, which involves the undetermined $\Re(\phi)$ and $\mat{Y}'$.

It is also worth noting that it is unclear how `complex' a metric is allowed to be to make good physical sense, as for example characterised by the Kontsevich--Segal--Witten (KSW) conditions~\cite{Kontsevich:2021dmb,witten2022notecomplexspacetimemetrics} that arise from QFT considerations for complex metrics. As we discuss later our condition above on the complex metric  is  weaker than that of KSW.

An important point is that a Lefschetz flow is not unique, and always requires a choice of metric on configuration space. The thimble that is constructed by such a flow depends on this choice, although the value of $e^{- \frac{1}{\hbar} S}$ integrated over it will not, at least for small deformations of that metric. The Einstein--Lefschetz flow is a very natural choice for the Lefschetz flow for gravity, but it is not the only one. The purpose of this work is to explore the properties of this canonical choice.

As we will see, depending on the stability of the saddle point, there may be a finite number of perturbations of the saddle point that grow but nonetheless are not captured by fixing the real class and imaginary part of the  conformal factor as final data, and they must be added to the data. Correspondingly there may also be a finite number of perturbations that decay in flow time,  and these  give a finite number of constraints on the final data.

Linearizing about a real saddle point we recover the GHP prescription -- the contour for the conformal class given by the transverse traceless perturbations remains real, whereas that for the conformal factor perturbations is rotated to be imaginary. However in detail, depending on the stability of the saddle, there may be a finite number of modes which must be treated differently. Interestingly, we will see that the prescription arising from the linearization of our Einstein--Lefschetz flow appears not to be the same as the perturbative prescription for Wick rotation of Marolf and Santos~\cite{Marolf_2022}. 

The key point of this work is that our construction is fully non-perturbative, and  allows us to understand how to continue the thimble into the non-linear regime, beyond that where GHP applies, and indeed beyond where the KSW condition holds.
To demonstrate that this construction can be solved and provides explicit results we numerically construct paths in the integration cycles for two cases; the large Euclidean Schwarzschild black hole and  Euclidean de Sitter. In both cases we show how non-perturbatively the conformal factor  forces the integration cycle to become complex.

A plan for this paper is as follows. In Section~\ref{sec:scalar} we briefly review Picard--Lefschetz theory and discuss the Lefschetz complex gradient flow  for  Euclidean real scalar $\phi^4$ field theory showing that this yields a `forward-backward' flow. We demonstrate that this may be numerically solved by prescribing the appropriate data at the two ends of the flow and then solving the entire flow all at once.
In Section~\ref{sec:Einstein} we then turn to Euclidean gravity and our warm-up exercise of finding a gradient ascent flow for the Einstein--Hilbert action for real metrics. We explore the Einstein flow, arguing that the conformal class and factor flow in opposite directions, and  demonstrate this may be explicitly numerically solved in the examples of metrics on the 4-sphere, so deformations of Euclidean de Sitter. We also construct the gradient ascent from the large to the small Euclidean black hole in a cavity.

In Section~\ref{sec:Lefschetz}  we outline how the Einstein--Lefschetz flow gives a path such that the Euclidean action has monotonically increasing real part and constant imaginary part, so that it may be used to construct Lefschetz thimbles.
We discuss the character of this  flow, describing how in a neighbourhood of real metrics, the real conformal factor and class flow in the same sense as for the real Einstein flow, and their imaginary counterparts flow in the opposite sense.
We show how to take final data for a flow that starts at a  saddle point, giving a path within the Lefschetz thimble of that saddle. We compute explicit numerical examples of such paths for the 4-sphere saddle of Euclidean de Sitter and for the large black hole saddle for Euclidean gravity in a cavity. We conclude with some outstanding questions in Section~\ref{sec:Outlook}.

\section{Forward-backward flows: Lefschetz flows for scalar $\phi^4$ theory}\label{sec:scalar}

We begin our discussion with a very brief review of Picard--Lefschetz theory~\cite{Pham:1983edi,ArnoldGuseinZadeVarchenko1988}.  This has received much attention in the theoretical physics literature in recent years. One avenue of exploration is the pragmatic one of tackling the sign problem that plagues certain lattice field theory computations, and in particular those for finite density thermal QCD -- see for example ~\cite{Cristoforetti:2012su,Alexandru:2020wrj}. Another direction is that of the current work, to define a path integral by deforming the integration contour of fields~\cite{Guralnik:2007rx,Witten:2010cx,witten2010newlookpathintegral,Tanizaki:2014tua}.
In the context of gravity Picard--Lefschetz theory has been employed to consider which saddle points are relevant in a quantum cosmological setting~\cite{Feldbrugge_2017,Feldbrugge_2018,DiazDorronsoro:2017hti} and a thermodynamic context~\cite{Marolf:2022ybi,Kolanowski:2026gii}. An excellent review of Picard--Lefschetz theory may be found in~\cite{Witten:2010cx}.

\subsection{A brief review of Picard--Lefschetz theory}

The realm of Picard--Lefschetz theory is naturally that of finite dimensional integrals. Given an integral with a real action $S(x)$ over a finite set of real variables $x^i$, $i = 1,\ldots,N$,
\be
I = \int \dd^N x \; e^{- \frac{1}{\hbar} S(x)}
\ee
we are interested in the $\hbar \to 0$ limit. 
If we can analytically continue this into the complex $z^i$ space $\mathbb{C}^N$ so that $S(z)$ is a holomorphic function, then we may attempt to  use complex analysis to evaluate the integral using a saddle point approximation. 
Let us suppose that this is the case, and furthermore that there is a single saddle point of this action at $z = z_\star$, where $\partial S / \partial z = 0$.

The basic idea of Picard--Lefschetz is to construct a new integration cycle associated to the saddle point that is complex, but still of real dimension $N$, that we may be able to deform our original real contour to.
This complex cycle should contain the point $z_\star$ and have the properties that the integrand $e^{- S(z)/\hbar}$ has constant phase everywhere on it and its magnitude is maximized at the saddle, and decreases as one moves away from it. Furthermore we want $e^{- S(z)/\hbar} \to 0$ at complex infinity suitably fast so that the integrand has no contribution there, and the integral over the complex cycle converges.
In the limit  $\hbar \to 0$ the integral along this cycle is then simply computed using the saddle point approximation associated to that complex saddle.

The cycle is called the Lefschetz thimble of that saddle point, and we denote it $\Gamma$, and its integral $I_\Gamma = \int_{\Gamma} e^{- \frac{1}{\hbar} S(z)}$.
The key idea is then that if we can deform our original real integration cycle to this Lefschetz thimble, possibly adding `arcs' at infinity in regions where the integral has no contribution, then $I$ is simply given by $I_\Gamma$. 

If there are multiple (complex) saddle points $A = 1, \ldots, m$, located at positions $z_{(A)}^i$, we denote the thimble associated to each as $\Gamma_A$. The integral over each of these $N$ real dimensional thimbles $\Gamma_A$ we denote,
\be
I_{\Gamma_A} = \int_{\Gamma_A}  e^{- \frac{1}{\hbar} S(z)} \; .
\ee
Picard--Lefschetz theory then explains that given reasonable behaviour of the action, the original integral with contour taken over the real variables can be deformed into a sum over the thimbles,
\be
\label{eq:thimbledecomp}
 I = \sum_A c_A   I_{\Gamma_A}
\ee
for integer coefficients $c_A$.\footnote{These coefficients may be computed by considering the reversed flow, $\frac{\dd z^i}{\dd\lambda} = - (\mathcal{G}^{-1})^{i\bar{j}}(z,\bar{z}) \frac{\partial \bar{S}}{\partial \bar{z}^{\bar{j}}}$, from the action saddle points, which generate `dual thimbles'. The coefficients are then determined by considering the intersection of these dual thimbles with the original  integration cycle.}

A key component of Picard--Lefschetz theory is an algorithm for constructing the thimble associated to a saddle point. This is given by a complex gradient ascent flow of the analytically continued action, $S(z)$. We call this the `Lefschetz flow' and it is given as,
\be
\label{eq:LF}
\frac{\dd z^i}{\dd\lambda} = (\mathcal{G}^{-1})^{i\bar{j}}(z,\bar{z}) \frac{\partial \bar{S}}{\partial \bar{z}^{\bar{j}}} \; .
\ee
Then $\frac{dS}{d\lambda} = \frac{\partial S}{\partial z^i} \frac{d z^i}{d \lambda}$ since $S$ is holomorphic so $\frac{\partial S}{\partial \bar{z}^{\bar{i}}} = 0$, and the action evolves in flow time $\lambda$ as
\be
\frac{\dd S}{\dd\lambda} =  (\mathcal{G}^{-1})^{i\bar{j}}(z,\bar{z}) \frac{\partial {S}}{\partial {z}^i} \frac{\partial \bar{S}}{\partial \bar{z}^{\bar{j}}},
\ee
and hence if $\mathcal{G}_{{i}\bar{j}}$ is a positive Hermitian metric, and $(\mathcal{G}^{-1})^{i\bar{j}}$ is its inverse, the righthand side above is real and positive, so the real part of the action monotonically increases along the flow, and the imaginary part remains constant. Hence the phase of $e^{- S(z)/\hbar}$ is constant along the flow, and its magnitude monotonically decreases with flow time.

A (non-degenerate) saddle point, $\partial S / \partial z^i = 0$ is a fixed point of the flow. For each complex variable the fixed point is a saddle, with some direction in the complex plane being unstable, and the orthogonal direction being stable.\footnote{
Consider a non-degenerate fixed point of the flow at $z = z_\star$. We write $\delta z^i =  O^{ij} ( z^j - z_\star ^j)$ for an appropriate complex coordinate transformation $O \in GL(N,\mathbb{C})$ that makes the Hermitian metric the Euclidean one at $z_\star$ so, $O \mathcal{G}^{-1} {O}^{\dagger} = 1$ there. Then, $S \simeq S_0 + \frac{1}{2} H^{ij} \delta z^i \delta z^j$ for a symmetric (complex) matrix $H^{ij}$ and the flow near the saddle is, $ \frac{d \delta z^i}{d\lambda} \simeq \overline{H^{ij} \delta {z}^j}$.
 We may write $H = U \Sigma U^T$, with unitary $U$ and $\Sigma$ being diagonal with positive entries on the diagonal $\sigma_i > 0$ (Takagi factorization). Defining $\delta z' = U^{T} \delta z$, and decomposing into real and imaginary parts, $\delta z'^i =  u^i + i  v^i$, then,
 \be
  \frac{d  u^i}{d\lambda} \simeq \sigma_i  u^i \; , \quad  \frac{d  v^i}{d\lambda} \simeq - \sigma_i  v^i
 \ee
 so the saddle is an unstable source for the flow in the $N$ $u^i$ directions, and a stable attractor in the $N$ $v^i$ directions.
}
Hence, given a saddle point in this $2N$-real dimensional space, there are $N$ orthogonal vectors corresponding to unstable flow directions. Perturbing the fixed point in an  unstable direction generates a flow in the complexified field space $\mathbb{C}^N$ originating from the fixed point as $\lambda \to - \infty$. The collection of all such flows defines the $N$ real dimensional Lefschetz thimble of the saddle point provided that the flows asymptote to infinity, $| z | \to \infty$, and the action is such that $e^{- S(z)/\hbar} \to 0$ suitably quickly so that the integral does not have a contribution there.\footnote{Generically only the single saddle point that generates a thimble will lie within the thimble. In cases where a symmetry implies multiple saddles lie in a thimble then one perturbs the action to break this symmetry~\cite{Witten:2010cx}.} 

Gradient flows such as~\eqref{eq:LF} require an inverse metric to convert the gradient covector, $\partial \bar{S}/\partial \bar{z}^{\bar{i}}$, into $dz^i/d\lambda$ which is a tangent vector on the configuration space.
An important point is that the individual flows will depend on the choice of this Hermitian metric $\mathcal{G}_{{i}\bar{j}}$. 
Hence the collection of all flows from a saddle point which generates the geometric embedding of the thimble of that saddle also depends on that choice. However, while the geometry of the thimble will deform under small changes in the metric $\mathcal{G}_{{i}\bar{j}}$, since the action is holomorphic, then the value of the integral of $e^{- \frac{1}{\hbar} S}$ will be invariant under a smooth deformation of the integration contour (unless the contour is deformed to cross a singularity or the behaviour at infinity is changed). Likewise the thimble decomposition in equation~\eqref{eq:thimbledecomp} will be unchanged too.

\subsection{Euclidean scalar $\phi^4$ theory}\label{sec:phi4}

Now let us consider the more subtle case of a field theory, so the integral in question is now infinite dimensional.
Suppose we have a field theory on a Riemannian manifold $\mathcal{M}$ defined by the real scalar field $\phi(x)$ and Euclidean action $S[\phi] =\int \dd^n x\; \sqrt{g} (  \frac{1}{2} (\partial \phi)^2 + V(\phi) )$ so that the equation of motion is $\nabla^2 \phi = V'(\phi)$. We note that if the manifold has boundaries, then boundary conditions and boundary terms must be added to ensure that $\delta S = \int \dd^n x\; \sqrt{g} ( - \nabla^2 \phi + V'(\phi) ) \delta \phi$ with boundary terms vanishing.
In order to evaluate the Euclidean path integral,
\be
I' = \int D \phi \; e^{- \frac{1}{\hbar} S[\phi] },
\ee
in the semiclassical limit $\hbar \to 0$, we analytically continue $\phi \mapsto \psi \in C^{\infty}(\mathcal{M},\mathbb{C})$, so complex fields over the Riemannian manifold $\mathcal{M}$, and correspondingly the action continues $S[\phi] \to S[\psi]$ to a holomorphic functional of $\psi(x)$. Then we define a complex gradient flow,
\be
\label{eq:scalarflow}
\frac{\partial }{\partial \lambda} \psi(x) & = & \frac{\mathcal{G}^{-1}(x, \psi, \bar{\psi}) }{\sqrt{g}} \frac{\delta \bar{S}}{\delta \bar{\psi}(x)} =  \mathcal{G}^{-1}(x, \psi, \bar{\psi}) \left( - \nabla^2 \bar{\psi} + V'(\bar{\psi}) \right)
\ee
where the local metric on field space $\mathcal{G}(x, \psi, \bar{\psi}) > 0$ is real and positive definite. The action evolves along the flow as,
\be
\frac{\dd S}{\dd\lambda} = \int \dd^n x\; \sqrt{g} ( - \nabla^2 \psi + V'(\psi) ) \frac{ \partial \psi}{\partial \lambda} =  \int \dd^n x \;\sqrt{g} \,  \mathcal{G}^{-1}(x, \psi, \bar{\psi}) \left| - \nabla^2 \psi + V'(\psi) \right|^2 \ge 0 \; .
\ee
Since the righthand side is real and positive, this flow is therefore a gradient ascent of the real part of the action, and leaves the imaginary part constant along the flow.
Thus as for the finite dimensional case, we could imagine using this flow to construct the thimbles associated to saddle points, $\delta S / \delta \psi(x) = 0$, by
perturbing the field away from the saddle and then flowing in the unstable directions to generate paths of complex fields that would lie in the thimble. 

However this na\"ive prescription fails immediately when we consider the well-posedness of the flow. Let us decompose the complex scalar into real and imaginary parts, $\psi(x) = u(x) + i v(x)$.
Then the flow~\eqref{eq:scalarflow} becomes
\be
\frac{\partial }{\partial \lambda}
\left(
 \begin{array}{c}
 u(x) \\
 v(x) 
 \end{array}
 \right)
& = & \mathcal{G}^{-1}(x, u,v) \left(
 \begin{array}{c}
- \nabla^2 u   \\
+ \nabla^2 v 
 \end{array}
 \right) + \mathrm{potential\;terms}
\ee
and hence we see that, while $v(x)$ obeys a parabolic flow equation forward in flow time, $u(x)$ does not, instead `anti-diffusing'. Unlike the finite dimensional case, where we may perturb a fixed point and the resulting flows generate the thimble, for this example of a scalar field we cannot flow to positive $\lambda$ away from a perturbed fixed point. Likewise, we cannot flow in the negative $\lambda$ sense, since the flow for $v$ is not parabolic and renders it ill-posed.

While this means we cannot think of~\eqref{eq:scalarflow} as a parabolic flow, we may still hope to solve it. Indeed, an illustrative example presents itself in the simple setting of the scalar theory on flat Euclidean space with a quadratic potential. 
Let us choose the simplest field space metric, $\mathcal{G}(x,u,v) = 1$, and  take $V(\psi) = \frac{1}{2} m^2 \psi^2$ with $m^2 > 0$. The resulting flow equations are then linear giving,
\be
\frac{\partial }{\partial \lambda}
\left(
 \begin{array}{c}
 u(x) \\
 v(x) 
 \end{array}
 \right)
& = & \left( - \partial^2 + m^2  \right) 
 \left(
 \begin{array}{c}
 u(x) \\
 - v(x) 
 \end{array}
 \right)
\ee
and so this decomposes into two decoupled parabolic flows: one for $v$ flowing forwards, and one for $u$ flowing backwards. 
Consider solving the flow from an initial time $\lambda_0$ to a final time $\lambda_1 > \lambda_0$.
Let the initial value of $u$ at $\lambda_0$ be $u_0(x)$, and the final value at $\lambda_1$ be $u_1(x)$, and likewise for $v$, so $v_0(x)$ and $v_1(x)$. Further let their Fourier transforms be $u_0(x) = \mathrm{Re} \int \dd^n k \;u_0(k) \,e^{- i k \cdot x}$ and similarly for $u_1(k)$ and $v_{0,1}(k)$.
We may write the general solution in terms of these data as,
\be
u(\lambda,x) &=& \mathrm{Re} \int \dd^n k \, u_1(k) \, e^{- i k \cdot x} \, e^{ - ( | k |^2 + m^2 ) (\lambda_1 - \lambda)} \nl
v(\lambda,x) &=& \mathrm{Re} \int \dd^n k \, v_0(k) \, e^{- i k \cdot x} \, e^{ - ( | k |^2 + m^2 ) ( \lambda - \lambda_0)} \; .
\ee
We expect that prescribing the initial value for $v$ and the final value for $u$, so $u_1(k)$ and $v_0(k)$, gives a well-posed PDE since these control the solution as,
\be
| u(\lambda,x ) | \le \int \dd^n k \, | u_1(k) | \; , \quad | v(\lambda,x ) | \le  \int \dd^n k \, | v_0(k) |
\ee
however long the flow time period is. 
While one could specify the initial value for $u$ too, instead of final data, now writing the solution as,
\be
u(\lambda,x) &=& \mathrm{Re} \int \dd^n k \, u_0(k) \, e^{- i k \cdot x} \, e^{ + ( | k |^2 + m^2 ) ( \lambda - \lambda_0)} 
\ee
this would not control the amplitude of the solution as for $\lambda > \lambda_0$ then the above integral may diverge for large $| k |$ even if $u_0(k)$ is bounded, signalling ill-posedness in the Hadamard sense.

A point we note for future reference is that we expect to have a well posed flow if we specify a linear combination of the $u$ and $v$ as initial and final data, specifically giving initial data $A_0(k)$ and final data   $A_1(k)$ where,
\be
A_0(k) = a(k) u_0(k) + b(k) v_0(k) \; , \quad A_1(k) = c(k) u_1(k) + d(k) v_1(k) \; .
\ee
Then since,
\be
u_0(k) = u_1(k) e^{ - ( | k |^2 + m^2 ) (\lambda_1 - \lambda_0)} \; , \quad v_1(k) &=& v_0(k) e^{ - ( | k |^2 + m^2 ) ( \lambda_1 - \lambda_0)} 
\ee
the data $A_{0,1}(k)$ determines $u_1(k)$ and $v_0(k)$ and hence the solution provided that,
\be
b(k) c(k) - a(k) d(k) e^{ - 2 ( | k |^2 + m^2 ) (\lambda_1 - \lambda_0)} \ne 0 \; .
\ee
In the short wavelength limit, $| k | \to \infty$, then
\be
A_0(k) \simeq b(k) v_0(k) \; , \quad A_1(k) \simeq c(k) u_1(k) 
\ee
and so specifying $A_0$ and $A_1$ is equivalent to giving $v_0$ and $u_1$ for large $k$ provided that $b(k)$ and $c(k)$ are non-vanishing in this limit, in which case we  expect the system to be well posed.

With this flow we may map out points on the thimble of the single fixed point $\psi(x) = 0$ by requiring the flow to start on the fixed point for $\lambda \to -\infty$, and then we may construct a flow up to a finite time $\lambda_1$ where we supply the data for $u_1(x)$ as above.
 We then flow $v$ forwards, obtaining simply the zero solution, and flow $u$ back, finding the solution,
\be
u(\lambda,x) &=& \mathrm{Re} \int \dd^n k \, u_1(k) \, e^{- i k \cdot x} \, e^{ - ( | k |^2 + m^2 ) ( \lambda_1 - \lambda) } 
\ee
In fact we may choose any $\lambda_1$, say taking $\lambda_1 = 0$, since a translation $\lambda_1 \to \lambda_1 + \mu$ is equivalent to a scaling $u_1(k) \to e^{ - ( | k |^2 + m^2 ) \mu}\, u_1(k)$. 
Different choices of $u_1(x)$ parametrize the different points in field space that lie on the thimble. Hence we see that the thimble generated is purely real and, since $u_1$ is free to specify, is just given by all real $u(x)$. Thus the thimble in this simple example represents a contour integration over real fields, and is indeed just the original Gaussian integration.

We term these \textit{forward-backward} flows; we should specify the imaginary part of $\psi$ (i.e. $v$) at the initial flow time since it flows forwards, and the real part (i.e. $u$) at the end of the flow as it flows backwards.
Forward-backward flows have been studied in other contexts in the mathematics literature, but not to our knowledge in the case of Lefschetz flows, or geometric flows more generally as we consider here. An example of previous studies are those of~\cite{Beals1,Beals2,Kruskal} which consider equations of the form $x \frac{\partial u}{\partial t} + A(x) u = f$, where $A$ is an elliptic second order operator depending on a coordinate $x$ and we see the diffusion constant varies linearly in space changing sign at $x=0$. 
Study of forward-backward diffusive systems also arises in control theory of diffusive PDEs and stochastic diffusion processes. These involve a forward diffusion problem, a backward problem, and a control condition that couples the two~\cite{Kolmogorov1931,FlemingRishel1975,Troltzsch2010}.

For a real fixed point, given that we have a real potential $V(\psi)$ for real $\psi$, and assuming that the fixed point is stable so the linear operator $-\nabla^2 + V''(\psi)$ is positive, then the flow starting at the fixed point will be purely real; at the fixed point we have $v = 0$, and any evolution away from it will keep $v$ fixed at zero.
More interesting examples arise if there are complex saddles of the action, but we will not pursue that here. Instead, it is instructive to consider how we might solve this flow more generally than starting at a fixed point.
\begin{figure}[t]
    \centering
    \subfigure[Purely real solution, $\Re\psi$, with $u_1 = \cos\theta$ and $v_0 = 0$.]{
        \includegraphics[width=0.47\textwidth]{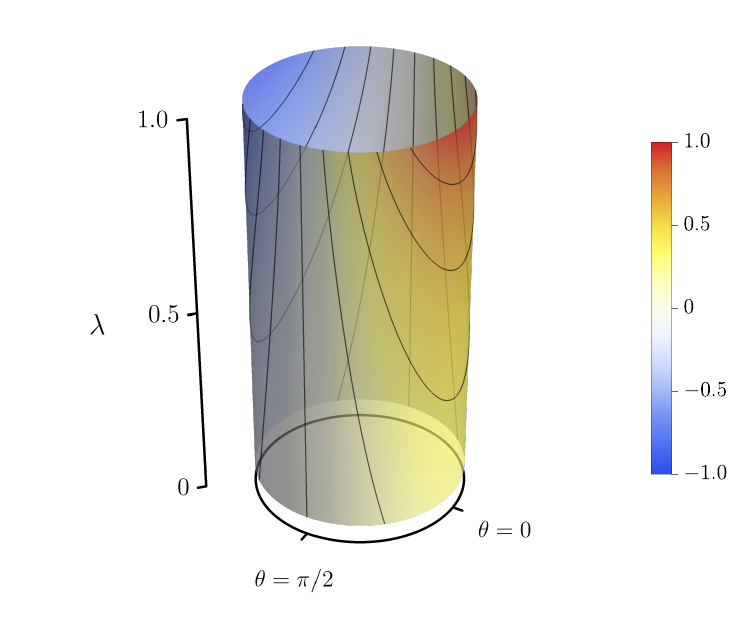}
        \label{fig:phi4-real}
    }
    \hfill
    \subfigure[Purely imaginary solution, $\Im\psi$, with $v_0 = \cos\theta$ and $u_1 = 0$.]{
        \includegraphics[width=0.47\textwidth]{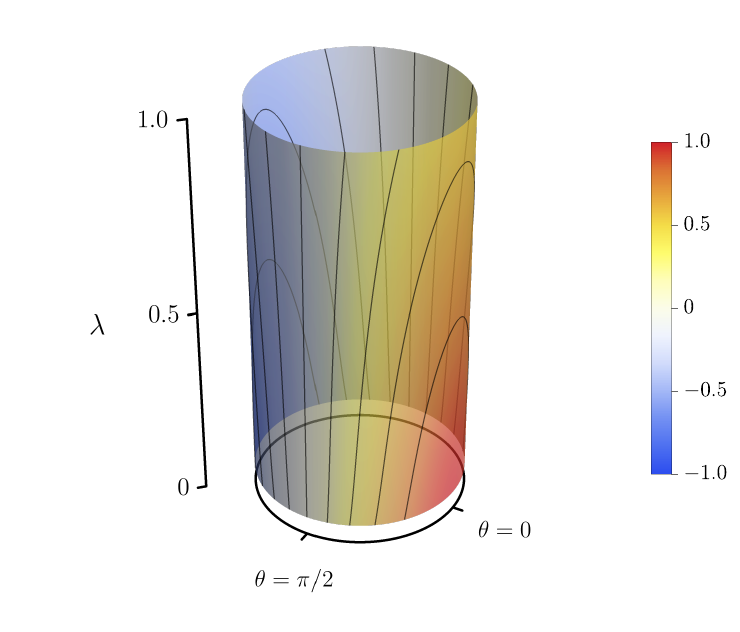}
        \label{fig:phi4-imag}
    }
    \caption{Numerical solutions of the forward-backward Lefschetz flow for $\phi^4$ theory on a circle. The purely real solution, $\Re\psi$, with $u_1 = \cos\theta$ and $v_0 = 0$ is on the left, and the purely imaginary solution, $\Im\psi$, with $v_0 = \cos\theta$ and $u_1 = 0$ is on the right.}
    \label{fig:phi4-pure}
\end{figure}

\begin{figure}[t]
    \centering
    \includegraphics[width=\textwidth]{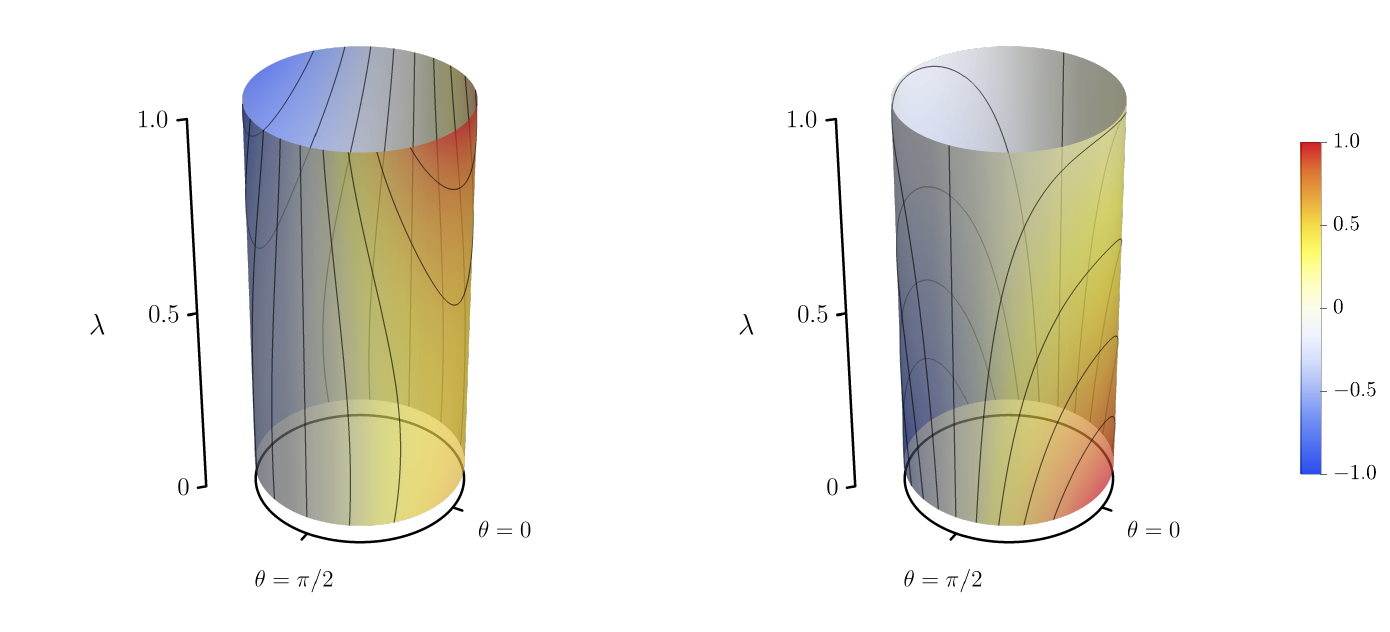}
    \caption{The real (left) and imaginary (right) parts of the mixed solution of the forward-backward Lefschetz flow for $\phi^4$ theory on a circle, with $u_1 = v_0 = \cos\theta$. The mixed solution is visibly not the sum of the purely real and purely imaginary solutions of Figure~\ref{fig:phi4-pure}, reflecting the non-linearity of the potential; in particular the $- v u^{2}$ term in the $\Im \psi$ equation makes the imaginary part decay faster.}
    \label{fig:phi4-mixed}
\end{figure}

To be concrete let us take a 1-dimensional scalar field on a circle, with coordinate $\theta$, and potential $V(\psi) = \frac{\mu}{4!} \psi^4$. The flow equations, decomposed into real and imaginary parts, are then
\be
\frac{\partial }{\partial \lambda}
\left(
 \begin{array}{c}
 u(\theta) \\
 v(\theta) 
 \end{array}
 \right)
& = &  - \partial^2_\theta 
 \left(
 \begin{array}{c}
 u(\theta) \\
 - v(\theta) 
 \end{array}
 \right)
 + \frac{\mu}{6}
 \left(
 \begin{array}{c}
 u(\theta)^3 - 3 u(\theta) v(\theta)^2 \\
 v(\theta)^3 - 3 v(\theta) u(\theta)^2
 \end{array}
 \right) \; .
\ee
Considering a flow from flow-time $\lambda_0$ to $\lambda_1 > \lambda_0$, we then define the forward-backward initial data as
\be
v_0(\theta) = \mathrm{Im}\left( \psi(\lambda_0,\theta) \right) \; , \quad u_1(\theta) = \mathrm{Re}\left( \psi(\lambda_1,\theta) \right).
\ee
We cannot solve this using the usual evolution method, and instead must solve it on the two-dimensional domain $(\lambda,\theta ) \in [ \lambda_0, \lambda_1 ] \times  [0, 2\pi] $, prescribing the data above.
We can do this numerically by discretising the flow equation on the domain, imposing the boundary conditions for $v_0$ and $u_1$, and then we employ Newton's method to find solutions to the resulting coupled system of equations.
Throughout this work we use pseudospectral differencing, employing a Chebyshev grid in the flow time and the non-trivial coordinate direction, which here is the angle $\theta$.

In Figures~\ref{fig:phi4-pure} and~\ref{fig:phi4-mixed} we show example flows where we take  $\lambda \in [0, 1]$ and have chosen the coupling $\mu = 5$. 
In Figure~\ref{fig:phi4-pure} we show a flow with $u_1 = \cos\theta$ and $v_0 = 0$, so the solution is purely real, and also a flow with $v_0 = \cos\theta$ and $u_1 = 0$ which yields a purely imaginary solution. Figure~\ref{fig:phi4-mixed}  shows the example of a mixed solution with $v_0 = u_1 = \cos{\theta}$. We can visually see the non-linearity due to the potential, as the mixed flow is not the sum of the first two. Furthermore, we observe that the real part relaxes backwards in flow time from its data at $\lambda = 1$, and the imaginary part forwards from its data at $\lambda = 0$.

Since the data we consider for these flows are even under the reflection $\theta \to -\theta$, so is the solution as we see in the figures. In practice this allows us to reduce the angular domain to $\theta \in [0, \pi]$, imposing Neumann conditions at $\theta = 0$ and $\theta = \pi$. The flows shown were computed on an $n_\lambda \times n_\theta = 60 \times 50$ Chebyshev grid, which yields solutions with good accuracy. We give some details of numerical convergence in Appendix~\ref{app:convergence}.

\section{Gradient ascent for Euclidean gravity and real metrics}\label{sec:Einstein}

We now turn to writing down a gradient ascent flow of the action for Euclidean gravity for  Riemannian metrics in $n\ge 3$ dimensions. Our motivation is that later we use this as the starting point to construct a flow in the space of complex metrics that is a Lefschetz flow which will construct a Lefschetz thimble associated to a saddle point. However in this section we study the simpler problem of understanding such a flow in the context of real metrics.

The canonical flow of real metrics is the famous Ricci flow. While this is a gradient flow of the action, it is not a gradient ascent/descent flow since the superspace metric (a DeWitt metric) governing the flow is not positive definite. Nonetheless it is useful to briefly review Ricci flow, which can be phrased as a parabolic flow and hence solved by integrating initial data, and we review this in the next Section~\ref{sec:ricci}. Following that we introduce the Einstein flow. While structurally this is a geometric flow that looks similar to the Ricci flow we shall see its features are quite different, and in particular while it does provide a gradient ascent of the action, it is a forward-backward flow like the Lefschetz scalar flow discussed earlier rather than being parabolic. Nonetheless we may still solve this following the strategy above of solving the entire flow at once. There is an elegant picture for the data for this flow; the conformal factor is specified at the initial flow time and the conformal class is specified at the final time. Having written the Einstein flow in Section~\ref{sec:EinsteinFlow} we then give explicit numerical examples where we solve it in the context of deformations of Euclidean de Sitter (Section~\ref{sec:EinsteinS4}) and for black holes in a cavity (Section~\ref{sec:bhflow}).

\subsection{A brief review of Ricci flow} \label{sec:ricci}

For an $n$-dimensional Riemannian manifold the Ricci flow~\cite{PhysRevLett.45.1057,hamilton1982three,perelman2002entropyformularicciflow} is defined by,
\be
\frac{\partial}{\partial \gamma} g_{\mu\nu} = - 2 R_{\mu\nu}
\ee
where $\gamma$ is the flow time, and for simplicity we  take the manifold to be compact.
The Ricci flow is a gradient flow of the Einstein--Hilbert action (taking units so $8 \pi G = 1$)~\cite{Headrick_2006,Tseytlin:2006ak},
\be
S = - \frac{1}{2} \int \dd^n x \,\sqrt{g}\, R
\ee
with respect to the local DeWitt metric;
\be
\label{eq:dewitt}
\frac{\partial}{\partial \gamma} g_{\mu\nu}(x) = - \mathcal{G}_{\mu\nu\alpha\beta}^{-1}(x) \frac{\delta S}{\delta g_{\alpha\beta}(x)} \; , \quad \mathcal{G}^{\mu\nu\alpha\beta}(x) = \frac{1}{4} \sqrt{g} \left( g^{\mu(\alpha} g^{\beta)\nu} - \frac{1}{2} g^{\mu\nu} g^{\alpha\beta} \right) \, .
\ee
Crucially, however, Ricci flow is not a gradient \textit{descent}, since the DeWitt metric above is not positive definite; the conformal perturbation of the metric has a negative norm with respect to it. 
Defining the inner product,
\be
\left( h_{\mu\nu} , h'_{\alpha\beta} \right) = \int \dd^n x \, \mathcal{G}^{\mu\nu\alpha\beta} \, h_{\mu\nu}\, h'_{\alpha\beta} 
\ee
we explicitly see that while a transverse traceless tensor has positive norm, and further is orthogonal to a conformal perturbation of the metric, the latter has negative norm; $(\phi g_{\mu\nu}, \phi g_{\alpha\beta} )= - \frac{n(n-2)}{8} \int \dd^n x \sqrt{g} \phi^2$.
The change in the action is given by,
\be
\frac{\dd S}{\dd \gamma}   = - \int \dd^n x \sqrt{g} \left( \hat{R}^{\mu\nu} \hat{R}_{\mu\nu} - \frac{n-2}{2n} R^2 \right),
\ee
where $\hat{R}_{\mu\nu} = R_{\mu\nu} - \frac{1}{n} g_{\mu\nu} R$ is the traceless part of the Ricci tensor. We see that for dimensions of interest $n \ge 3$, this traceless part contributes oppositely in sign to the trace part, the Ricci scalar, for the change of the action in flow time.

The flow is geometric in the sense that if one performs a flow-time dependent diffeomorphism on a solution, it remains a solution. Adding the infinitesimal diffeomorphism
$\delta g_{\mu\nu}(\gamma) = \delta \gamma \, \nabla_{(\mu} \xi_{\nu)}$ with the covector field $\xi_\mu(\gamma)$ being a function of flow time,
then the flow becomes
\be
\frac{\partial}{\partial \gamma} g_{\mu\nu} = - 2 R_{\mu\nu} + \nabla_{(\mu} \xi_{\nu)} \; .
\ee
This is geometrically equivalent to the original flow, differing only by a flow dependent diffeomorphism.
For an appropriate choice of $\xi_\mu$, this flow becomes parabolic. Choosing some smooth connection $\tilde{\Gamma}$ on the manifold, one takes $\xi^\mu = 2 g^{\alpha\beta} \left( \Gamma^\mu_{~\alpha\beta} - \tilde{\Gamma}^\mu_{~\alpha\beta} \right)$ which, being a difference of connections, is a globally defined vector field. This is the Ricci--DeTurck flow~\cite{DeTurck1983} whose principal part is, 
\be
\frac{\partial}{\partial \gamma} g_{\mu\nu} = g^{\alpha\beta} \partial_\alpha \partial_\beta g_{\mu\nu} + \text{lower order terms},
\ee
and hence for Riemannian $g_{\mu\nu}$ this is parabolic --- locally a diffusion in each metric component.

We may modify the flow to include a cosmological constant in the obvious manner; $\frac{\partial}{\partial \gamma} g_{\mu\nu} = - 2 \left( R_{\mu\nu} - \Lambda g_{\mu\nu} \right) + \nabla_{(\mu} \xi_{\nu)}$. 
The fixed points of this flow include solutions of the Einstein equation, $R_{\mu\nu} = \Lambda g_{\mu\nu}$ with $\xi_\mu = 0$, which are the saddle points of interest in the path integral.
However Ricci flow may also admit Ricci solitons -- these are fixed points of the flow that continuously evolve by a diffeomorphism and are not Einstein, solving instead $R_{\mu\nu} = \Lambda g_{\mu\nu} + \frac{1}{2} \nabla_{(\mu} \xi_{\nu)}$. For example the 4-dimensional Koiso–Cao solution is a Ricci soliton with positive cosmological term~\cite{Koiso1990,Cao1996}.

The character of Ricci flow is to flow the metric towards fixed points where they are stable, so have a positive spectrum of the Lichnerowicz operator (including cosmological contribution). In this sense the flow is naturally in the `opposite' sense to a Lefschetz flow, namely flowing towards rather than away from fixed points.
In the case that the fixed point is unstable, with a finite number of negative modes, then the metric data to flow towards a fixed point must be tuned in order to reach it. Conversely these finitely many unstable modes generate flows away from the fixed point.
A nice gravitational example is large Euclidean black holes with cavity boundary conditions which are stable fixed points of Ricci flow~\cite{Headrick_2006}. Starting with an initial geometry which is a deformation of a large black hole, then provided the amplitude of the deformation is not too large, the Ricci flow asymptotically returns to the large black hole.
On the other hand, the small black hole in a cavity has a single unstable mode~\cite{Gregory:2001bd}. One can flow to it if one suitably tunes the initial data~\cite{Headrick:2009pv,Wiseman:2011by}, and the negative mode generates a flow away from the small black hole in two directions that asymptotes either to the stable large black hole, or to hot flat space after a finite flow time singularity that must be resolved~\cite{Headrick_2006}.

Famously Ricci flow also exhibits a non-linear behaviour where positive curvature regions are driven to singularities in finite flow time --- for example a sphere has a linear decrease in the square of its radius in flow time. A consequence of this is that when a positive cosmological term is introduced, the round sphere --- so Euclidean--de Sitter --- which is the fixed point of that flow is unstable to the homogeneous mode that changes its radius --- it either grows forever, or shrinks to a singularity. 

\subsection{Towards a gradient ascent for real metrics} \label{sec:ascent}

We now wish to find a gradient ascent of the Einstein--Hilbert action for real Riemannian metrics as a warm-up for a Lefschetz flow of complex metrics. 
We would like this to flow in the opposite sense to Ricci flow, so away from fixed points rather than towards them.
A natural choice is to modify the Ricci flow by its trace and reverse the sign of the flow,
\be
\frac{\partial}{\partial\lambda} g_{\mu\nu} = 2 R_{\mu\nu} - \alpha  g_{\mu\nu} R  + \nabla_{(\mu} \xi_{\nu)} \; ,
\ee
for some constant $\alpha$. 
We again include a diffeomorphism term along the flow.
Up to the reversed sign, this is the Ricci--Bourguignon flow~\cite{aubin1970metriques,Bourguignon1981Ricci,catino2016riccibourguignonflow}. To get a better idea of how this new flow differs from the Ricci flow, we will examine its principal symbol.
To do this we first linearize the perturbation $\delta g_{\mu\nu}$ about the background metric ${g}_{\mu\nu}$. We also take a vanishing background vector $\xi_\mu$.
To study the character of the flow we focus on the short distance behaviour of the perturbation, and hence we `freeze the coefficients' in the linearization, so take the background metric to be constant.
We then look at the highest derivative terms for the linearization which dominate at short distance;
\be
\frac{\partial}{\partial\lambda} \delta g_{\mu\nu} - {\partial}_{(\mu} \delta \xi_{\nu)} + \ldots = P[  \delta g ]_{\mu\nu} =  P_{\mu\nu}^{\alpha\beta\rho\sigma}( {g} ) \partial_\alpha \partial_\beta \delta g_{\rho\sigma} + \ldots
 \ee
where $\ldots$ are lower derivative terms which we ignore. Working in Fourier space on this constant background then $\partial_\mu \to i k_\mu$, and the principal part on the righthand side above becomes a function of the wavevector $k_\mu$ and is given explicitly by;
\be
P[ \delta g ]_{\mu\nu} &= & {g}^{\alpha\beta} k_\alpha k_\beta \delta g_{\mu\nu} + {g}^{\alpha\beta} k_\mu k_\nu \delta g_{\alpha\beta} - 2 {g}^{\alpha\beta} k_\alpha k_{(\mu} \delta g_{\nu)\beta} \nl
&& \qquad
- \alpha {g}_{\mu\nu} \left(  
{g}^{\alpha\beta} {g}^{\rho\sigma}  k_\alpha k_\beta \delta g_{\rho\sigma} - {g}^{\alpha\rho} {g}^{\beta\sigma}  k_\alpha k_\beta \delta g_{\rho\sigma} 
 \right) \; . 
 \ee
 We then look for  $\delta g_{\mu\nu}$ which are eigenvectors of this up to a vector term parameterized by $u_\mu$, with eigenvalue $\omega$ satisfying,
 \be
 P[ \delta g ]_{\mu\nu} = \omega \, \delta g_{\mu\nu} + k_{(\mu} u_{\nu)} \; .
 \ee
Then this yields $\frac{\partial}{\partial\lambda} \delta g_{\mu\nu} =  \omega \, \delta g_{\mu\nu} + \ldots$ with the vector part being absorbed by the diffeomorphism $\delta \xi_\mu$.
One  finds the eigenvectors (up to diffeomorphism) are given by a scalar mode $\delta g^{(S)}_{\mu\nu}$,
\be
\omega_{(S)} = (1 - (n-1) \alpha) k^2 \; , \qquad 
\delta g^{(S)}_{\mu\nu}  = \phi \, {g}_{\mu\nu}  \; , \quad u^{(S)}_\mu = (n-2) \phi \, k_\mu
\ee
and transverse-traceless modes $\delta g^{(TT)}_{\mu\nu}$ with,
\be
\omega_{(TT)} = k^2 \; , \quad k^\mu \delta g^{(TT)}_{\mu\nu}  = {g}^{\mu\nu} \delta g^{(TT)}_{\mu\nu} = 0
\ee
and finally the vector modes which have vanishing eigenvalue,
\be
\omega_{(V)} = 0 \; , \quad  \delta g^{(V)}_{\mu\nu} = k_{(\mu} v_{\nu)}
\ee
 and  arise due to the diffeomorphism invariance of the flow.
Then given that ${g}_{\mu\nu}$ is a Riemannian metric, the transverse-traceless perturbation {\it on its own} has a parabolic flow backwards in flow time so we would specify final data rather than initial data. 
However the mode $\delta g^{(S)}$ differs from Ricci flow. We see the flow is parabolic backwards in flow time if $\alpha < 1/(n-1)$, and note that $\alpha = 0$ is the value corresponding to  Ricci flow. Then one could specify a final metric and flow the metric backwards in time \cite{catino2016riccibourguignonflow}. 
However for  $\alpha > 1/(n-1)$ the conformal mode itself obeys a parabolic flow equation forwards in flow time. Hence for $\alpha > 1/(n-1)$ the principal symbol shows that the flow cannot be parabolic in character (even ignoring vector gauge modes) since then the transverse-traceless modes propagate backwards in flow-time $\lambda$, but the conformal mode flows forwards.\footnote{For the marginal value $\alpha = 1/(n-1)$, the righthand side of the flow equation is the Schouten tensor, and the conformal mode does not flow at leading derivative order.} 

The flow can be written as a gradient flow with respect to the local DeWitt metric $\mathcal{G}^{\mu\nu\rho\sigma}(x;\alpha)$;
\be
\frac{\partial}{\partial\lambda} g_{\mu\nu}(x) = \mathcal{G}_{\mu\nu\rho\sigma}^{-1}(x; \alpha) \frac{\delta S}{\delta g_{\rho\sigma}(x)} \;,
\ee
where,
\be
\mathcal{G}^{\mu\nu\rho\sigma}(x;\alpha) = \frac{\sqrt{g} }{4} \left( g^{\mu\rho} g^{\nu\sigma} +  \frac{\alpha - 1}{2-  \alpha n}\, g^{\mu\nu} g^{\rho\sigma} \right)  , \; \mathcal{G}_{\mu\nu\rho\sigma}^{-1}(x; \alpha) = \frac{4}{\sqrt{g}}\left(g_{\mu\rho}\, g_{\nu\sigma} - \frac{1-\alpha}{n-2}\,g_{\mu\nu}\, g_{\rho\sigma}\right)  \; .\nl
\ee
Note that we have chosen a positive sign on the righthand side of the flow equations above, unlike for Ricci flow, as this will yield a gradient ascent of the action for appropriate $\alpha$ as we will now show.
Since the associated inner product on transverse traceless tensors is independent of $\alpha$ they always have positive norm. For conformal modes we find the norm given by this DeWitt metric to be,
 \be
( \phi  g_{\mu\nu} , \,\phi g_{\rho\sigma} ) =  \frac{n}{4}\frac{n-2}{\alpha n - 2} \int \dd^n x \; \sqrt{g}\, \phi^2 \; .
\ee
Hence for dimensions $n \ge 3$ we see that this metric is positive for both conformal and transverse-traceless modes for $\alpha > 2/n$. The action varies along the flow as,
\be
\frac{\dd S}{\dd \lambda} & = &  \int \dd^n x \sqrt{g} \left( \hat{R}^{\mu\nu} \hat{R}_{\mu\nu} + \frac{1}{4n}(n-2)(\alpha n-2) R^2 \right) 
\ee
and so monotonically increases provided $\alpha > 2/n$.

For dimensions $n \ge 3$ the two conditions of gradient ascent of the action, $\alpha > 2/n$, and parabolicity of the flow $\alpha < 1/(n-1)$ cannot be mutually compatible. 
For $\alpha < \frac{1}{n-1}$ we may have a (backward) parabolic flow that is not an ascent, as for Ricci flow which corresponds to vanishing $\alpha$. For the intermediate range $\frac{2}{n} > \alpha > \frac{1}{n-1}$ the flow is neither parabolic, being of forward-backward character, nor is the flow of the action monotonic. Finally for $\alpha > \frac{2}{n}$ the flow is a gradient ascent of the action, but is a forward-backward flow where
 transverse-traceless perturbations flow backwards in the flow time $\lambda$, and the conformal mode flows forwards.

\subsection{Einstein flow} \label{sec:EinsteinFlow}

Perhaps the most natural choice of flow which yields monotonicity in any dimension is $\alpha = 1$ given in equation~\eqref{eq:EinsteinFlow} of the Introduction, which we term the `Einstein flow'. 
Let us now consider this flow, including a diffeomorphism term, cosmological term and also allowing our Riemannian manifold to potentially have boundaries where appropriate boundary conditions for the flow should be imposed.
The DeWitt metric controlling the flow takes the elegant form;
\be
\label{eq:Einsteinmetric}
\mathcal{G}^{\mu\nu\rho\sigma}(x;1) = \frac{\sqrt{g} }{4}  g^{\mu\rho} g^{\nu\sigma} 
\; , \quad \mathcal{G}_{\mu\nu\rho\sigma}^{-1}(x; 1) = \frac{4}{\sqrt{g}} g_{\mu\rho}\, g_{\nu\sigma} 
\; .
\ee
Including the cosmological term, the Euclidean gravitational action is,
\be
\label{eq:EHaction}
S = - \frac{1}{2} \int \dd^n x \,\sqrt{g}\, \left( R - (n-2) \Lambda \right) + \mathrm{boundary\;terms}
\ee
where we take the boundary terms so that upon variation the action is stationary when the bulk Einstein equation $G_{\mu\nu} = - \frac{n-2}{2}  \Lambda g_{\mu\nu}$ together with the boundary conditions are imposed. Hence,
\be
\delta S = \frac{1}{2} \int \dd^n x \,\sqrt{g}\, \left( G^{\mu\nu} + \frac{n-2}{2} \Lambda g^{\mu\nu} \right) \delta g_{\mu\nu}
\ee
with no boundary contribution when $\delta g_{\mu\nu}$ is restricted to comply with the boundary conditions. 
Then the gradient ascent of this action, together with a flow dependent diffeomorphism, gives the expression in~\eqref{eq:EinsteinFlow},
\be
\frac{\partial}{\partial\lambda} g_{\mu\nu} = \mathcal{G}_{\mu\nu\rho\sigma}^{-1}(x; 1) \frac{\delta S}{\delta g_{\rho\sigma}(x)}  + \nabla_{(\mu} \xi_{\nu)}  = 2 \left( G_{\mu\nu} + \frac{n-2}{2}  \Lambda g_{\mu\nu} \right) + \nabla_{(\mu} \xi_{\nu)} 
\ee
and we require that the flow time dependent diffeomorphism, generated by the vector field $\xi^\mu$ vanishes on any boundaries.
The corresponding change in the action is,
\be
\label{eq:actionreal}
\frac{\dd S}{\dd \lambda} 
& = & \int \dd^n x \; \sqrt{g}\,  \left( G_{\mu\nu} + \frac{n-2}{2}  \Lambda g_{\mu\nu} \right)^2 \ge 0 \; , 
\ee
where to obtain this we have used the vanishing of $\xi^\mu$  on  boundaries so that integrating by parts we see the term $ \int \dd^n x \; \sqrt{g}\,  \left( G_{\mu\nu} + \frac{n-2}{2}  \Lambda g_{\mu\nu} \right) \nabla^\mu \xi^\nu$ vanishes.

At a fixed point of the flow, $\frac{\partial}{\partial\lambda} g_{\mu\nu} = 0$, the action is constant so $\dd S/\dd\lambda=0$. From equation~\eqref{eq:actionreal} this implies that the metric must satisfy the Einstein condition $R_{\mu\nu} = \Lambda g_{\mu\nu}$, and hence from the flow equation we see $\nabla_{(\mu} \xi_{\nu)} = 0$. Then either $\xi^\mu$ vanishes, or it is a Killing vector, and hence the diffeomorphism it generates vanishes. 
Thus we see the nice result that for the Einstein flow the only fixed points are Einstein metrics --- there can be no  Ricci solitons. This is in contrast to Ricci flow, where for positive cosmological term there are known Ricci solitons.\footnote{Indeed this simple absence of solitons follows directly from the flow of the action being monotonic, and so holds for any $\alpha > \frac{2}{n}$.}

Let us now assume that the diffeomorphism vector field, $\xi_\mu$, does not alter the principal part of the equation, unlike the case of the DeTurck choice. Hence it does not alter the flow characteristics of the Einstein flow.
We have seen above that for the Einstein flow the  conformal perturbations flow forwards in flow time $\lambda$ and the transverse traceless perturbations flow backwards. Since it is a geometric flow, the vector gauge perturbations do not flow.
Thus this is a forward-backward flow -- how should we think of the data for it?
We claim that provided the metric is Riemannian, then we should prescribe data such that at the initial time it constrains the  conformal perturbation, and at the final time it constrains the transverse-traceless perturbation. If the manifold has boundaries we should also prescribe appropriate data there too along the flow.
We then expect the flow equation will be well-posed, although we do not provide a rigorous proof of this, and it would clearly be interesting to do so.

Conformal perturbations correspond to perturbative Weyl scalings of the metric and the transverse-traceless perturbations correspond to perturbations of the conformal class of the metric. One possible boundary condition is that we specify the conformal factor of the metric at the initial flow-time $\lambda_0$, and the conformal class  at the final flow-time $\lambda_1$. We may decompose the metric as
\be
\label{eq:decomp}
g_{\mu\nu} = e^\Phi \hat{g}_{\mu\nu} \;,
\ee
and choose to fix $\mathrm{det}\, \hat{g}_{\mu\nu}$ to define the separation into the conformal factor $e^\Phi$ and conformal class with representative $\hat{g}_{\mu\nu}$.\footnote{For example, given a fixed reference metric $g^{(ref)}_{\mu\nu}$ we could require $\mathrm{det}\, \hat{g}_{\mu\nu} = \mathrm{det}\, {g}^{(ref)}_{\mu\nu}$.}
Then about a background ${g}_{\mu\nu}$ this gives a perturbation,
\be
\label{eq:metpert1}
\delta g_{\mu\nu} = \delta \Phi \, {g}_{\mu\nu} + \delta g^{(T)}_{\mu\nu}
\ee
where $\delta g^{(T)}_{\mu\nu} = e^\Phi \delta \hat{g}_{\mu\nu}$ is traceless with respect to the background ${g}_{\mu\nu}$ due to the determinant condition. We should compare this to a metric perturbation which we may decompose as,
\be
\label{eq:metpert2}
\delta g_{\mu\nu} =  \left( \phi + \frac{1}{n} {\nabla}^2 u \right) {g}_{\mu\nu} + a^{(TT)}_{\mu\nu} + {\nabla}_{(\mu} v^{(T)}_{\nu)} + \left( {\nabla}_{\mu} \partial_{\nu} u - \frac{1}{n} {\nabla}^2 u {g}_{\mu\nu} \right)
\ee
where we have from above that $\phi$ flows forwards, the transverse traceless $a^{(TT)}_{\mu\nu}$ (so that ${g}^{\mu\nu} a^{(TT)}_{\mu\nu} = 0$ and ${\nabla}^\mu a^{(TT)}_{\mu\nu} = 0$) flows backwards, and the diffeomorphism terms generated by the covector $v_{\mu} = v^{(T)}_\mu + \partial_\mu u$ do not flow, and we have decomposed this into a divergence free part ${\nabla} \cdot v^{(T)} = 0$ and a scalar part $u$.
This York decomposition is unique up to global issues.

Consider a flow on a manifold $\mathcal{M}$,  with initial and final flow-times $\lambda_0$ and $\lambda_1$. To solve the flow we could specify an initial conformal factor $\Phi^{(0)}(x)$ over $\mathcal{M}$, and final conformal class with representative $\hat{g}^{(1)}_{\mu\nu}(x)$, so
\be
\Phi^{(0)}(x) = \Phi(\lambda_0, x) \; , \quad \hat{g}^{(1)}_{\mu\nu}(x) = \hat{g}_{\mu\nu}(\lambda_1,x).
\ee
These boundary conditions would then ensure that $\delta g^{(T)}_{\mu\nu} = 0$ at the final flow time and $\delta \Phi = 0$ at the initial flow time  about any background. At the final flow time $\delta g^{(T)}_{\mu\nu} = 0$ then matching~\eqref{eq:metpert1} and~\eqref{eq:metpert2} we simply see (ignoring global uniqueness issues),
\be
 \delta \Phi  = \phi \; , \quad a^{(TT)}_{\mu\nu} = 0 \; , \quad v^{(T)}_{\mu} = 0 \; , \quad u = 0
\ee
and hence this indeed provides a final boundary condition for $a^{(TT)}_{\mu\nu}$ and no constraint on $\phi$, as the character of our flow requires.

Now consider the initial flow time where $\delta \Phi =0 $. This imposes,
\be
\phi + \frac{1}{n} {\nabla}^2 u  = 0 \; , \quad \delta g^{(T)}_{\mu\nu} = a^{(TT)}_{\mu\nu} + {\nabla}_{(\mu} v^{(T)}_{\nu)} + \left( {\nabla}_{\mu} \partial_{\nu} u - \frac{1}{n} {\nabla}^2 u {g}_{\mu\nu} \right) \; .
\ee
A subtlety is that the scalar part of the diffeomorphism, $u$, mixes with the conformal mode $\phi$ in the conformal factor perturbation.
Our flow requires a boundary condition for $\phi$, but we see the mixing with the scalar diffeomorphism complicates this.
Ideally we might impose a coordinate choice on $\hat{g}_{\mu\nu}$ in such a way that $u = 0$; then we simply find that $\phi = 0$ and $a^{(TT)}_{\mu\nu}$ is unconstrained, as required by the character of the flow. However in practice it may be inconvenient to impose such coordinates. Instead any coordinate choice will work provided that $u$ is fixed in terms of $a^{(TT)}_{\mu\nu}$ and not in terms of $\phi$. Schematically this will then impose a boundary condition $\phi + f( a^{(TT)} ) = 0$, where $f$ is a function involving $a^{(TT)}_{\mu\nu}$ but not $\phi$, derived from the gauge condition. This is a mixed boundary condition for the forward flowing $\phi$ and backward flowing $a^{(TT)}_{\mu\nu}$, but definitely overlaps with $\phi$ and hence will be a good boundary condition given the character of the flow. At short wavelengths perturbations of $a^{(TT)}$ decay to the past, while perturbations of $\phi$ grow, so for short wavelengths this initial condition effectively controls those $\phi$ perturbations, as required.
Hence, provided we are careful with our gauge choices, then with this boundary condition of fixing the conformal factor $\Phi$ initially, and the class representative $\hat{g}_{\mu\nu}$ finally, we can then expect to solve the flow equation over the domain $\mathcal{M} \times [\lambda_0, \lambda_1 ]$.

A simple example of a flow is given by that of a round sphere. Writing $ds^2 = \frac{n-1}{\Lambda} S(\lambda) d\Omega^2$, where $d\Omega^2$ is the unit round sphere metric, then this is a truncation only to the conformal factor $S$, and the flow equation is simply given by,
\be
\label{eq:homogflow}
\frac{dS}{d\lambda} = (n-2) \Lambda ( S - 1 ) \; ,
\ee
so that $S = 1 + (S_0 - 1) e^{(n-2) \Lambda (\lambda - \lambda_0)}$ for a constant $S_0$ so that $S(\lambda_0) = S_0$.
Hence we see that the round sphere Einstein metric is a fixed point of the flow. However it is unstable in the forward direction so that under a perturbation the sphere either expands forever or collapses to a point in finite flow time. This is exactly the same flow equation (and hence flow behaviour) as for Ricci flow with a cosmological term when we make the replacement of Ricci flow time $\gamma = \frac{n-2}{2} \lambda$.

Unless we work in a highly symmetric setting such as the above example, we cannot expect to find analytic solutions to the flow. Nonetheless, to be of use in practice, we should at least expect to be able to solve the flow equation using numerical constructions.
However without fixing a gauge, there is no unique solution --- any solution can be transformed to another by a flow dependent diffeomorphism. Explicitly, given a solution $g_{\mu\nu}$, then we may perturb this to a new solution given by $g_{\mu\nu} \to g_{\mu\nu} + \epsilon \nabla_{(\mu} v_{\nu)}$ for any smooth vector field $v_\mu$ for infinitesimal $\epsilon$. Integrating this in $\epsilon$ relates solutions by flow dependent diffeomorphisms.
 In practice this lack of unique solution would likely prohibit solving the flow numerically and we should expect to have to eliminate this local gauge invariance in order to obtain explicit numerical solutions. 
One option could be to use the DeTurck choice of vector field as discussed above for Ricci flow. However, while this does eliminate the gauge invariance the DeTurck choice is somewhat complicated and is specifically designed to make Ricci flow parabolic. Given that we are working with the Einstein flow, which is manifestly not parabolic, it is not at all clear that this is any longer advantageous. Thus a simpler choice is to choose a local gauge condition for the metric, and then solve for the diffeomorphism vector field $\xi_\mu$ that yields this local gauge condition. If there are boundaries on the manifold we should also ensure $\xi$ vanishes on these.

\subsection{Einstein flows to and from fixed points} \label{sec:EinsteinFixed}

We are particularly interested in flows that asymptote from fixed points since these are saddle points of the Euclidean action. Later when we generalize the Einstein flow to give a Lefschetz flow for complex metrics, it will be the flows from a saddle point that generate the thimble associated to that saddle.

For flows in the vicinity of a fixed point we may linearize the flow about the fixed point Einstein metric. Then we may write the perturbation to the fixed point ${g}_{\mu\nu}$ up to diffeomorphisms as above, so $\delta g_{\mu\nu} =  \phi \, {g}_{\mu\nu} + a^{(TT)}_{\mu\nu}$. The flow equations for these perturbations decouple in the interior of the manifold, and choosing an appropriate diffeomorphism vector $\xi_\mu = (n-2) \partial_\mu \phi$, we find,
\be
\label{eq:linEinstein}
\frac{\partial}{\partial \lambda}  \phi &=& (n-2) \left( {\nabla}^2 + \Lambda \right) \phi  \; , \nl
\frac{\partial}{\partial \lambda} a^{(TT)}_{\mu\nu} &=& - {\nabla}^2  a^{(TT)}_{\mu\nu} - 2 {R}_{\mu~~\nu}^{~~\alpha~\beta} a^{(TT)}_{\alpha \beta} 
\ee 
 and we note that the flow preserves the transverse and traceless condition on $a^{(TT)}_{\mu\nu}$.
In many cases the conformal modes and transverse-traceless modes decouple; for example for a compact manifold, or an asymptotically flat one. However in other situations, such as manifolds with boundary, they do not, as emphasized in~\cite{Santos_2022}.

For perturbations of flat space, the behaviour is given precisely by our characteristic analysis above; conformal mode perturbations decay and transverse traceless perturbations grow.
More generally we introduce the notion of stability of a fixed point geometry. We assume for simplicity that there are no marginal perturbations; if they did exist we assume they have been fixed  by appropriate boundary conditions or constraints. We then define stability as follows;
\begin{quote}
 A fixed point is {\it stable} if any conformal perturbation at an initial flow time then decays to the future, and any transverse-traceless perturbation at a final flow time decays to the past. 
\end{quote}
In the simplest setting where the two sets of modes in~\eqref{eq:linEinstein} fully decouple, then stability simply corresponds to the operators $\left( - {\nabla}^2 - \Lambda \right)$ and $\left( - \delta_\mu^\alpha \delta_\nu^\beta {\nabla}^2   - 2 {R}_{\mu~~\nu}^{~~\alpha~\beta} \right)$ being positive.

For a flow to a stable fixed point, the boundary conditions for the flow are specified by giving an initial conformal factor at some finite flow time, and requiring in the future the flow asymptotes to the fixed point geometry.
As with a flow for a finite time, the initial conformal class cannot be specified as data, and instead must be read off from the flow solution. Thus the data for flows to a fixed point is the initial conformal factor, and provided this is a sufficiently small deformation of that of the fixed point, we expect flows to exist.
Conversely, for a flow from a stable fixed point the data for the flow is given by specifying the final conformal class. The final conformal factor cannot be specified as data, and again is recovered from the flow solution.

The asymptotically flat Schwarzschild solution is an unstable fixed point as the Gross--Perry--Yaffe (GPY) mode is a transverse-traceless mode that grows in the past~\cite{GPY}.
Likewise a small black hole in a cavity is unstable as a generic transverse-traceless perturbation will excite the GPY mode, now not a pure transverse-traceless mode, which will grow to the past.

Consider a general unstable fixed point. Let there be an $N$ dimensional space of `unstable' conformal factor perturbations which grow rather than decay. Then considering a flow towards the fixed point we must avoid these perturbations -- so only a codimension-$N$ set of conformal factor deformations flow to the fixed point. 
However since this $N$ dimensional space of modes grow, it represents a finite dimensional set of flows that originate from the fixed point in the past that do not overlap with the class at some finite flow time. Hence they are not constrained by fixing the conformal class at a final flow time.

Now consider the conformal class deformations. Let there be an $M$ dimensional space of `unstable' conformal class perturbations which grow to the past, and hence  do not give final data that originates from the fixed point. Then only a codimension-$M$ space of conformal class deformations flow from the fixed point in the past. However the $M$-dimensional space of unstable perturbations instead yields a set of flows that asymptotes to the fixed point in the future, and are not fixed by specifying the initial conformal factor deformation.

Hence the deformations that flow to the fixed point are given by the codimension-$N$ set of suitably small `stable'  deformations of the conformal factor, together with the $M$ dimensional set of `unstable' conformal class deformations.
Similarly the unstable manifold of deformations that flow from the fixed point is given by the codimension-$M$ stable class deformations, together with the $N$ unstable conformal mode perturbations.

The question then arises; how do we {\it non-linearly} construct flows from a fixed point  if there are unstable conformal factor or class perturbations given that for finite time flows the initial data for the flow is specified only by the conformal factor, and the final data by the class?
Unstable conformal class deformations are easy to understand -- this just constrains the set of final conformal class deformations that can flow from the fixed point. Given $M$ unstable directions, we would have to tune the conformal class final data in $M$ parameters to `emerge' from the fixed point. But how do we see the $N$ dimensional space of unstable conformal factor deformations that should also give flows asymptoting from the fixed point?
There are two obvious methods. Firstly instead of starting the flow at the fixed point as $\lambda \to - \infty$, one can start the flow at a very early, but finite flow time, and impose initial data so that the conformal factor is fixed to include these $N$ unstable perturbative deformations. All the stable conformal factor deformations decay, so it is irrelevant if they are included or not. Only the $N$ unstable perturbations will grow and influence the subsequent flow solution.

A second method is to realize that while for a finite flow we should not expect to give the conformal factor as final data, if there is an $N$ dimensional space of unstable conformal factor deformations and we are flowing from the fixed point, then we can specify an $N$ dimensional subset of the conformal factor as final data. Thus we can choose to impose $N$ constraints on the final conformal factor to move in the space of flow solutions from the fixed point. 

The converse is true for flows to a fixed point; only a codimension-$N$ space of conformal factor deformations can be used as initial data. One would have to tune generic conformal factor data in $N$ parameters to ensure the flow does asymptote to the fixed point. And likewise the $M$-dimensional space of `unstable' class deformations can either be included by putting $M$ constraints on the initial conformal class, which otherwise we are not free to specify, or alternatively one can end the flow at a very large, but finite flow time in the future, and set the final data to be that dictated by this $M$ dimensional unstable space of conformal class deformations.

\subsection{Example: Einstein flow of deformations of $S^4$}\label{sec:EinsteinS4}

We now consider explicit examples of the forward-backward Einstein flow. Our first example is for deformations of 4-dimensional Euclidean de Sitter, and hence we consider the flow with a positive cosmological term. We shall begin by solving the flow for linear perturbations of the round $S^4$ solution which is a fixed point of the flow. We then proceed to numerically solve the non-linear flow for certain axisymmetric deformations that may become large.

\subsubsection{Linear Flow}
In 4-dimensions our flow equation is  given by
\be\label{eq: Einstein Flow}
\frac{\partial}{\partial\lambda} g_{\mu\nu} =  2\left( G_{\mu\nu} +\Lambda g_{\mu\nu}\right) + \nabla_{(\mu} \xi_{\nu)} \, ,
\ee
with $\Lambda>0$. 
Then 4-dimensional Euclidean de Sitter is the round sphere $ds^2 = {g}_{\mu\nu} dx^\mu dx^\nu = \frac{3}{\Lambda} d\Omega^2$ which is a fixed point of this flow provided the diffeomorphism term vanishes.
Perturbing about this round sphere background, we decompose our perturbation in the same manner as~\eqref{eq:linEinstein} and choosing an appropriate gauge, our linearized flow equations become
\be
 \frac{\partial}{\partial\lambda} \phi =  2 \left( {\nabla}^2 +  \Lambda \right) \phi \; , \quad
 \frac{\partial}{\partial\lambda} a^{TT}_{\mu\nu} = - \left( {\nabla}^2 - \frac{2}{3}\Lambda\right) a^{TT}_{\mu\nu}\, .
\ee
We may expand  our trace and transverse-traceless perturbations in terms of eigenmodes of the $S^4$ Laplace operator; $Y^{(l, m)}$ are the scalar spherical harmonics on $S^4$ with integer $l \ge 0$, and multiplicity $(l+1)(l+2)(2 l+3)/6$, and $\Upsilon^{(l,m)}_{\mu\nu}$ are the transverse traceless tensor harmonics labelled by integer $l \ge 2$ with multiplicity $5(l-1)(l+4)(2 l +3)/6$, which obey~\cite{Allen:1986tt},
\be
\nabla^2 Y^{(l,m)} = - \frac{\Lambda}{3} l(l+3) Y^{(l,m)} \; , \quad
\nabla^{2}\Upsilon^{(l,m)}_{\mu\nu} = -\frac{\Lambda}{3} (l(l+3)-2) \Upsilon^{(l,m)}_{\mu\nu} \; .
\ee
Then we may write a solution to the linearized flow as a sum over these,
\begin{align}
\phi &= \sum_{l \ge 0,m} \phi_{l, m} Y^{(l, m)} e^{- \frac{2\Lambda}{3}\left(l(l+3)-3\right)  \lambda  } \, ,\nl
a^{TT}_{\mu\nu} &= \sum_{l \ge 2,m} a_{l,m} \, \Upsilon^{(l,m)}_{\mu\nu} e^{+ \frac{\Lambda}{3}  l(l+3)   \lambda  } \; .
\end{align}
We see that the large eigenvalue modes of $\phi$ decay to future flow times, corresponding to the conformal factor having the characteristic that it flows in the forward direction. However the lowest single mode, given by $\phi_{0,0}$, which is simply a constant mode on the sphere, has an exponentially growing behaviour in forward flow time. This homogeneous growing mode corresponds to the single unstable mode of the flow. It is precisely this single mode that generates the unstable homogeneous non-linear flow discussed above in equation~\eqref{eq:homogflow}. In the language of the previous section, the codimension of the set of stable conformal deformations in the space of all conformal deformations for the forward flow is $N=1$.
In the transverse-traceless sector all modes decay to the past, reflecting that these modes obey a diffusive flow that is parabolic backwards in time; hence this corresponds to $M=0$, so there are no unstable class deformations.

The set of coefficients $\{ \phi_{l,m}, a_{l+2,m} | l \ge 0 \}$ specifies a flow. From the perspective of a flow from a finite flow time $\lambda_0$ to the later time $\lambda_1$ we may think of $\{ \phi_{l,m} \}$ as giving the initial data for the conformal factor, $\phi(\lambda_0,x) = \sum_{l,m} \phi_{l, m} Y^{(l, m)} e^{- \frac{2\Lambda}{3}\left(l(l+3)-3\right)  \lambda_0  }$ and $\{ a_{l,m} \}$ gives the final class data, $a^{TT}_{\mu\nu}(\lambda_1,x)  = \sum_{l,m} a_{l,m} \, \Upsilon^{(l,m)}_{\mu\nu} e^{+ \frac{\Lambda}{3} l(l+3) \lambda_1  }$.

However we are particularly interested in flows from or to fixed points. 
For the flow from the fixed point, the final data at $\lambda_1$ is given by all class perturbations, together with the single unstable conformal mode -- the homogeneous one -- $\{ a_{l,m} , \phi_{0,0} | l \ge 2 \}$. All other modes of the conformal factor are set to zero in the far past at the fixed point, $\lambda_0 \to - \infty$, and hence the conformal factor at  $\lambda_1$ is given by the single homogeneous perturbation $\phi(\lambda_1,x) =  \phi_{0, 0} Y^{(0,0)} e^{ 2\Lambda \lambda_1 }$.
Conversely if we consider a flow forward towards the round sphere fixed point then the initial data at $\lambda_0$ is given by the stable conformal modes, $\{ \phi_{l,m} , l \ge 1 \}$.

\subsubsection{Non-linear numerical flow of deformations of $S^4$}\label{sec:EinsteinS4nonlin}

We now consider the Einstein flow of non-linear deformations of $S^4$. Going beyond the homogeneous sector we have no way to solve the flow except numerically. The purpose of this section is to demonstrate that we may explicitly do this, showing how to solve the flow both for finite flow times, and also for flows that asymptote from a fixed point in the past or to one in the future.

We will consider a specific sector of axisymmetric deformations that preserve an $SO(3) \times SO(2)$ subgroup of the round sphere $SO(5)$ isometries, hence reducing the flow to a PDE in two variables; the flow time $\lambda$ and a polar angle on the sphere. We write the round sphere as a fibration of $S^2$ and $S^1$ over an interval, so,
\be
ds^2 = \frac{3}{\Lambda} \left( d\psi^2 + \sin^2(\psi) d\Omega_{(2)}^2 + \cos^2(\psi) d\chi^2 \right)
\ee
where $\psi \in [0, \frac{\pi}{2}]$, $d\Omega_{(2)}^2$ is the metric on a unit radius round 2-sphere, and $\chi$ is the angle coordinate of the $S^1$. Then we deform in the most general way that preserves the $SO(3) \times SO(2)$  isometries taking,
\be
    d s^2 = \frac{3}{\Lambda}  \left( e^{2A}\,\dd \psi^2 +  e^{2B}  \sin^2(\psi) d\Omega_{(2)}^2  + e^{2C }\cos^2(\psi) d\chi^2 \right) 
\ee
where $A, B, C$ are smooth functions of the angle $\psi \in [0, \frac{\pi}{2} ]$ as well as flow time.
Regularity at $\psi = 0$ where the $S^2$ shrinks requires the functions are even at $\psi = 0$ and further that $A = B$ there. Given that $\chi$ is an angle coordinate with period $2 \pi$ then regularity at $\psi = \frac{\pi}{2}$ where the $S^1$ shrinks implies the functions are even about $\psi = \frac{\pi}{2}$ and also satisfy $A = C$ there.

There remains the coordinate freedom to take $\psi \to \theta = f(\psi)$ for any smooth monotonic function $f$ such that $f(0) = 0$ and $f(\pi/2) = \frac{\pi}{2}$. As a consequence, we may use this freedom to choose to fix $g_{\theta\theta}$ to some prescribed function. We make the following choice,
\be
\label{eq:spherecoordchoice}
    d s^2 = \frac{3}{\Lambda}  \left( e^{A} \left( \dd \theta^2 + \sin^2(\theta) d\Omega_{(2)}^2 \right) + e^{C}\cos^2(\theta) d\chi^2 \right)
\ee
which is convenient as the regularity condition at $\theta = 0$ is then automatically satisfied provided $A$ and $C$ are smooth even functions. This transformation is achieved by defining the new angular coordinate $\theta$ to be related to $\psi$ as $\psi = \psi(\theta)$ which satisfies the first order ODE,
\be
\sin( \theta )  \psi'(\theta) =  \sin( \psi(\theta) ) e^{- A( \psi(\theta) ) + B( \psi(\theta) ) }  \; .
\ee
Given a solution, we may invert it to find $\theta = \theta(\psi)$. 
Provided $A$ and $B$ are smooth functions of $\psi^2$ with $A = B$ at $\psi =0$, as required by regularity, then there is a one-parameter family of regular solutions whose power series goes as $\psi(\theta) = k \theta + O(\theta^3)$ for any value of the parameter $k$, so that $\psi = 0$ corresponds to $\theta = 0$. Given a value for the constant of integration $\psi'(0)$, we integrate to  obtain $\psi(\theta)$, which is a smooth function of $\theta$ that is odd under $\theta \to - \theta$. The initial data $\psi'(0)$ yields a one parameter family of such solutions. However, if we wish to have the location where the $S^1$ shrinks at the coordinate location $\theta = \pi/2$ in the new coordinate, then we must also impose the condition $\psi(\pi/2) = \pi/2$, which fixes this one parameter. This yields a unique solution for the coordinate transformation where $\psi$ is monotonically increasing. 
Such a coordinate transform can be performed at each flow time along the Einstein flow. 

Now we wish to write the metric above in~\eqref{eq:spherecoordchoice} as in the decomposition in equation~\eqref{eq:decomp}. Hence we choose to define $A = \Phi + T$ and $C = \Phi - 3 T$ so,
\be
\label{eq:spheremetricform}
    d s^2 = e^{\Phi} \hat{g}_{\mu\nu} dx^\mu dx^\nu = \frac{3}{\Lambda} e^{\Phi} \left( e^{T} \left( \dd \theta^2 + \sin^2(\theta) d\Omega_{(2)}^2 \right) + e^{ -3T}\cos^2(\theta) d\chi^2 \right) 
\ee
where now $\Phi = \Phi(\lambda,\theta)$ and $T = T(\lambda,\theta)$ are functions of flow time and the polar coordinate, so we see $\det( \hat{g}_{\mu\nu} )$ is  fixed in flow time to be the same as that of the round sphere. 
Regularity requires $\Phi, T$ are smooth even functions of  the  angle $\theta$. Regularity implies that $T = 0$ where the circle shrinks at $\theta = \pi/2$. In practice we solve for the functions on the domain $\theta \in [0, \pi/2]$ and impose Neumann boundary conditions on $\Phi$ and $T$ at the domain boundaries. Since the Einstein flow is a geometric flow and preserves the regularity of the metric, one finds that for smooth $\Phi$ and $T$ with these boundary conditions then $\left. \frac{\partial T}{\partial \lambda} \right|_{\theta = \pi/2} = 0$. Hence if we provide data for $T$ at some flow time such that $T = 0$ at $\theta = \pi/2$ at that time, then $T$ will vanish there for all flow times.

We see that ${\Phi}$ is a conformal factor deformation of the sphere, and ${T}$ determines $\hat{g}$, the representative of the conformal class. A perturbation ${\Phi} \to {\Phi} + \delta {\Phi}$ with $\delta T = 0$ generates a conformal perturbation, while a perturbation,
\be
\label{eq:TTpertSphere}
{T} \to {T} + \frac{3}{4} \delta {X} \; , \quad \Phi \to \Phi + \frac{1}{4} \delta {X}
\ee
so that,
\be
\frac{\Lambda}{3} \delta g_{\mu\nu}
=  e^{\Phi} \left( \delta X e^T \sin^2(\theta) d\Omega_{(2)}^2   - 2 \delta X e^{-3 T} \cos^2(\theta) d\chi^2 \right) + {\delta X} e^{\Phi + T} \dd \theta^2 
\ee
generates a traceless perturbation that is transverse with respect to the vector $\partial_\theta$, given by the first term (in brackets), together with a diffeomorphism in $\theta$, the last term.

Now if one computes the Einstein tensor of the metric~\eqref{eq:spheremetricform}, it gives a tensor of similar form with the $SO(3) \times SO(2)$ isometry; however it does not have the relation between its $\dd \theta^2$ and $d\Omega_{(2)}^2$ component  that the metric has with our chosen coordinates. Hence the Einstein flow with no diffeomorphism term is not compatible with this gauge fixing. However we may appropriately choose the diffeomorphism term to ensure the flow has the above form for all time. Since the coordinate transformation to put the metric in this form is a redefinition of the polar angle only, then the required flow diffeomorphism is given by a vector field $\xi^\mu$ with vanishing components except for its angular component,
\be
\label{eq:diffvecsphere}
\xi^\theta = \sin( \theta )\cos(\theta) F(\lambda,\theta) \; .
\ee
We require that $\xi^\theta(0) = 0$ and $\xi^\theta(\pi/2) = 0$ to preserve the positions where the $S^2$ and $S^1$ shrink, and due to the trigonometric factors included in the ansatz, this simply implies that $F$ is a smooth function in $\theta$ which is even about $\theta = 0$ and about $\theta = \pi/2$, similar to the requirement for $\Phi$ and $T$. Hence, as for $\Phi$ and $T$, this is imposed by specifying Neumann boundary conditions at the angle domain boundaries $\theta = 0$ and $\pi/2$. 

With the above choice, the flow equations themselves determine $F$. They give equations for $\Phi$ and $T$ with the highest derivative terms going as,
\be
\frac{\partial}{\partial \lambda} \Phi = g^{\theta\theta} \partial^2_\theta \left( 2\Phi -  T \right) + \ldots \; , \quad \frac{\partial}{\partial \lambda} T = - g^{\theta\theta} \partial^2_\theta  T  + \ldots \; .
\ee
From this we see the perturbation $\delta \Phi$ with $\delta T =0$ flows forwards, as we expect. 
Diagonalizing we see the perturbation flowing backwards is $(\delta T, \delta \Phi) \propto \left(1, \frac{1}{3} \right)$. This is precisely as we expect from~\eqref{eq:TTpertSphere}, namely $\delta T = 3 \delta \Phi = \frac{3}{4} \delta X$, with $\delta X$ the transverse-traceless mode.
We obtain an ODE for $F$ where $\frac{\partial}{\partial \theta} ( \cos(\theta) F )$ is given by an expression composed only of $\Phi$, $T$ and their $\theta$ derivatives; it does not contain $F$ or any flow derivatives. As discussed above, the coordinate transformation to put the metric in this form, with the further requirement that the positions where the $S^2$ and $S^1$ shrink are at $\theta = 0$ and $\pi/2$ respectively, has no free continuous data. We then expect a unique solution.

Having dealt with fixing the coordinate freedom, we may now discuss the geometric data for the flow. Here, and for all future examples with a positive cosmological constant, we choose units so that $\Lambda = 1$.
Let us first consider  a flow for a finite flow time. Then at the initial flow time $\lambda_0$ we should prescribe the conformal factor giving the data $\Phi_0(\theta) = \Phi(\lambda_0, \theta)$, and at the final flow time $\lambda_1$ we fix the conformal class giving the data $T_1(\theta) = T(\lambda_1,\theta)$, ensuring that $T_1(\pi/2)= 0$ to ensure regularity where the $S^1$ shrinks.
Fixing $T_1$ fixes the transverse-traceless perturbation $\delta X$ to vanish at the final time (since $\delta T = \frac{3}{4}\delta X)$. 
As discussed above, fixing $\Phi_0$ fixes the conformal perturbation $\delta \Phi$ at the initial time.
\begin{figure}[h]
    \centering
    \subfigure[Flow of $T(\lambda,\theta)$]{
        \includegraphics[width=0.47\textwidth]{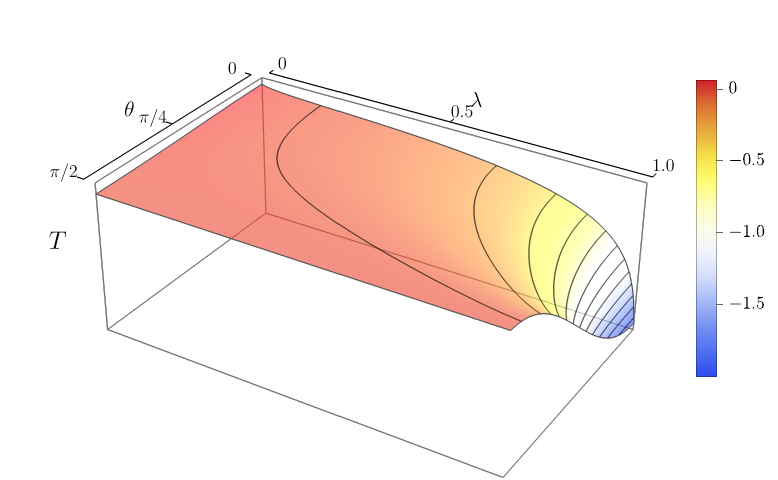}
        \label{fig:Finite_T_S4}
    }
    \hfill
    \subfigure[Flow of $\Phi(\lambda,\theta)$]{
        \includegraphics[width=0.47\textwidth]{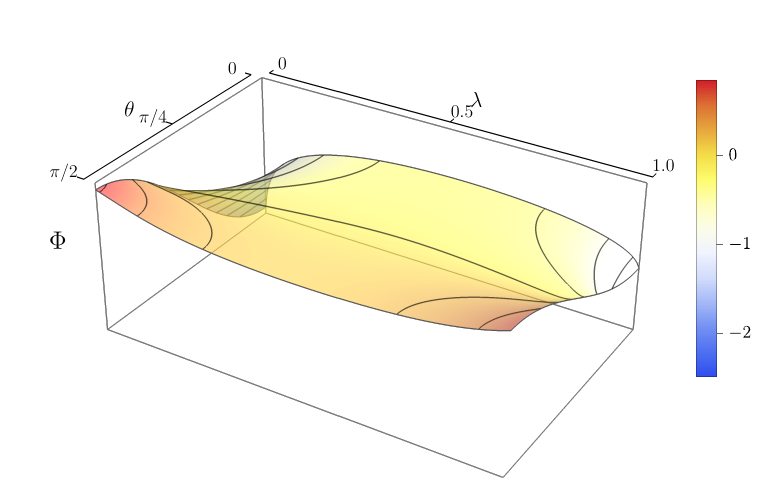}
        \label{fig:Finite_Phi_S4}
    }
    \caption{Numerical solutions of the Einstein flow near the $S^4$ fixed point for finite flow time period $\lambda \in [0,1]$. The initial conditions for the conformal factor are $\Phi_0(\theta) = \frac{2}{3}(5\sin^2\theta - 3)$ at $\lambda = 0$. The final condition at $\lambda =1$ is $T_1(\theta) = -2\cos^2\theta$ fixing the class. $T$ relaxes from its final data backwards in flow time towards the initial flow time $\lambda = 0$. The inhomogeneous profile of $\Phi$ initially relaxes away from the initial flow time, but towards the final flow time $\lambda = 1$ the coupling to $T$ keeps it appreciably deformed there.  }    \label{fig:Finite_Flow_S4}
\end{figure}
In Figure~\ref{fig:Finite_Flow_S4} we show an example of a finite time flow solution from $\lambda = 0$ to $1$, where we prescribe the initial data $\Phi_0(\theta) = \frac{2}{3}(5\sin^2\theta - 3)$ and the final data
 $T_1(\theta) = -2\cos^2\theta$.
We plot the amplitudes of $\Phi$ and $T$ as functions of the flow time, $\lambda$, and the angle coordinate $\theta$. The computational domain is then given by $(\lambda,\theta) \in [0,1]\times[0,\frac{\pi}{2}]$, and as for the scalar field example earlier, the data shown is for a Chebyshev discretization of the domain using $n_\lambda = 60$ and $n_\theta = 50$ points. Boundary conditions are imposed for regularity and to fix the data $\Phi_0$ and $T_1$, and then solutions are found using a Newton method to solve the resulting discrete system.
Here we can see from the figure quite clearly that $T$ relaxes backwards in flow time from its data at $\lambda_1$, whereas $\Phi$ relaxes forwards from its data at $\lambda_0$. We note that relatively small deformations in $T$ induce quite strong deformations in $\Phi$, but we do not observe the converse behaviour, finding that $T$ remains small at $\lambda = 0$, despite considerable deformations in $\Phi(0,\theta)$. 

Let us now consider a flow to the round sphere fixed point in the future, so $\lambda_1 \to \infty$. Without loss of generality we may then set $\lambda_0 = 0$. In order to compute such a flow we compactify the flow time domain, introducing a computational flow coordinate $y$ as,
\be
\label{eq:compflowtime}
\lambda(y) = \frac{1}{a} \frac{y}{1-y}  
\ee
so that $y \in [0, 1]$ corresponds to $\lambda \in [0, \infty)$. 
 The parameter $a$ allows us to increase grid sampling around regions of interest, so that we may see the perturbation decay in greater resolution. In all flows towards a fixed point presented we have chosen the value $a = 10$ unless otherwise specified. Our computational domain is then $(y, \theta) \in [0,1]\times[0,\frac{\pi}{2}]$, and for the data presented we  discretize this with a Chebyshev grid of $n_y = 60$ and $n_\theta = 50$ points. 

We impose that as $\lambda \to \infty$ the flow asymptotes to the round sphere fixed point, and hence take $\Phi = T = F = 0$ at $y = 1$. Then we impose boundary conditions for regularity and also specify the data for $\Phi_0$ at the initial flow time.
However, we have seen above that the conformal mode has one unstable mode about the round sphere fixed point.  Thus the initial data for the flow, given by the initial conformal factor $\Phi_0$, is a codimension one set in the space of all conformal factors --- we must tune one parameter for this initial conformal factor to ensure we reach the fixed point to the future. 

\begin{figure}[h]
    \centering
    \subfigure[Flow of $T(\lambda,\theta)$]{%
        \includegraphics[width=0.485\textwidth]{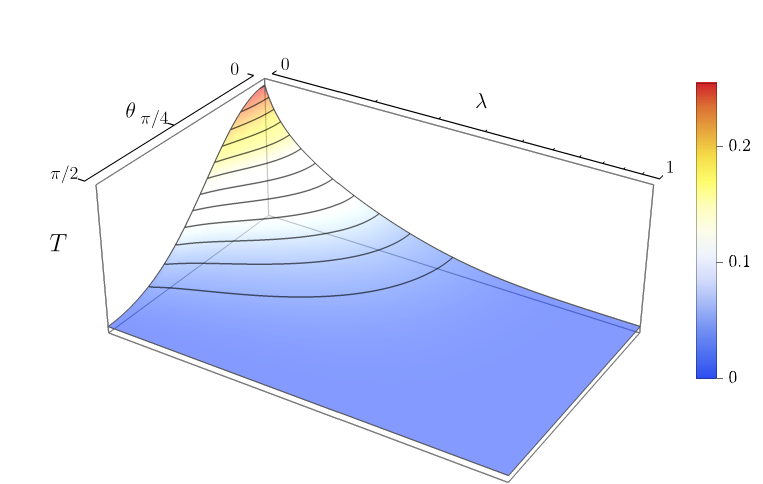}%
        \label{fig:Einstein_towards_T_S4}
    }
    \hfill
    \subfigure[Flow of $\Phi(\lambda,\theta)$]{%
        \includegraphics[width=0.485\textwidth]{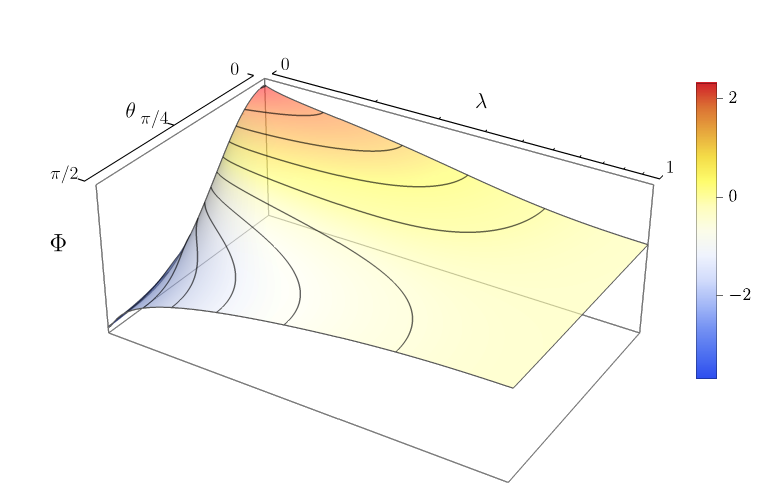}%
        \label{fig:Einstein_towards_Phi_S4}
    }
    \\
    \subfigure[Evolution of the action]{
    \includegraphics[width=0.48\textwidth]{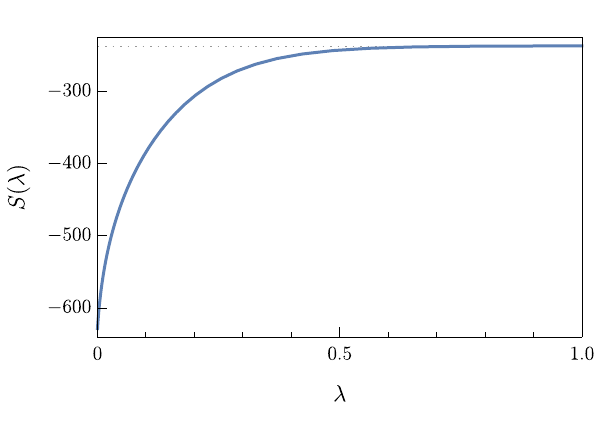}
  \hspace{1cm}          \label{fig:Einstein_towards_Action_S4}
    }
    \caption{Einstein flow towards the $S^4$ fixed point. The data for the flow is the large deformation $\Phi_\perp(\theta) = 6\left(\cos^4\theta - \tfrac{8}{35}\right)$, so this is far from the linear regime and the  response of $T$ due to non-linearity can be seen. While the flow is computed for $\lambda \in [0, \infty)$, only the portion $\lambda \in [0,1]$ is shown since the geometry quickly settles to that of the fixed point where $\Phi$ and $T$ vanish. Indeed for clarity we have used a log scale to depict the $\lambda$ dependence. We see that the action increases monotonically towards its value, $- 24 \pi^2$ (dashed line), at the fixed point.}    \label{fig:Einstein_Towards_S4}
\end{figure}

For small deformations the unstable mode is the constant component of the conformal factor. Hence we decompose the initial data into a homogeneous part, $\Phi_h$, and its complement $\Phi_{\perp}(\theta)$ with respect to the volume measure on the sphere, so that,
 \be
\Phi_0(\theta) = \Phi_h + \Phi_\perp(\theta) \; ,\quad \; \Phi_h = 3\int_0^{\pi/2} \dd\theta \sin^2\theta \cos\theta \, \Phi_0(\theta)
 \label{eq:Phidecomp}
\ee
where we have chosen to normalize so that $\Phi_h = \Phi_0$ for a constant function.
Having done so we may freely specify the smooth function $\Phi_\perp(\theta)$ at the initial flow time, so $y=0$. Crucially, we do \textit{not} fix $\Phi_h$ at that initial flow time, and its value is determined entirely by the flow equations there, and hence implicitly by the condition at the asymptotic end of the flow $\lambda \to \infty$ that we reach the round sphere fixed point.
Figure~\ref{fig:Einstein_Towards_S4} provides an example of such a flow. We show only range of flow time $\lambda \in [0, 1]$, since the fields relax to the fixed point within the early stages of the flow, and so after $\lambda \simeq 1$ the flow is very close to the fixed point solution where $\Phi$ and $T$ vanish. The data chosen for the example is the relatively large deformation, $\Phi_\perp(\theta) = 6 \left(\cos^4\theta - \tfrac{8}{35}\right)$, so we are far from the linear regime. We can see that the function $T$ is deformed from zero at the initial flow time in response to this strong deformation $\Phi_\perp(\theta)$.  Given the size of the deformation we impose as data for $\Phi_\perp$ which is strongly non-linear, the response of $T$ is relatively modest. By construction the Einstein--Hilbert action should increase along the flow. In the figure we show the action as a function of flow time $\lambda$, and indeed see that it quickly and monotonically increases from its initial value at $\lambda = 0$ over the period to $\lambda \simeq 1$ where it tends to the fixed point value $S = - 24 \pi^2$.

Finally we consider a flow away from the round sphere fixed point in the past, so $\lambda_0 \to - \infty$ and we fix vanishing $\Phi, T$ and $F$ there. Without loss of generality we may take the final flow time $\lambda_1 = 0$. 
Then we specify the class $T_1(\theta)$ at that final flow time, and since perturbations of the sphere by the class decay to the past, we don't expect to have to perform any tuning --- any smooth function obeying the regularity conditions would be expected to yield a flow solution, provided that it is not too large a deformation of the round sphere class. 
As discussed earlier, we now expect extra data due to the unstable homogeneous conformal mode which decays in the past. 
In order to specify this data we decompose the final conformal deformation $\Phi(0,\theta)$ in the same manner as above in equation~\eqref{eq:Phidecomp}, so that,
 \be
 \label{eq:PhiDecompFinal}
\Phi(0,\theta) = \Phi_h + \Phi_\perp(\theta) \; ,\quad \; \Phi_h = 3\int_0^{\pi/2} \dd\theta \sin^2\theta \cos\theta \, \Phi(0,\theta)
\ee
and then impose data by fixing $\Phi_h$ at the final flow time, whereas $\Phi_\perp$ is determined by the flow equations there.
We compactify the flow time similarly to the flow towards the fixed point, choosing a computational time variable $y$ as,
\be
\lambda(y) = \frac{1}{a} \frac{y-1}{y}
\label{eq:away compact flow time}
\ee
so that we compactify $\lambda \in (-\infty, 0]$, and we fix vanishing $\Phi, T$ and $F$ at $y = 0$. For the flows away from the sphere presented here we have chosen the scaling parameter to have the value  $a=0.5$, but for all later flows away from a fixed point we have used $a = 10$.\footnote{The computational domain $(y,\theta) \in [0,1] \times [0, \frac{\pi}{2}]$ is discretised as for the flow to the fixed point, and the data shown is for the same resolution $60 \times 50$.}

\begin{figure}[h]
\centering
    \subfigure[Flow of $T(\lambda,\theta)$]{%
        \includegraphics[width=0.48\textwidth]{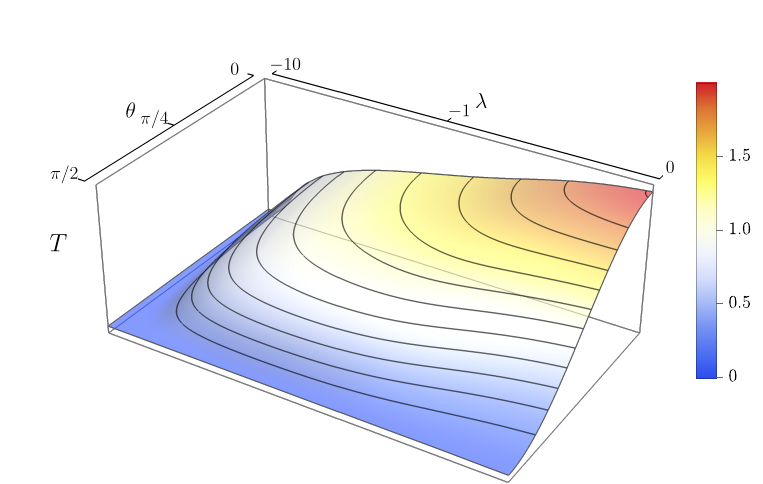}%
        \label{fig:Einstein_Away_T_S4}
    }
    \hfill
    \subfigure[Flow of $\Phi(\lambda,\theta)$]{%
        \includegraphics[width=0.48\textwidth]{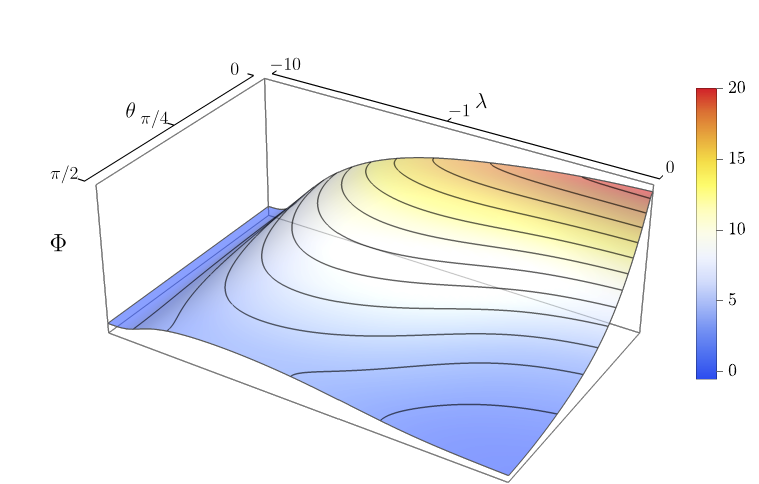}
        \label{fig:Einstein_Away_S_S4}
        }
    \subfigure[Evolution of the action]{
    \includegraphics[width=0.48\textwidth]{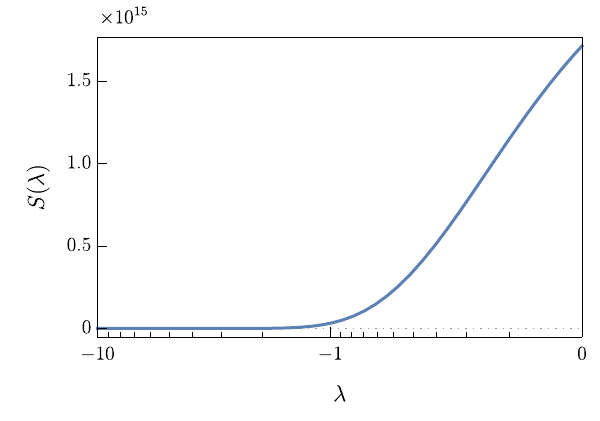}
  \hspace{1cm}
        \label{fig:Einstein_Away_Action_S4}
    }
    \caption{An example solution for the Einstein flow away from the $S^4$ fixed point. We show only the portion of the flow $\lambda \in [-10, 0]$ and employ a log scale for the flow time. The final data are $T_1(\theta) = 2 \cos^2\theta$ for the class and $\Phi_h = 2$ for the homogeneous mode of the conformal factor. We see that the action monotonically increases from the round sphere value, $-24 \pi^2$, in the far past.}
    \label{fig:Einstein_Away_S4}
\end{figure}

In Figure~\ref{fig:Einstein_Away_S4} we show an example of a flow from the round sphere where we have prescribed the data $T_1(\theta) = 2 \cos^2\theta$ and $\Phi_h = 2$ at the final flow time. We see that the deformation in $T$ induces a significant deformation in $\Phi$. The figure also shows the action which monotonically increases from the round sphere value asymptotically in the past as the flow proceeds. Only the range $\lambda \in [-10,0]$ is shown in the figures as the flow is very close to the fixed point geometry for earlier times.

\subsection{Example: Einstein flow of large to small black holes in a cavity}\label{sec:bhflow}

As our second example for Einstein flow, we consider the case of 4-dimensional Euclidean Schwarzschild black hole saddle points, taking $\Lambda = 0$, in a cavity with boundary $S^1_{(\beta)} \times S^2_{(R)}$ so that the Euclidean time $S^1$ factor has length $\beta$ and the spatial round sphere $S^2$ factor has radius $R$. This setting computes the canonical ensemble for gravity in a round cavity of (areal) radius $R$ and temperature $1/\beta$, and we choose units so that $R = 1$.
At sufficiently high temperature, $\beta^{-1} > 3 \sqrt{3} / 8 \pi$, there exist two Schwarzschild saddle point solutions, the small and large black holes.  

Firstly let us review Ricci flow for these solutions. The stability properties of the saddle points are  controlled by the linearization of the Ricci tensor, the Lichnerowicz operator.
The spectrum of perturbations of the Lichnerowicz operator for the small black hole famously has a single negative mode. The conformal perturbation modes and transverse-traceless modes are coupled together by the cavity boundary conditions and cannot be decoupled, so this negative mode is a mix of both conformal and transverse-traceless.
For a very small black hole this is approximately the Gross--Perry--Yaffe (GPY) unstable mode of the asymptotically flat Euclidean Schwarzschild solution~\cite{GPY} which is purely transverse-traceless. The unstable mode persists as one moves along the small black hole branch, increasing the black hole size, until at the minimum temperature black hole it becomes a zero mode, which is precisely the mode given by the tangent to the space of solutions there. For the large black holes the mode continues to become positive, and the spectrum of the Lichnerowicz operator for large black holes is positive.
As a consequence for Ricci flow the small black hole is unstable to perturbation by the negative mode, whereas the large black holes are a stable attractor. The unstable mode preserves the static and spherical $U(1) \times SO(3)$ isometry, and hence this flow shares this isometry of the Schwarzschild solution. 
There are two flows generated by the unstable mode of the small Schwarzschild black hole in the cavity~\cite{Headrick_2006}.  In the sense where the unstable mode increases the horizon size, the flow asymptotes to the large black hole which is stable under Ricci flow. 
We will not be concerned here with the flow in the other sense, where the flow goes to hot flat space via a finite flow time singularity which must be resolved. 

Now let us turn to the Einstein flow. We first consider the case of a Schwarzschild fixed point which is asymptotically flat, so not in a cavity. 
Then the conformal and transverse-traceless perturbations decouple from each other.
The stability of fixed points for the flow is given by the linearization of the Einstein tensor. 
Eigenmodes of the linearized Ricci tensor which are transverse-traceless are also eigenmodes of the linearization of the Einstein tensor with the same eigenvalue. Conformal eigenmodes are also eigenmodes of the linearized Einstein tensor, but with the opposite sign for their eigenvalue which is also rescaled by $(n-2)$. 
These eigenmodes are given by solutions to the equations~\eqref{eq:linEinstein} where $\phi(\lambda,x) = \phi(x) e^{\omega \lambda}$ and $a^{(TT)}_{\mu\nu}(\lambda,x) = a^{(TT)}_{\mu\nu}(x) e^{\omega \lambda}$ so that $\omega$ is the corresponding eigenvalue.
All the conformal eigenmodes are stable, so have $\omega < 0$, and hence decay in Einstein flow time. The transverse-traceless modes are all stable in the backward flow sense, so have $\omega > 0$ and grow in flow time, except for the single GPY mode $h^{*}_{\mu\nu}$
which preserves the static and spherical $U(1) \times SO(3)$ isometry
and decays as,
\be
\label{eq:negmodeflow}
a^{(TT)}_{\mu\nu}(\lambda,x) =  a e^{- \mu^{2} \lambda} \, h^{*}_{\mu\nu}(x)
\ee
where $a$ is a constant giving the amplitude of the deformation by the unstable mode. Note that since we have reversed the `sense' of the flow for our Einstein flow relative to Ricci flow, what was an exponentially growing mode for Ricci flow becomes exponentially decaying for our Einstein flow.
Hence we see asymptotically flat Euclidean Schwarzschild is an unstable fixed point with $N =0$ and $M = 1$ in the conventions of Section~\ref{sec:EinsteinFixed}.

Flowing from the Schwarzschild fixed point then the data for the flow is given by a codimension one subset of all conformal classes at some final flow time $\lambda_1$.
If the flow is everywhere a perturbation of Schwarzschild then this tuning is simply that the perturbation contains no unstable mode, and hence neither does the class data. If the class data is a larger non-perturbative deformation, a one parameter family of data will have to be tuned in order that the flow asymptotes correctly to the fixed point in the past. The tuning is such that near the fixed point in the past, the perturbation has no unstable mode component.

For a flow that asymptotes to the fixed point in the future, so $\lambda_1 \to + \infty$, then taking any conformal factor data near Schwarzschild at the initial finite flow time $\lambda_0$ we expect the flow to exist. However, exactly as with Euclidean de Sitter, there is the additional data of tuning the GPY mode in the asymptotic future. Since this is orthogonal to the conformal factor modes, it may be tuned either by prescribing one constraint on the conformal class at $\lambda_0$, or alternatively truncating the flow  at a large flow time $\lambda_1$ and imposing the class to be that of Schwarzschild including the GPY mode, as in~\eqref{eq:negmodeflow}. Then tuning the constant $a$ amplitude of the GPY mode, and taking $\lambda_1 \to \infty$ will give this extra data.

Now in the setting of black holes in a cavity the two classes of modes, conformal and transverse-traceless, become coupled by the cavity boundary conditions. Hence the eigenmodes of the linearization of the Ricci tensor will not generally be eigenmodes of the linearization of the Einstein tensor, and vice versa.
However for very small black holes, we expect the unstable mode to be approximately transverse-traceless, since it localizes near the horizon and so does not `see' the cavity. By continuity, we then expect that for flows that asymptote to the small black hole in the past we will need a one-parameter tuning of the conformal class data at the final flow time $\lambda_1$. Conversely, for flows that tend to the small black hole in the future, we expect there is one extra parameter that governs the flow as well as the usual initial conformal factor data. 

For large black holes we expect no such unstable mode, by which we mean that if we flow from a large black hole in the past, we can choose any conformal class in the future near to the black hole and expect a flow to exist with no tuning. Likewise for a flow towards a large black hole in the future, any conformal factor data near the black hole should determine a unique flow to the fixed point. A natural expectation is that all the small black holes are unstable, in the sense they require a one parameter tuning of the class to flow from them in the far past. 
Indeed, in Appendix~\ref{app:negative} we numerically confirm that for small black holes there is an additional unstable mode of the linearized Einstein tensor, analogous to the GPY mode of the linearized Ricci tensor. For very small black holes this is the usual GPY mode.

We construct flows with $U(1) \times SO(3)$ symmetry in a similar manner as for the Euclidean sphere. Writing Euclidean Schwarzschild in a unit radius cavity as,
\be
ds^2 =  r_0^2 \left( r^2 d\tau^2 + 4 ( 1 - r_0) g(r)^2 dr^2 + g(r) d\Omega^2\right) \; , \quad g(r) = \frac{1}{(1 - (1-r_0) r^2 )^2} 
\label{eq: BH metric}
\ee
then the radial coordinate $r \in [0,1]$, with the unit radius cavity at $r = 1$, and the horizon at $r = 0$ having radius $r_0$.
The small black holes have radius $r_0 < \frac{2}{3}$, and the large black holes have $\frac{2}{3} < r_0 < 1$. We then deform this as,
\be
ds^2 =  r_0^2 e^{\Phi} \left( r^2 e^{ A} d\tau^2 + 4 ( 1 - r_0) g(r)^2 e^{ T}  dr^2 + g(r) e^{- T}  d\Omega^2\right) 
\label{eq: BH metric with T and Phi}
\ee
where $\Phi, T, A$ are functions of radius and flow time. 
$A$ may be chosen by prescribing the radial coordinate. We choose to fix the gauge $A = T$. 
Then the variation $\delta \Phi$ is a conformal perturbation and $\delta T$ gives a traceless perturbation. Regularity at the horizon implies $\Phi$ and $T$ are even functions of $r$. We impose a Dirichlet cavity boundary condition setting $\Phi = T = 0$ at the cavity wall, $r=1$.\footnote{
While the status of Dirichlet boundary conditions in Euclidean gravity is subtle~\cite{Avramidi:1997sh,Avramidi:1997hy} in general, in this setting of high symmetry we do not expect a problem with well-posedness. Another possible boundary condition that should be well-posed for Euclidean signature is the conformal or generalized conformal one~\cite{Anderson:2006lqb,Figueras:2011va,Adam:2011dn,Witten:2018lgb,Anninos:2023epi,Liu:2024ymn}.
}
Having fixed the gauge as above we take the diffeomorphism vector to have non-vanishing component,
\be
\xi^r = r F(\lambda,r) 
\ee
and then the condition that the coordinates preserve the horizon being at $r = 0$ is built into the above ansatz provided $F$ is smooth at $r = 0$, and the cavity remaining at $r = 1$ requires $F$ vanishes there. With the above choice, the flow equations themselves determine $F$, and the boundary condition at the cavity wall  fixes the constants of integration in the solution for $F$.
The flow equations determine the highest derivative terms for the conformal factor $\Phi$ and class $T$ and go as,
\be
\frac{\partial}{\partial \lambda} \Phi = g^{rr} \partial^2_r \left( 2 \Phi - T \right) + \ldots \; , \quad \frac{\partial}{\partial \lambda} T = - g^{rr} \partial^2_r T + \ldots
\ee
with $\ldots$ being lower derivatives, and hence the conformal factor $\Phi$ flows forwards and the class $T$ back as we expect. We solve the system for $\Phi, T, F$ in a similar manner as described above for the sphere.
In Appendix~\ref{App:Sch} we give numerical examples of finite flows near Schwarzschild, both for small and large black holes, where we prescribe $\Phi$ at the initial flow time and $T$ at the final time.  Further we consider flows to and from a Schwarzschild fixed point, and show that for flows that  asymptote from small black hole fixed point in the past, the future conformal class given by $T$ must be tuned as discussed above.

Here we consider whether there exist  flows that connect a small black hole to a large black hole, as there are for Ricci flow. 
Since the action increases along the Einstein flow, if such a flow exists it must be in the direction of flowing from the large black hole to the small one.
If such a flow exists, then since one can always translate the flow in flow time, there should be a one parameter family of flows.

Suppose we attempt to construct such a flow by taking a finite flow from time $\lambda_0$ to $\lambda_1$ and taking the initial data as the conformal factor for the large black hole, and the final data as the class for the small black hole, and then take $\lambda_1 = - \lambda_0 \to \infty$ --- where would that one parameter that we expect arise?
From the discussion above there should precisely be one more piece of data to fix, namely the GPY mode at the future small black hole fixed point. Hence giving this data, the constant $a$ above in equation~\eqref{eq:negmodeflow} where now the background is the small black hole, will give the one parameter family of flows. Indeed in the far future, scaling $a$ can be reabsorbed into a translation of flow time. As with the Ricci flow case, the existence of such a flow from the large to the small black hole is again due to the GPY mode.
\begin{figure}[t]
    \centering
    \subfigure[Evolution of $r_T$ and $r_A$]{%
        \includegraphics[width=0.7\textwidth]{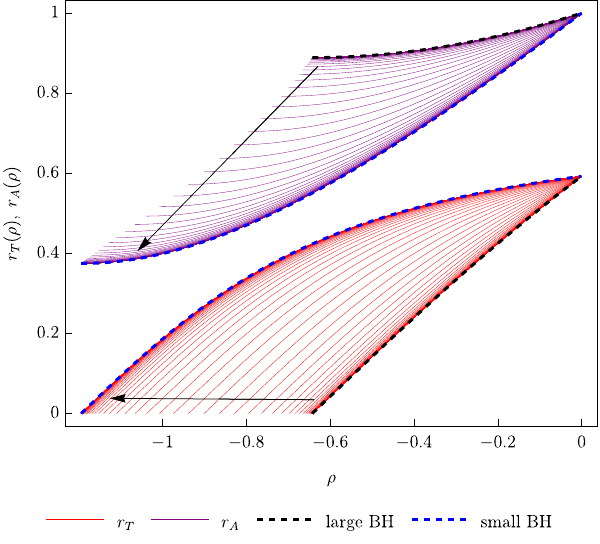}%
        \label{fig:Einstein_BH_Transition_radii}%
    }
    \hfill
    \subfigure[Evolution of the action]{%
        \includegraphics[width=0.6\textwidth]{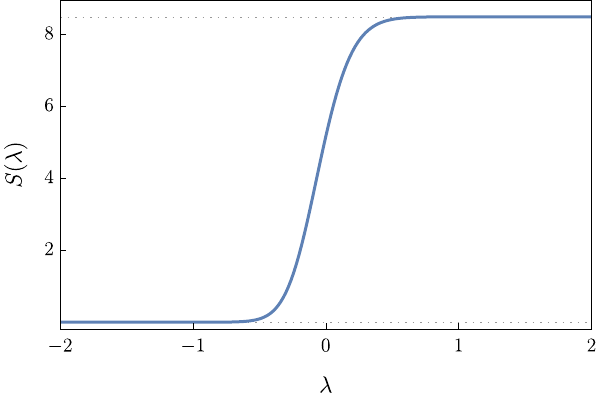}%
        \label{fig:Einstein_BH_Transition_action}%
    }
    \caption{The Einstein flow from the large to the small black hole for Schwarzschild in a cavity of radius $R=1$ at the Hawking--Page temperature $\beta = 32\pi/27$. In (a) we plot the Euclidean time circle radius $r_T$ and the sphere radius $r_A$ against the proper distance $\rho$ from the cavity wall; in (b) the action along the flow. The thin curves are equally spaced in flow time, every $\Delta\lambda = 0.05$ over the range $-2 < \lambda < 2$.}
    \label{fig:Einstein_BH_Transition}
\end{figure}
Numerically we are able to find such flows, starting with the data described above. In Figure~\ref{fig:Einstein_BH_Transition} we present an example flow for the choice of $\beta = 32 \pi/27$, which is the Hawking--Page transition temperature for our choice of units $R=1$. The small black hole to  which the flow asymptotes then corresponds to the Schwarzschild solution above in~\eqref{eq: BH metric} with $r_0 = \frac{1}{18} (1 + \sqrt{33}) \simeq 0.375$. We wish to choose this as the asymptotic final data for the flow, corresponding to taking this value of $r_0$ and setting $\Phi = T = F =0$ as $\lambda \to + \infty$. The  large black hole of the same temperature which gives the  fixed point from which the flow originates has a non-trivial profile for $\Phi$ and $T$ in our ansatz~\eqref{eq: BH metric with T and Phi} with that value of $r_0$ and this depends on the choice of gauge through the function $F$.  We have chosen to present the flow for this temperature as this matches the Ricci flow shown in the work~\cite{Headrick_2006}. We have also presented the figures in a similar style to those of that work, namely plotting the size of the Euclidean time circle and sphere against proper distance from the cavity wall.

In order to construct this flow we compactify the flow time, choosing our computational flow time $y$ such that,     
\be
\lambda(y) = \frac{c\, y}{1 - y^2}
\ee
so that  taking  $y \in [-1,1]$ corresponds to $\lambda \in (-\infty, \infty)$. The constant $c$ controls the resolution around $y \sim 0$ where the transition will occur, and for the data presented we have chosen $c = 1.2$.
As described above, having chosen $r_0 = \frac{1}{18} (1 + \sqrt{33})$ in our metric ansatz then at $y = +1$ we set $\Phi$, $T$ and $F$ to vanish, imposing the geometry to be the small black hole at the Hawking--Page temperature.
For the choice of gauge $F = 0$ the corresponding large black hole  does not have an obvious analytic form, and we deduce $\Phi$ and $T$ for it  by numerically solving the fixed point condition in this gauge. 
Hence at $y = -1$ we impose this data.

Since the flow asymptotes to fixed points in the past and the future, and the equations are invariant under translations of flow time, any translation of a solution is again a solution. To obtain a unique solution we fix this translation freedom with a single condition in the interior of the flow: the $T$ flow equation at one grid point $(\lambda_*, r_*)$ is replaced by the condition $T(\lambda_*, r_*) = T_*$, with $T_*$ lying between its values at the two black holes, which places the transition near $\lambda \simeq 0$. 
This plays the role of the constant $a$ above. One could instead fix the GPY amplitude at a large but finite flow time, as in the truncated flow described above; this is equivalent in principle, but numerically poorly conditioned, since that amplitude is exponentially small and so only weakly determines where the transition occurs.
In practice we choose to fix $T$ at the horizon, so $r_* = 0$, at the intermediate flow time $\lambda_* = 0$, replacing the Neumann condition.\footnote{We do this at a grid point closest to $\lambda_* = 0$ if one does not lie exactly at that value.} 
In the past the value of $T$ at the horizon is given by the large black hole solution, $T \simeq -2.07$, and in the future it asymptotes to zero for the small black hole. Thus we select an intermediate value for $T_*$, and for the data presented we have chosen $T_* = -1$. Having replaced the Neumann condition at this point, we subsequently verify that it is accurately satisfied in the resulting solution to the same precision as we are solving the flow equations to.
The computational domain $(y,r) \in [-1,1]\times[0,1]$ is discretised in the same way as for the de Sitter flows above, and the flow is solved in the same manner, with a solution at resolution $60 \times 50$ presented here. Since the approach to both fixed points is exponential in flow time, the transition is confined to a narrow window around $\lambda = 0$, and we plot the flow over $-2 \le \lambda \le 2$.

\begin{figure}[t]
    \centering
    \includegraphics[width=0.75\textwidth]{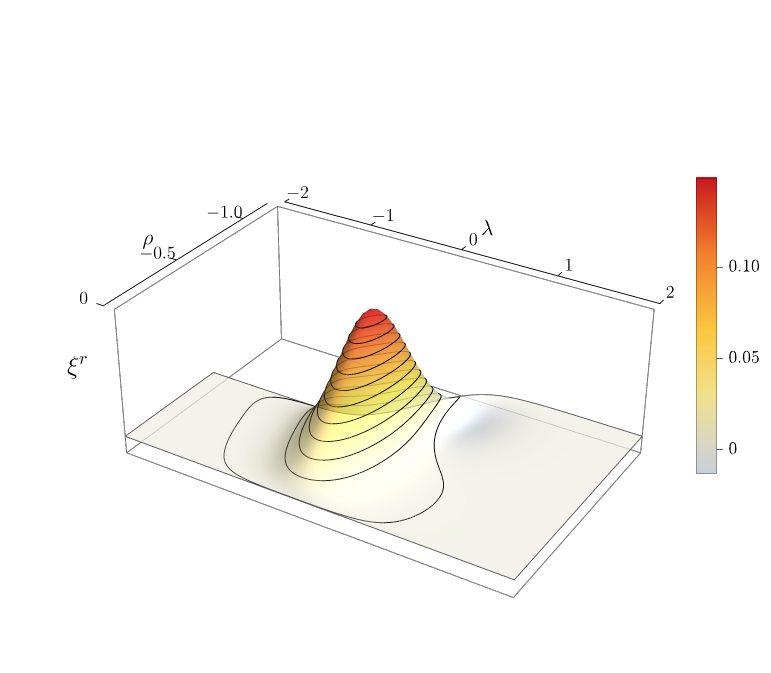}
    \caption{The radial component of the diffeomorphism vector, $\xi^r = r F$, along the flow from the large to the small black hole of Figure~\ref{fig:Einstein_BH_Transition}. Notably $\xi^r$ vanishes at the cavity wall and at both $\lambda$ endpoints.}
    \label{fig:Einstein_BH_Transition_Diffeo}
\end{figure}

In Figure~\ref{fig:Einstein_BH_Transition} we show the flow of the Euclidean time circle radius $r_T = 2\sqrt{1 - r_0} r_0 r e^{(\Phi + T)/2}$ and the sphere radius $r_A = \frac{r_0 e^{(\Phi - T)/2}}{1 - (1 - r_0)r^2}$ against the proper distance $\rho(r) = - \int_r^1 dr \sqrt{g_{rr}}$ measured relative to the position of the cavity wall, and increasing towards it. We find the flow from the large black hole solution to the small which is qualitatively similar to the Ricci flow solution found in~\cite{Headrick_2006}. We also plot the action along the flow and we see that this asymptotes in the past to zero, since we have chosen the Hawking--Page temperature so the large black hole has the same vanishing on-shell action as hot flat space, and then increases monotonically to the value for the small black hole asymptotically in the future of the flow.
In Figure~\ref{fig:Einstein_BH_Transition_Diffeo} we show $\xi^{r} = r F$ plotted against the proper radius $\rho(r)$ along the flow. Indeed, we see that this vanishes at both the initial and final flow times as we have required for the large and small black hole fixed points. Further we see it also vanishes at the cavity wall as imposed by our boundary conditions.

\section{A Lefschetz flow for Euclidean gravity}\label{sec:Lefschetz}

We now turn to the main aim of this paper, namely to give a Lefschetz flow for  Euclidean semiclassical gravity. 
This is a flow of complex metrics that is a gradient ascent of the real part of the Einstein--Hilbert action, and leaves the imaginary part unchanged. 
Starting the flow at a fixed point, a saddle point of the Einstein--Hilbert action, we will specify final data at a finite flow time. The flow solution then gives a path in the space of complex metrics which lies in the Lefschetz thimble of that saddle point. The collection of all such flows gives the Lefschetz thimble of that saddle.

The underlying manifold remains real. The metric tensor is extended to have complex components, which locally are functions of the real coordinates, and the Einstein--Hilbert action is then analytically continued as a function of this complex metric. We will discuss the first such Lefschetz flow for a general complex metric which was given earlier in equation~\eqref{eq:ELflow} in the Introduction. This flow reduces to the Einstein flow we have just discussed in the case that the metric is real. Hence we term this the `Einstein--Lefschetz' flow.

A subtlety is that the Einstein--Hilbert action in equation~\eqref{eq:EHaction} must be analytically continued as we extend the metric $g_{\mu\nu}$ to be complex. 
While the measure $\sqrt{\det{g_{\mu\nu}}}$ has a branch cut when thought of as a complex function of the components  $g_{\mu\nu}$, 
since we work with the exponential parameterization of the metric  in~\eqref{eq:ELmetric} given in the Introduction, this defines the analytic extension.
More generally it is unclear whether we should allow all complex metrics or rather place some restrictions on how `complex' they can be. The KSW condition arises from the requirement that QFT for fundamental fields which are $p$-forms can be defined for a complex metric on a real manifold. Considering `$0$-forms' yields the condition $\Re\left( \sqrt{\det(g_{\mu\nu})} \right) > 0$ which in fact implies that $\det{g_{\mu\nu}} \notin (-\infty,0]$ so KSW would not allow the square root branch cut in the measure to be crossed.

Since the Einstein--Lefschetz flow reduces to the Einstein flow for real metrics, then if the metric is real, the condition on the metric that the flow acts as a forward-backward flow for the conformal factor and class is simply that the metric is Riemannian. We shall later derive the condition that the flow remains such a forward-backward one when the metric becomes complex. As we shall see, the flow requires boundary conditions, and  we have described these in the Introduction, but intuitively starting at a fixed point one fixes the real conformal class and imaginary conformal factor at the final flow time.
These boundary conditions require the complex metric to satisfy the condition in~\eqref{eq:ELconditions}, namely that $\Re\left( \left( \sqrt{\det(g_{\mu\nu})} g^{\mu\nu} q_\mu q_\nu \right)^{1/2} \right) > 0$ for all real non-zero $q_\mu$. 
This is a weaker condition than the KSW $1$-form condition which requires $\Re\left( \sqrt{\det(g_{\mu\nu})} g^{\mu\nu} q_\mu q_\nu \right) > 0$ for all such $q_\mu$.
Thus thimbles constructed by our flow will generally leave the space of KSW allowable metrics although the interpretation of this is unclear as these are not on-shell geometries~\cite{witten2022notecomplexspacetimemetrics}.

\subsection{An Einstein--Lefschetz flow for complex metrics}

On our real manifold we now continue our metric to be a complex symmetric tensor, $g_{\mu\nu}$, such that its complex conjugate components are $\bar{g}_{\mu\nu}$. We analytically continue the Einstein--Hilbert action in~\eqref{eq:EHaction} so that, given we satisfy~\eqref{eq:ELconditions}, it is a holomorphic function of the complex metric, and hence does not depend on the conjugate metric; $\partial S/\partial \bar{g}_{\mu\nu} = 0$. 
The generalization of the Lefschetz flow for a scalar field, as discussed in Section~\ref{sec:phi4}, to a complex metric is the following,
\be
\frac{\partial}{\partial \lambda} g_{\mu\nu} = \mathcal{G}_{\mu\nu\rho\sigma}^{-1}(g, \bar{g}) \,\overline{\frac{\delta S}{\delta g_{\rho\sigma}}} + \nabla_{(\mu} \xi_{\nu)}
\ee
where we have added a complex vector field $\xi_\mu$ that generates a flow dependent complex diffeomorphism along the flow. If there are boundaries to the manifold we should require this to vanish there.
Then the action varies along the flow as,
\be
\frac{\dd S}{\dd \lambda} =  \int \dd^n x\, \mathcal{G}_{\mu\nu\rho\sigma}^{-1}(g, \bar{g})  \frac{\delta S}{\delta g_{\mu\nu}} \overline{\frac{\delta S}{\delta g_{\rho\sigma}}} 
\ee
where we note that the diffeomorphism term gives no contribution.
Then provided  $\mathcal{G}_{\mu\nu\rho\sigma}^{-1}$ gives a positive definite inner product on complex symmetric tensors,
the integrand will be real and positive. Hence  the  real part of the action increases and its imaginary part remains constant, as we require for a Lefschetz flow.

Saddle points of the Einstein--Hilbert action will be fixed points of this flow.
The choice of superspace metric, $\mathcal{G}_{\mu\nu\rho\sigma}^{-1}(g, \bar{g})$, will yield different flows from a fixed point, and correspondingly will construct different Lefschetz thimbles for a saddle point.
The choice we make here is similar to that for the Einstein flow, where we replace the metric in~\eqref{eq:Einsteinmetric} with a real Riemannian metric $\mathfrak{g}_{\mu\nu}$, which we term the real reference metric, giving,\footnote{
Just as for the real flows considered earlier, we could modify the superspace metric by a trace and obtain flows with similar properties. Referring back to Section~\ref{sec:ascent}, then the Einstein flow is given by taking $\alpha = 1$, but flows with $\alpha$ close to one will have similar gradient ascent properties, and remain forward-backward flows for the conformal factor and class.
In our complex setting we might take,
\be
\mathcal{G}_{\mu\nu\rho\sigma}^{-1}(x; \beta) =  \frac{4}{\sqrt{\det{\mathfrak{g}}}} \left( \mathfrak{g}_{\mu\rho}\,\mathfrak{g}_{\nu\sigma} - \beta \mathfrak{g}_{\mu\nu}\, \mathfrak{g}_{\rho\sigma}\right)
\ee
for some real $\beta$ close to zero. Provided this superspace metric is positive definite then this will preserve that $d S/d\lambda > 0$ along the flow, and we can expect the flow to have similar properties to the Einstein--Lefschetz case. We note this freedom but will not explore it further here. 
}
\be
\label{eq:PLmetric}
\mathcal{G}^{-1}_{\mu\nu\rho\sigma}( g, \bar{g} ) =   \frac{4}{\sqrt{\det{\mathfrak{g}}}} \, \mathfrak{g}_{\mu\rho}\,\mathfrak{g}_{\nu\sigma} 
\ee
which yields the Einstein--Lefschetz flow equation given in equation~\eqref{eq:ELflow} of the Introduction.
Computing the variation of the action along the flow we find no contribution from the diffeomorphism term, and obtain,
\be
\frac{\dd S}{\dd \lambda} = 
 \int \dd^n x \sqrt{ \frac{ \det{g} \det{\bar{g}}  }{ \det{\mathfrak{g}}} }\, \mathfrak{g}_{\mu\rho}\,\mathfrak{g}_{\nu\sigma}  \left( G^{\mu\nu} + \frac{(n-2) \Lambda}{2}  g^{\mu\nu} \right)  \overline{  \left( G^{\rho\sigma} + \frac{(n-2)\Lambda}{2}  g^{\rho\sigma} \right) } \ge 0
\ee
which is indeed real and positive given that $\mathfrak{g}_{\mu\nu}$ is Riemannian, and we note the measure, $| \det{g} | / \sqrt{ \det{\mathfrak{g}} }$, is real and positive and also correctly transforms as a tensor density with the correct weight. Hence this indeed defines a Lefschetz flow for complex metrics. We note that the same argument that there can exist no stationary soliton solutions for the Einstein flow applies here too; at a fixed point $dS/d\lambda = 0 $, and then the  above equation implies the Einstein condition $G^{\mu\nu} + \frac{(n-2) \Lambda}{2}  g^{\mu\nu}  = 0$ must hold.

Taking $\xi_\mu = 0$, so setting the diffeomorphism term to zero, then the flow is geometric being covariant under the real manifold diffeomorphisms. Given a flow solution, then a real diffeomorphism of this will yield another flow solution. 
While the action is invariant under an infinitesimal complex diffeomorphism, $g_{\mu\nu} \to g_{\mu\nu} + (\mathcal{L}_v g)_{\mu\nu}$, for complex covector $v_\mu$, the superspace metric does not transform covariantly, since it involves the real reference metric, and nor does $\overline{\delta S/\delta g_{\mu\nu}}$, due to the complex conjugation.
Thus a complex diffeomorphism of a flow solution does not yield another flow solution in general. The exception to this is  for linear perturbations of a fixed point, where the terms involving the real reference metric that do not transform covariantly do not appear as the background is Einstein, so the flow is invariant under linear complex diffeomorphisms.  

Our complex vector field $\xi^\mu$ allows us to construct flows in different complex coordinates. For real $\xi^\mu$ these are the same flow up to a real coordinate transformation. However in general for complex $\xi^\mu$ these flow solutions do not differ by a complex diffeomorphism, but are geometrically different. Thus non-perturbatively the flow equation is not covariant under complex diffeomorphisms. Different complex gauge choices will lead to different trajectories in the space of complex metrics away from a fixed point.

If we require a unique solution, as for example in a numerical analysis, then we must fix the complex gauge, not just the real one. One way to do this would be to use a vector field $\xi_\mu$ analogous to the DeTurck vector to break the complex diffeomorphism invariance. However another method, similar to that discussed above for the real Einstein flow, is simply to fix a complex gauge in the metric, and then solve the flow equation for the necessary complex vector $\xi_\mu$ that achieves this gauge. That is the approach we take later in our examples. 

What should we pick for the real reference metric? In principle any Riemannian metric will do, and different choices will lead to different flows from a saddle point, and hence will construct different Lefschetz thimbles. A key requirement that we impose is:
 \be
 \label{eq:requirement}
 \mathfrak{g}_{\mu\nu} \to g_{\mu\nu} \quad \mathrm{as} \quad \Im\left( g_{\mu\nu} \right) \to 0
 \ee
 so that for real metrics this flow is simply the Einstein flow of the previous section. 
  One choice which is the simplest, remaining linear in the metric is $\mathfrak{g}_{\mu\nu} = \Re\left( g_{\mu\nu} \right)$.
However we will shortly find that there is a more natural choice which we will employ and which we have detailed in equation~\eqref{eq:ELreference} of the Introduction.\footnote{We note however that if one is exploring the thimble of a real saddle point, another natural candidate for $\mathfrak{g}$ is to take the saddle metric; however this lacks the nice property that for real metrics the flow simplifies to the Einstein flow, and we do not consider this further here.} 

\subsection{Character and boundary conditions near real metrics}

We now consider the character of the Einstein--Lefschetz flow.
When the metric is real, since the flow reduces to the Einstein one, then for real perturbations of the metric the conformal factor flows forwards and the class backwards as discussed above. 
To understand how the imaginary part of the metric flows, we 
linearize about a real Riemannian metric $g_{\mu\nu}$ with $\delta g_{\mu\nu}$ being a complex metric perturbation. Then to leading order we have that $\mathfrak{g}_{\mu\nu} = {g}_{\mu\nu}$ so that the additional factors involving the real reference metric drop out in the highest derivative terms, and then the complex linearized flow is simply,
\be
\label{eq:pertELflow}
\frac{\partial \delta g_{\mu\nu}}{\partial \lambda} = 2 \, \overline{ \delta G_{\mu\nu}  } + \ldots
\ee
where $\ldots$ are terms of lower derivative order, including the diffeomorphism term. We further decompose the perturbation as,
\be
\label{eq:complexpert}
\delta g_{\mu\nu} = \delta \phi {g}_{\mu\nu} + \delta g^{(T)}_{\mu\nu} + \mathrm{diffeomorphism\,terms}
\ee
with $\delta g^{(T)}_{\mu\nu}$ traceless as for when we considered the real Einstein flow, although now $\delta \phi$ and $\delta g^{(T)}_{\mu\nu}$  are complex. Then to understand the character of the flow we consider the most relevant terms, those with the highest derivatives, and obtain,
\be
\frac{\partial \delta \phi}{\partial \lambda} =  (n-2) \overline{{\partial}^2 \delta \phi} + \ldots \; , \quad \frac{\partial \delta g^{(T)}_{\mu\nu}}{\partial \lambda} = - \overline{{\partial}^2 \delta g^{(T)}_{\mu\nu}} + \ldots 
\ee
where $\ldots$ represent terms of lower order in derivatives.
While a real perturbation evolves as for the real Einstein flow, we see the linearized operator is antilinear, and this implies that for an imaginary perturbation the flow is in the opposite sense. Explicitly decomposing into real and imaginary parts we obtain the real flows,
\be
\frac{\partial \Re(\delta \phi)}{\partial \lambda} &=& + (n-2) \partial^2 \Re(\delta \phi) + \ldots \; , \quad \frac{\partial \Im(\delta \phi)}{\partial \lambda} =  - (n-2) \partial^2 \Im(\delta \phi) + \ldots\nl
\frac{\partial \Re(\delta g^{(T)}_{\mu\nu})}{\partial \lambda} &=& - \partial^2 \Re(\delta g^{(T)}_{\mu\nu})+ \ldots \; , \quad \frac{\partial \Im(\delta g^{(T)}_{\mu\nu})}{\partial \lambda} =  + \partial^2 \Im(\delta g^{(T)}_{\mu\nu}) + \ldots\; .
\ee
Thus the imaginary conformal factor flows backwards and the imaginary class flows forwards. 

Moving away from a real Euclidean metric the perturbations become more complicated, and we will describe them in detail in what follows. However for metrics sufficiently close to real ones, we see that we require a choice of boundary conditions for the flow that fix the real conformal factor perturbation at the initial flow time and the real class perturbations at the final time -- just as for the real Einstein flow -- and conversely fix the imaginary conformal class perturbation at the initial time and imaginary conformal factor perturbation at the final time.

\subsection{Perturbative flows away from real fixed points and the GHP prescription}

We are now in a position to discuss integration cycles for the Euclidean path integral. Given a real fixed point, so a real Einstein geometry, the flows away from that fixed point taken together construct the Lefschetz thimble for it. We consider perturbative Einstein--Lefschetz flows emanating from a real fixed point and argue that they correctly reproduce the GHP prescription.

For a real fixed point geometry we have that complex perturbations flow as in equation~\eqref{eq:pertELflow}. We decompose the complex perturbation as $\delta g_{\mu\nu} = \phi {g}_{\mu\nu} + a^{(TT)}_{\mu\nu}$ on the real background fixed point geometry ${g}_{\mu\nu}$, and $\phi$ and $a^{(TT)}_{\mu\nu}$ are now complex, with the latter being transverse and traceless again as $\nabla^\mu a^{(TT)}_{\mu\nu} = 0$ and $g^{\mu\nu} a^{(TT)}_{\mu\nu} = 0$. For an appropriate complex diffeomorphism term, $\xi_\mu = (n-2) \partial_\mu \overline{\phi}$, we obtain,
\be
\label{eq:linEinsteinComplex}
\frac{\partial}{\partial \lambda}  \phi &=& (n-2)  \left( {\nabla}^2 + \Lambda \right) \overline{\phi  } \; , \nl
\frac{\partial}{\partial \lambda} a^{(TT)}_{\mu\nu} &=&    - {\nabla}^2  \overline{a^{(TT)}_{\mu\nu}}  - 2 {R}_{\mu~~\nu}^{~~\alpha~\beta} \overline{a^{(TT)}_{\alpha \beta}  } \; .
\ee 
For now let us suppose for simplicity that we have a manifold with no boundary -- for example that of the Euclidean sphere or asymptotically flat Euclidean Schwarzschild. The flow solutions for the conformal perturbation, $\phi$, and for the transverse-traceless perturbation, $a^{(TT)}_{\mu\nu}$, decouple.
We decompose these complex solutions into real and imaginary parts as,
\be
\phi = \phi^{(+)} + i \phi^{(-)} \; , \quad a^{(TT)}_{\mu\nu} = a^{(+)}_{\mu\nu} + i a^{(-)}_{\mu\nu}
\ee
to yield a pair of real flows for the real and imaginary parts of the conformal mode,
\be
\frac{\partial}{\partial \lambda}  \phi^{(\pm)} &=& \pm (n-2)  \left( {\nabla}^2 + \Lambda \right) \phi^{(\pm)} = \mp (n-2) \mathcal{O} \phi^{(\pm)}
\ee
and a real pair of flows for the real and imaginary parts of the transverse-traceless perturbation,
\be
\frac{\partial}{\partial \lambda} a^{(\pm)}_{\mu\nu} &=&  \pm \left(  - {\nabla}^2 a^{(\pm)}_{\mu\nu}  - 2 {R}_{\mu~~\nu}^{~~\alpha~\beta} a^{(\pm)}_{\alpha\beta} \right) = \pm \mathcal{O}_{\mu\nu}^{~~~\alpha\beta} a^{(\pm)}_{\alpha\beta} \; .
\ee 
For the real parts ($+$) this is simply the real Einstein flow  in~\eqref{eq:linEinstein}. For the imaginary parts the flow is the backwards Einstein flow. 

Assume for the moment that the fixed point is stable in the sense that the operators $ \mathcal{O} = - {\nabla}^2 - \Lambda$ and $\mathcal{O}_{\mu\nu}^{~~~\alpha\beta}$ are positive, and assume there are no zero modes. Since we want the flow to emanate from the fixed point, we must have $\delta g_{\mu\nu} \to 0$ as $\lambda \to - \infty$ and this implies that $ \phi^{(+)} = a^{(-)}_{\mu\nu} = 0$ for all flow time -- if they are non-zero at any time, they would become larger to the past. Thus the flow of a perturbation from a stable fixed point must take the form;
\be
\label{eq:pertthimble}
\phi(\lambda) = i \phi^{(-)}(\lambda) \; , \quad a^{(TT)}_{\mu\nu}(\lambda) = a^{(+)}_{\mu\nu}(\lambda)
\ee
having only a real class perturbation and imaginary conformal factor. Both of these flow backward, and hence we should specify the solution by giving the data for these at the final flow time $\lambda_1$, so $\phi^{(-)}(\lambda_1)$, $a^{(+)}_{\mu\nu}(\lambda_1)$. This data will then decay to the past, and approach zero as $\lambda \to - \infty$, and hence the flow tends to the fixed point as we require. There is no restriction on this final data, except that it is small enough to ensure it may be treated as a perturbation. The collection of all these flows then generates the portion of the thimble for the fixed point which is perturbatively close to the fixed point, and is given by all real class perturbations, $a^{(+)}_{\mu\nu}$, and all imaginary conformal factor perturbations, $i \phi^{(-)}$. This is precisely the same as the GHP prescription,
namely that to obtain the integration cycle one should Wick rotate the conformal perturbation modes to be imaginary.

Following the discussion in Section~\ref{sec:EinsteinFixed}, a real fixed point may be unstable so that it possesses  growing modes that are not controlled by the final boundary conditions, namely here, fixing the final real conformal class perturbation and imaginary conformal factor perturbation. In this case there may be an additional finite set of data to specify too. 
In the setting where the conformal mode and transverse-traceless sector decouple, then this will occur if either of the operators $ \mathcal{O}$ and $\mathcal{O}_{\mu\nu}^{~~~\alpha\beta}$ has negative eigenmodes, meaning that the real part of their eigenvalue is negative. Let us focus on the simplest case where these negative mode eigenvalues and eigenvectors are real.\footnote{We note that there may be situations where eigenvalues and hence eigenmodes are complex~\cite{Marolf_2022}.}
If $\mathcal{O}$ has negative modes, then there will be real conformal modes that grow, and imaginary conformal modes that decay. Such unstable imaginary conformal modes grow to the past and will destabilize the flow there. Hence the final imaginary conformal mode perturbation must be tuned to ensure these unstable modes are not present in the far past of the flow. Conversely the unstable growing real conformal modes will not be fixed by the final boundary condition which fixes instead the imaginary conformal mode. Hence, as described in Section~\ref{sec:EinsteinFixed}, we must supply the extra data at the fixed point to dictate how much of these growing real conformal modes are excited in the flow.
The situation for negative modes of $\mathcal{O}_{\mu\nu}^{~~~\alpha\beta}$ is analogous. They imply that there are real transverse-traceless modes that decay, and these imply the final real conformal class must be tuned so that the flow can tend to the fixed point in the far past. They also imply there are imaginary  transverse-traceless modes that grow, so that these must be specified as additional data associated to the fixed point to fully specify the flow.

As a concrete example let us consider the asymptotically flat Euclidean Schwarzschild fixed point with $\Lambda = 0$. Perturbatively there is the real GPY negative mode  of the Lichnerowicz operator $\mathcal{O}_{\mu\nu}^{~~~\alpha\beta}$ with Einstein flow $a^{(GPY)}_{\mu\nu}(\lambda)$ such that it exponentially grows in the past. Then the flow must have no negative mode component, which implies a codimension one constraint on the final data for the real class. On the other hand, we may now excite a single imaginary class mode, given by the flow time reverse of the real $a^{(GPY)}_{\mu\nu}(\lambda)$ multiplied by $i$. Hence for Schwarzschild, perturbatively near the fixed point we will have purely imaginary $\phi(\lambda)$ as above, but,
\be
\label{eq:pertthimbleSch}
a^{(TT)}_{\mu\nu}(\lambda) = a'^{(+)}_{\mu\nu}(\lambda) + i a^{(GPY)}_{\mu\nu}(-\lambda) \; ,
\ee
where $a'^{(+)}_{\mu\nu}(\lambda)$ is a general solution to the real Einstein flow with no GPY component so that it decays. Since $a^{(GPY)}_{\mu\nu}(-\lambda)$ also decays to the past, this flow indeed tends to the Schwarzschild fixed point as $\lambda \to - \infty$.
Hence for Schwarzschild our construction yields the GHP prescription extended to Wick rotate the conformal factor {\it and} the single transverse-traceless GPY negative mode.
Another example is the Euclidean sphere solution. As discussed above there is an unstable conformal perturbation which is homogeneous. This results in an integration cycle where one Wick rotates the conformal factor {\it except} for the homogeneous mode,
\be
\label{eq:pertthimbleSphere}
\phi(\lambda) = i \phi'^{(-)}(\lambda) + c(\lambda) \; ,
\ee
where $c(\lambda)$ is the real homogeneous mode, which decays to the past, and $\phi'^{(-)}$ has no homogeneous component.

Thus we see that broadly our flow agrees with the perturbative prescription of GHP. 
However as Marolf and Santos have emphasized, on manifolds with boundary the situation is more subtle as the conformal and transverse-traceless perturbations may be coupled together by boundary conditions. An example is the case of black holes in a cavity with Dirichlet boundary conditions~\cite{Marolf_2022}. They propose to modify the GHP prescription so that given an eigenmode, $h_{\mu\nu}$, of the Lichnerowicz operator -- the linearization of the Ricci tensor condition for a fixed point, so $\delta (R_{\mu\nu} - \Lambda g_{\mu\nu} )$ -- then if the norm is negative with respect to the local DeWitt metric in~\eqref{eq:dewitt},  it is Wick rotated. In the simple setting where there is no coupling between the conformal modes and transverse-traceless modes, this would mean the conformal modes are rotated as for GHP. 

Our prescription appears to differ from that of Marolf and Santos as it is not phrased directly in terms of the norm of a mode, and does not refer to the Lichnerowicz operator. 
Perturbatively near a real fixed point we may simply summarize our rule for Wick rotation. Given the linearization of the Einstein condition, 
\be
\label{eq:linEinsteinOp}
\mathcal{L}_G \delta g_{\mu\nu}= \delta \left( G_{\mu\nu} + \frac{n-2}{2} \Lambda g_{\mu\nu} \right) \; , 
\ee
which defines the real operator $\mathcal{L}_G$, then perturbatively near a fixed point the Einstein--Lefschetz flow is
\be
\frac{\partial \delta g_{\mu\nu}}{\partial \lambda} = 2 \mathcal{L}_G ( \overline{ \delta g_{\mu\nu}} ) \; .
\ee
Given a real eigenfunction, $h_{\mu\nu}$ of $\mathcal{L}_G$, up to a diffeomorphism that can be absorbed by $\delta \xi_\mu$, with non-zero real eigenvalue $\alpha \ne 0$, so that $\mathcal{L}_G h_{\mu\nu} = \alpha h_{\mu\nu} + \nabla_{(\mu} v_{\nu)}$ for some $v_\mu$,
then we may construct the perturbative flow solution,
\be
\delta g_{\mu\nu} =  \left( a e^{2 \alpha \lambda} + i b e^{- 2 \alpha \lambda} \right) h_{\mu\nu} 
\ee
where $a$ and $b$ are real constants. If $\alpha > 0$ then the flow emanating from the fixed point is given by $b = 0$, and hence perturbations $\delta g_{\mu\nu} =  a h_{\mu\nu} $ for all $a$ lie in the thimble. This corresponds to no Wick rotation of the mode $h_{\mu\nu}$. On the other hand for $\alpha < 0$ then the flow from the fixed point is $a = 0$, and consequently the perturbations $\delta g_{\mu\nu} =  i b h_{\mu\nu} $ for all $b$ lie in the thimble, and we see that the mode $h_{\mu\nu}$ is Wick rotated by a factor of $i$ to lie in the thimble.
Thus whether a (real) perturbation mode, which is an eigenfunction of the linearized Einstein equation, is Wick rotated or not depends only on its growth or decay under the Einstein flow about the fixed point, and hence its eigenvalue. Its norm plays no direct role, unlike for the Marolf--Santos prescription.

The eigenspectrum of the linearized Einstein equation~\eqref{eq:linEinsteinOp} is in general different to that of the linearized Ricci condition $\delta (R_{\mu\nu} - \Lambda g_{\mu\nu} )$. Since these differ by a trace, in the simple setting where conformal modes and transverse-traceless modes decouple, the transverse-traceless eigenfunctions are the same for both linearized operators. The conformal modes correspond too, although with modified eigenvalues. However in the more interesting case in~\cite{Marolf_2022} where the boundary conditions couple these modes, the eigenfunctions of the two operators are different in general, and hence there is no simple relation between their eigenvalues. This gives another difference with Marolf and Santos. This is illustrated explicitly in Appendix~\ref{app:negative} where we compute the unstable mode eigenvalue for small black holes for the linearized Einstein tensor, and compare it to that for the linearized Ricci tensor.

Having seen that our flow reproduces the expectations from the GHP prescription for perturbations of a real fixed point, we now wish to understand its non-perturbative behaviour.
Beyond perturbation theory, but still sufficiently close to a real fixed point so the metric is not `too' complex, we expect the flows starting at the fixed point to exist with suitable final flow time boundary conditions. In the far past the flow geometry is set to that of the fixed point. At the end point of the flow, we should fix boundary conditions that suitably  fix the real conformal class and imaginary conformal factor, and we will discuss these in detail shortly. While for small perturbations of a real fixed point the real and imaginary parts of the perturbation may decouple, this is not true non-linearly. Hence going beyond the perturbative regime where GHP applies, and taking an imaginary conformal deformation as final data for the flow, but no real class deformation, will result in a  complex flow. Non-linearity will excite both the real and imaginary parts of both the conformal factor and class at intermediate flow times.

\subsection{Non-perturbative character of the flow}
\label{sec: complex flow character}

We now consider in detail the character of the flow for complex backgrounds that may be far from a real geometry. 
In order to do this we must choose a prescription for the reference metric $\mathfrak{g}_{\mu\nu}$.

We now make a key restriction on our complex metric. Firstly we require it is invertible, and then we require that for all real non-zero $q_\mu$ we have,
\be
\label{eq:restriction}
\Re\left( \sqrt{ q_\mu H^{\mu\nu} q_\nu } \right) > 0 \; , \quad H^{\mu\nu} =  \left( \sqrt{\det(g_{\alpha\beta})} \right) g^{\mu\nu} 
\ee
which is the condition given in equation~\eqref{eq:ELconditions} in the Introduction.
We note that this is weaker than the KSW $1$-form condition which instead requires $\Re\left(  q_\mu H^{\mu\nu} q_\nu  \right) > 0$.

Let us take a fixed real  frame on our manifold $\hat{\theta}^\mu_{a}$ and coframe $\hat{\theta}_\mu^{a}$  so that $\hat{\theta}^{a} \cdot \hat{\theta}_{b} = \delta^a_b$. Then we can write the metric in this coframe as,
\be
g_{\mu\nu} = \hat{\theta}^a_\mu g_{ab} \hat{\theta}^b_\nu  =  \vect{\hat{\theta}}_\mu \, \mat{g} \, \vect{\hat{\theta}}_\nu 
\ee
where we have introduced matrix notation for the frame indices, so for example $g_{ab} \leftrightarrow \mat{g}$, $ \hat{\theta}^a_\mu  \leftrightarrow   \vect{\hat{\theta}}_\mu$. 
Then given the restriction~\eqref{eq:restriction} on the complex metric, we may find real symmetric matrix valued functions $\mat{X}$ and $\mat{Y}$ such that,
\be
\label{eq:metricform}
\mat{g} = \mat{A} \mat{A}^T = e^{\frac{1}{2} \mat{X}} e^{i \mat{Y}}  e^{\frac{1}{2} \mat{X}}  \; , \quad \mat{A} = e^{\frac{1}{2} \mat{X}} e^{\frac{i}{2} \mat{Y}} \; .
\ee
Furthermore, given $\mat{g}$ then the decomposition returns a unique $\mat{X}$.
The argument that the metric may be written in this manner is detailed in Appendix~\ref{app:metricform} and relies on the restriction~\eqref{eq:restriction}.
We then choose the reference metric $\mathfrak{g}_{\mu\nu} =  \vect{\hat{\theta}}_\mu \, \mat{\mathfrak{g}} \, \vect{\hat{\theta}}_\nu$  to be,
\be
\mat{\mathfrak{g}} = \mat{A} \overline{ \mat{A} }^T = e^{\mat{X}} \; .
\ee
We emphasize that since $\mat{X}$ is real and symmetric, $\mathfrak{g}_{\mu\nu}$ is then a Riemannian metric as we previously required, and furthermore it correctly equals $g_{\mu\nu}$ when $\mat{Y}$ vanishes so that the metric is real. Given a complex metric $\mat{g}$ satisfying~\eqref{eq:restriction}, while $\mat{Y}$ is not unique, $\mat{X}$ is, and hence there is no ambiguity in this definition of the real reference metric, or indeed the resulting flow. We note that while giving $\mat{g}$ there is an ambiguity in the corresponding $\mat{Y}$, since the flow equation only depends on $\mat{g}$ and its curvature, and the real reference metric $\mat{\mathfrak{g}}$, the flow is not sensitive to this.

Linearizing about a complex background, to examine the character we will retain the highest derivative terms in the flow and coordinates, and also the diffeomorphism term. 
We again freeze the coefficients in the background, so we may regard the complex metric $g_{\mu\nu}$ and diffeomorphism vector field $\xi_\mu$, together with the real reference metric $\mathfrak{g}_{\mu\nu}$, as all being constant.
This linearized flow is,
\be
\frac{\partial}{\partial \lambda} \delta g_{\mu\nu} - {\partial}_{(\mu} \delta \xi_{\nu)} =  2  \sqrt{\frac{\det{\bar{g}}}{\det{\mathfrak{g}}}} \, \mathfrak{g}_{\mu\rho}\,\mathfrak{g}_{\nu\sigma}  \delta \overline{G}^{\rho\sigma} + \ldots
\ee
where $\ldots$ are lower derivative terms we discard and $\delta G^{\mu\nu}$ is the linearization of the Einstein tensor with indices raised. 
On this frozen background we decompose into Fourier modes, both in flow time and coordinates as,
\be
\delta g_{\mu\nu}(\lambda, x) = \delta g_{\mu\nu} \, \Re\left( e^{\omega \lambda} e^{i k_\mu x^\mu} \right)
\ee
with $k_\mu$ real and then consider a single mode.
Thus we replace $\partial_\mu \partial_\nu \to - k_\mu k_\nu$, and then,
\be
2 \delta {G}_{\mu\nu} &=&  {g}^{\alpha\beta} k_\alpha k_\beta \delta g_{\mu\nu} + {g}^{\alpha\beta} k_\mu k_\nu \delta g_{\alpha\beta} - 2 {g}^{\alpha\beta} k_\alpha k_{(\mu} \delta g_{\nu)\beta} \nl
&& \quad +  {g}_{\mu\nu} \left(  {g}^{\alpha\rho} {g}^{\beta\sigma}  k_\alpha k_\beta \delta g_{\rho\sigma}  -
{g}^{\alpha\beta} {g}^{\rho\sigma}  k_\alpha k_\beta \delta g_{\rho\sigma} 
 \right) \; .
 \ee
 Similarly we replace  $\partial / \partial \lambda \to \omega$.
Now defining the projector orthogonal to $k_\mu$, so,
\be
P^{\mu\nu} = g^{\mu\nu} - \frac{1}{k^2} {k}^\mu {k}^\nu 
\ee
where we emphasize that while $k_\mu$ is real, ${k}^\mu$ is generally complex since $k^2$ is, then we can write the linearized flow neatly as,
\be
\label{eq:linflow1}
\omega \delta g_{\mu\nu} - i k_{(\mu} \delta \xi_{\nu)} =  \overline{Q} \mathfrak{g}_{\mu\rho}\,\mathfrak{g}_{\nu\sigma} \overline{ \left( P^{\rho\alpha} P^{\sigma\beta}  - P^{\rho\sigma} P^{\alpha\beta}   \right) \delta g_{\alpha\beta} } + \ldots
\ee
where we have defined,
\be
\label{eq:Q}
Q = {k^2}  \sqrt{\frac{\det{{g}}}{\det{\mathfrak{g}}}} \; .
\ee
These projectors ensure that perturbations that are diffeomorphisms, so of the form $\delta g_{\mu\nu} = k_{(\mu} v_{\nu)}$, are projected out of the righthand side and hence evolve only due to the diffeomorphism term  given by $\delta \xi_{\mu}$.
A point we note for later is that since $Q \propto (\sqrt{g})  g^{\mu\nu} k_\mu k_\nu$, with $k_\mu$ and the other factors being real, then our restriction on complex metrics earlier in~\eqref{eq:restriction} implies that,
\be
\label{eq:Qcondition}
\Re(\sqrt{Q}) > 0 \; .
\ee
We have ${k}_\mu = {\vect{k}} \vect{\hat{\theta}}_\mu$ so ${\vect{k}}$ is real. We now introduce a covector $\tilde{k}_\mu = \tilde{\vect{k}} \vect{\hat{\theta}}_\mu$ such that,
\be
\tilde{\vect{k}} = \frac{1}{\sqrt{k^2}} \mat{A}^{-1} \vect{k}
\ee
so that $\tilde{\vect{k}} \cdot \tilde{\vect{k}} = 1$, noting that while $\vect{k}$ is real, $\tilde{\vect{k}}$ is not. Then using that $\mat{g}^{-1} = (\mat{A}^T)^{-1} \mat{A}^{-1}$ then we can write the components of $P^{\mu\nu} = \vect{\hat{\theta}}^\mu \mat{P} \vect{\hat{\theta}}^\nu$ in the frame as, 
\be
 \mat{P} = (\mat{A}^{T})^{-1} \left( 1  - \tilde{\vect{k}} \tilde{\vect{k}}^T \right) \mat{A}^{-1} \; .
\ee
We now construct a natural complex basis which is orthogonal with respect to the Euclidean dot product;
\be
\vect{n}^{(A)} = ( \tilde{\vect{k}}, \vect{v}, \vect{n}^{(2)}, \ldots, \vect{n}^{(n-1)} ) = ( \tilde{\vect{k}} , \vect{n}^{(I)} )
\ee
so that $\vect{n}^{(A)} \cdot \vect{n}^{(B)} = \delta^{AB}$, with the property that $\vect{n}^{(2)} , \ldots, \vect{n}^{(n-1)}$ are real vectors, whereas $\tilde{\vect{k}}$ and $\vect{v}$ are complex. In the last equality above we have written $\vect{n}^{(1)} = \vect{v}$ and introduced indices $I = 1,2,\ldots,(n-1)$.
Let us now detail this construction. Generally $\tilde{\vect{k}}$ is complex, so not real or imaginary. Then the real and imaginary parts of $\tilde{\vect{k}}$ define vectors,
\be
\label{eq:xydefn}
\vect{x} = \Re(\tilde{\vect{k}}) \; , \quad \vect{y} = \Im(\tilde{\vect{k}})
\ee
and we take the unit norm versions of these, $\hat{\vect{x}} = \vect{x} / \sqrt{\vect{x} \cdot \vect{x}}$ and similarly for $\hat{\vect{y}}$. If it is pure real, then it defines one of $\vect{x}$ or $\vect{y}$ and we may simply choose the other to be any real orthogonal vector.
Now we can write,
\be
\tilde{\vect{k}} = \alpha \hat{\vect{x}} + i \beta \hat{\vect{y}} 
\ee
with real $\alpha, \beta$ and since $\tilde{\vect{k}} \cdot \tilde{\vect{k}} = 1$, then $\hat{\vect{x}} \cdot \hat{\vect{y}} = 0$ and $1 = \alpha^2 - \beta^2$. Now taking,
\be
\vect{v} = - i \beta \hat{\vect{x}} + \alpha \hat{\vect{y}}
\ee
which has unit norm, this is orthogonal to $\tilde{\vect{k}}$. Together $\tilde{\vect{k}}$ and $\vect{v}$ span a two complex dimensional subspace of $n$ complex dimensions -- hence there are $(n-2)$ complementary directions whose basis can be chosen to be the real $\vect{n}^{(2)}, \ldots, \vect{n}^{(n-1)}$. 

Using this complex basis we now write our metric perturbation in our frame, in a general gauge, as,
\be
\label{eq:linflowmet1}
\delta \mat{g} =  \mat{A} \left(  c^{(A)} ( \tilde{\vect{k}} \vect{n}^{(A) T} + \vect{n}^{(A)} \tilde{\vect{k}}^T ) + c^{(IJ)} \vect{n}^{(I)}  \vect{n}^{(J) T}  \right) \mat{A}^T 
\ee
so that $c^{(A)}$ parameterizes the gauge freedom, which is projected out of the righthand side of~\eqref{eq:linflow1}, 
and we include a diffeomorphism $\delta \xi_\mu = \vect{\delta \xi} \vect{\hat{\theta}}_\mu$, with,
\be
\delta \vect{\xi} =  \xi^{(A)} \mat{A} {\vect{n}}^{(A)} \; .
\ee
Given our chosen form for the reference metric in equation~\eqref{eq:ELreference} then we have $\mat{A} = \mat{\mathfrak{g}} \overline{ (\mat{A}^T)^{-1} }$, and consequently after a little computation we find the flow can be written as,
\be
 \omega c^{{(IJ)}} \vect{n}^{(I)}  \vect{n}^{(J) T}  =  \overline{Q} \overline{\left(  c^{(IJ)}  - c^{(KK)} \delta^{IJ} \right)}  \bar{\vect{n}}^{(I)}  (\bar{\vect{n}}^{(J)})^T  + \frac{1}{2}  u^{(A)} \left( \vect{\tilde{k}} \vect{n}^{(A) T} + \vect{n}^{(A)} \vect{\tilde{k}}^T \right)
\ee
where the components $u^{(A)}$ depend on only $c^{(A)}$ and the diffeomorphism vector components $\xi^{(A)}$.

Projecting the equation above with $\vect{\tilde{k}}$ then determines the values of $u^{(A)}$. We note that while $\vect{\tilde{k}} \cdot \vect{n}^{(I)} = 0$ and $\vect{\tilde{k}} \cdot \overline{\vect{n}}^{(2,\ldots,n-1)} = 0$ we have $\vect{\tilde{k}} \cdot \overline{\vect{n}}^{(1)} \ne 0$. Having found $u^{(A)}$, this then determines the vector $\xi^{(A)}$ in terms of the data $c^{(IJ)}$ and the gauge choice $c^{(A)}$. 
We then consider the projection in the orthogonal directions to $\tilde{\vect{k}}$, projecting with $\vect{n}^{(I)}$. Our basis obeys the simple relations,
\be
\vect{n}^{(I)} \cdot \bar{\vect{n}}^{(J)} = a^{(I)} \delta^{IJ} \; , \quad a^{(I)} = ( 1 + 2 \beta^2, 1, 1, \ldots, 1) \; .
\ee
Defining the rescaled constants $\tilde{c}^{(IJ)}$ related to ${c}^{(IJ)}$ as,
\be
{c}^{(IJ)} =  \sqrt{\bar{Q}} \tilde{c}^{(IJ)} 
\ee
then we can write the remaining components of the flow equation as,
\be
\omega \, \tilde{c}^{(IJ)}  = a^{(I)} a^{(J)}  | Q |   \overline{\left( \tilde{c}^{(IJ)}  - \tilde{c}^{(KK)} \delta^{IJ}  \right)} 
\ee
and we emphasize that now $a^{(I)} a^{(J)}  | Q | $ are real.
Then decomposing $ \tilde{c}^{(IJ)} = p^{(IJ)} + i q^{(IJ)}$ into real and imaginary parts we can write the mode flow equation as an eigenvalue problem in the $n(n-1)$ dimensional vector $(p^{(IJ)} ,  q^{(IJ)})$;
\be
 \omega p^{(IJ)}  &=& + a^{(I)} a^{(J)}  | Q |   \left(  p^{(IJ)}  - p^{(KK)}  \delta^{IJ}  \right) \nl
 \omega q^{(IJ)}  &=& - a^{(I)} a^{(J)}  | Q |   \left( q^{(IJ)}  - q^{(KK)}  \delta^{IJ}    \right) \; .
\ee
This is straightforwardly solved for the $n(n-1)$ eigenvectors. These are either purely real with $q^{(IJ)} = 0$ or purely imaginary with $p^{(IJ)} = 0$, and they are paired so that a real mode with $\tilde{c}^{(IJ)} = p^{(IJ)}$ and eigenvalue $\omega = \Omega$ is paired with an imaginary mode $\tilde{c}^{(IJ)} = i p^{(IJ)}$ with opposite sign eigenvalue $\omega = -\Omega$.
\\

\noindent
{\bf 4-dimensions}\\
Let us now explicitly give the modes in the case of $n=4$ dimensions. We firstly enumerate the 6 growing modes and these are grouped into 3 off-diagonal and 3 diagonal modes. 
The 3 growing off-diagonal modes are real and we label them as modes A, B and C;
\be
\mathrm{A\;and\;B:} \qquad  \omega &=& (1 + 2 \beta^2) | Q | \; , \quad \tilde{c}^{(IJ)} = \left(
\begin{array}{ccc}
0 & 1 & 0 \\
\cdot & 0 & 0 \\
\cdot & \cdot & 0
\end{array}
\right) 
\; \mathrm{and} \; \left(
\begin{array}{ccc}
0 & 0 & 1 \\
\cdot & 0 & 0 \\
\cdot & \cdot & 0
\end{array}
\right) 
\nl
\mathrm{C:} \qquad  \omega &=&  | Q | \; , \quad \tilde{c}^{(IJ)} = \left(
\begin{array}{ccc}
0 & 0 & 0 \\
\cdot & 0 & 1 \\
\cdot & \cdot & 0
\end{array}
\right) \; .
\ee
The diagonal modes follow from solving the cubic for their eigenvalues,
\be
\omega^3 - ( ( a^{(1)} a^{(2)} )^2 + ( a^{(2)} a^{(3)} )^2 + ( a^{(1)} a^{(3)} )^2 ) |Q|^2 \omega + 2 ( a^{(1)} a^{(2)} a^{(3)} )^2 |Q|^3 = 0 \; .
\ee
Given the form of the $a^{(I)}$ above the roots have a simple form,
\be
\omega =  |Q| \, , \;  |Q| (1 + \gamma) \, , \; -2 |Q| (1 + \frac{1}{2} \gamma) 
\ee
where we have defined the non-negative real constant $\gamma$,
\be
\label{eq:gamma}
\gamma = \frac{1}{2} \left( \sqrt{ 1 + 8 ( 1 + 2 \beta^2 )^2 }  - 3   \right) \ge 0 \; .
\ee
These then translate into 2 growing real diagonal modes which we label as D and E;
\be
\mathrm{D:}\qquad \omega &=&  |Q| \; , \quad \tilde{c}^{(IJ)} = \left(
\begin{array}{ccc}
0 & 0 & 0 \\
\cdot & -1 & 0 \\
\cdot & \cdot & 1
\end{array}
\right) \nl
\mathrm{E:}\qquad \omega &=&  |Q| (1 + \gamma) \; , \quad \tilde{c}^{(IJ)} = \left(
\begin{array}{ccc}
-2 - \gamma & 0 & 0 \\
\cdot & 1 & 0 \\
\cdot & \cdot & 1
\end{array}
\right) 
\ee
and finally one growing imaginary mode, which we label F,
\be
\mathrm{F:} \qquad \omega &=&   2 |Q| (1 + \frac{1}{2} \gamma) \; , \quad \tilde{c}^{(IJ)} = i \left(
\begin{array}{ccc}
1 + \gamma & 0 & 0 \\
\cdot & 1 & 0 \\
\cdot & \cdot & 1
\end{array}
\right) \; .
\ee
The remaining modes are decaying in flow time and have opposite sign eigenvalue $\omega$, and $\tilde{c}^{(IJ)}$ as above but multiplied by $i$. We label these A', B', C', D', E' and F'.

We now make a convenient choice of gauge in~\eqref{eq:linflowmet1} using the freedom $c^{(A)}$ which we write as $c^{(A)} = \sqrt{\bar{Q}}  \tilde{c}^{(A)}$ so that the perturbation takes the form,
\be
\label{eq:linflowmet2}
\delta \mat{g} =   \sqrt{\bar{Q}}  \mat{A} \left(  \tilde{c}^{(A)} ( \tilde{\vect{k}} \vect{n}^{(A) T} + \vect{n}^{(A)} \tilde{\vect{k}}^T ) + \tilde{c}^{(IJ)} \vect{n}^{(I)}  \vect{n}^{(J) T}  \right) \mat{A}^T \; .
\ee
Recall our previous basis, defined starting with~\eqref{eq:xydefn}, which in 4 dimensions introduces the orthogonal real $\vect{n}^{(2)}$ and $\vect{n}^{(3)}$. These define a \emph{real} orthonormal basis $\vect{N}^{(A)} = \{ \hat{\vect{x}} , \hat{\vect{y}} , \vect{n}^{(2)}, \vect{n}^{(3)} \}$ so that $\vect{N}^{(A)} \cdot \vect{N}^{(B)} = \delta^{AB}$.
We then choose the gauge for the perturbation to be, 
\be
\label{eq:metpert}
\delta \mat{g} =   \sqrt{\bar{Q}}  \mat{A} \left(  \psi \mat{1} + a^{(IJ)} \vect{N}^{(I)}  \vect{N}^{(J) T}  \right) \mat{A}^T 
\ee
where $a^{(IJ)}$ is tracefree, so $\sum_I a^{(II)} = 0$.
The explicit transformation is given by,
\be
\psi & = & \frac{1}{3 + 2 \beta^2} \left( \tilde{c}^{(11)} + (1+\beta^2) (\tilde{c}^{(22)} +\tilde{c}^{(33)}) \right) \nl
a^{(1b)} &=& \frac{1}{\alpha} \tilde{c}^{(1b)} \; , \quad a^{(23)} = \tilde{c}^{(23)} \; , \quad a^{(bb)} = \tilde{c}^{(bb)}  - \psi 
\ee
for $b = 2,3$, and this gauge can always be chosen, with the $\tilde{c}^{(A)}$ required being,
\be
\tilde{c}^{(0)} &=& \frac{1}{2(3 + 2 \beta^2)} \left( \tilde{c}^{(11)} (1 - 2 \beta^2) + \left( \tilde{c}^{(22)} + \tilde{c}^{(33)} \right) ( 1 + 2 \beta^2) \right) \nl
\tilde{c}^{(1)} &=& \frac{i \alpha \beta}{(3 + 2 \beta^2)} \left( 2 \tilde{c}^{(11)} - \tilde{c}^{(22)} - \tilde{c}^{(33)} \right) \; , \quad \tilde{c}^{(b)} = \frac{i \beta}{\alpha} \tilde{c}^{(1b)}
\ee
and we note that $\alpha = \sqrt{1 + \beta^2}$ is always strictly positive.
Thus in this gauge the complex perturbation for the principal part of the flow equation takes the form,
\be
\delta {\mat{g}} = \delta \phi\, \mat{g} +   \delta {\mat{g}}^{(TT)}  
\ee
so that the trace and traceless parts of the perturbation in our frame are,
\be
\delta \phi = \sqrt{\bar{Q}} \psi \; , \quad \delta {\mat{g}}^{(TT)} =  \sqrt{\bar{Q}} a^{(IJ)} \mat{A} \vect{N}^{(I)}  \vect{N}^{(J) T} \mat{A}^T
\ee
and we note that $\mathrm{Tr}( \mat{g}^{-1} \delta {\mat{g}}^{(TT)} ) = 0$ so indeed $\delta {{g}}^{(TT)}_{\mu\nu}$ is traceless. It is also transverse in the following sense,
\be
\Re(\tilde{\vect{k}}) \mat{A}^{-1} \cdot \delta {\mat{g}}^{(TT)} + \ldots = 0 
\ee
where $\ldots$ represent terms that are lower order in the wavevector, and hence are irrelevant in the principal part of the flow.
Thus in this short wavelength limit, we may view $a^{(IJ)}$ as yielding a perturbation to the conformal class, and $\psi$ to the conformal factor.

In summary, in $4$-dimensions, implementing the gauge transform for our above modes given in terms of $\tilde{c}^{(IJ)}$, then writing these in terms of the trace, $\psi$, and traceless part $a^{(IJ)}$ for the perturbation in our gauge~\eqref{eq:metpert}, and after suitable rescaling by a real constant, we find growing modes;
\be
\begin{array}{c|c|c|c}
\mathrm{mode} & \omega & \psi & a^{(IJ)} \\
\hline 
\hline
A\,\mathrm{and}\,B & (1 + 2 \beta^2) | Q |  & 0 & \left(
\begin{array}{ccc}
0 & 1 & 0 \\
\cdot & 0 & 0 \\
\cdot & \cdot & 0
\end{array}
\right) 
\; \mathrm{and} \; \left(
\begin{array}{ccc}
0 & 0 & 1 \\
\cdot & 0 & 0 \\
\cdot & \cdot & 0
\end{array}
\right) 
\\
\hline
C & | Q | & 0 & \left(
\begin{array}{ccc}
0 & 0 & 0 \\
\cdot & 0 & 1 \\
\cdot & \cdot & 0
\end{array}
\right)
\\
\hline
D & | Q | & 0 & \left(
\begin{array}{ccc}
0 & 0 & 0 \\
\cdot & -1 & 0 \\
\cdot & \cdot & 1
\end{array}
\right)
\\
\hline
E &  |Q| (1 + \gamma) & \frac{1}{3} ( 2 \beta^2 - \gamma ) &  \left(
\begin{array}{ccc}
-2 - \frac{2}{3} \gamma & 0 & 0 \\
\cdot & 1 + \frac{\gamma}{3} & 0 \\
\cdot & \cdot & 1 + \frac{\gamma}{3} 
\end{array}
\right)
\\
\hline
F & 2 |Q| (1 + \frac{1}{2} \gamma) & \frac{i}{3} \left( 2 \beta^2  + 3 + \gamma \right) &  i \left(
\begin{array}{ccc}
\frac{2}{3} \gamma & 0 & 0 \\
\cdot & -\frac{1}{3} \gamma & 0 \\
\cdot & \cdot & -\frac{1}{3} \gamma 
\end{array}
\right) 
\end{array}
\ee
We recognize A,B,C,D as the transverse traceless modes. Mode E is the continuation of a transverse traceless mode when the background is real, so $\gamma = 0$. Finally F is the continuation of  the conformal mode.
\\

\noindent
{\bf 3-dimensions}\\
We may repeat the same analysis in the case of $3$ dimensions where it is somewhat simpler. The off-diagonal mode takes a very similar form to mode A in $4$ dimensions, and the two diagonal modes are determined by a simple quadratic relation, $\omega^2 = (a^{(1)} a^{(2)} | Q |)^2$, rather than a cubic eigenvalue equation. Then the growing modes  are all degenerate with the same eigenvalue $\omega = (1 + 2 \beta^2) | Q |$ and we find,
\be
\begin{array}{c|c|c}
\mathrm{mode}  & \psi & a^{(IJ)} \\
\hline 
\hline
A &  0 & \left(
\begin{array}{cc}
0 & 1  \\
\cdot & 0 
\end{array}
\right)
\\
\hline
E & - \frac{1}{2} \beta^2 & \left(
\begin{array}{cc}
 -(1+ \beta^2) & 0 \\
 \cdot & (1+ \beta^2)
\end{array}
\right)
\\
\hline
F &  i \left( 1+ \frac{3}{2} \beta^2 \right) &  i \left(
\begin{array}{cc}
 \beta^2 & 0 \\
 \cdot & -\beta^2 
\end{array}
\right) 
\end{array}
\ee
Again A and E are continuations of the transverse traceless modes of a real background, and F is the continuation of the conformal mode.

Thus we see that non-perturbatively, both in $3$ and $4$ dimensions, given the restriction in equation~\eqref{eq:restriction} on the complex metric, the Einstein--Lefschetz flow has precisely the same forward-backward character as it does near real metrics.

\subsection{Boundary conditions for flows from a fixed point}
\label{sec:complexBC}

If we construct a flow starting at a fixed point as $\lambda \to - \infty$, then we require boundary conditions at the final finite flow time $\lambda_1$. In order for our forward-backward flow to be well-posed these should be such that they constrain the short distance growing modes we found in the previous section. Our intuition is that we should somehow fix the real conformal class and imaginary conformal factor.

A naive choice would be to again write as before $g_{\mu\nu} = e^\phi \hat{g}_{\mu\nu}$, where $\phi$ is now complex and the complex $\hat{g}$ again has $\det(\hat{g}) = $ fixed, and describes the conformal class, with the determinant condition defining the choice of class representative.  Then the perturbations decompose as in~\eqref{eq:complexpert} with $\delta g^{(T)}_{\mu\nu} = e^\phi  \delta \hat{g}_{\mu\nu} $ being traceless.
The natural guess is then to fix the real class, $\Re( \hat{g}_{\mu\nu} )$, at the final time. However, the conditions,
\be
\Re( \hat{g}_{\mu\nu} ) = \mathrm{fixed} \; , \quad \det(\hat{g}) = \mathrm{fixed}
\ee
then impose a further restriction on $\Im( \hat{g}_{\mu\nu} )$ at the final flow time. This follows as the determinant is inherently a non-linear function of the class representative $\hat{g}_{\mu\nu}$, whereas the reality condition is linear. This leads to an inconsistency as we expect $\Im( \hat{g}_{\mu\nu} )$  should not be constrained at the final flow time (except perhaps for a finite number of constraints arising from unstable modes as we have previously discussed).

Instead we use our complex metric decomposition from above, writing the components in our fixed frame as,
\be
\mat{g} = e^\phi \hat{\mat{g}} \; , \quad \hat{\mat{g}} = e^{\frac{1}{2} \mat{X}'} e^{i \mat{Y}'}  e^{\frac{1}{2} \mat{X}'}  
\ee
where we include a conformal factor and $\hat{\mat{g}}$ is the conformal class representative, as in equation~\eqref{eq:ELmetric} of the Introduction. The real reference metric then takes the form in equation~\eqref{eq:ELreference} there.
In order to define this split we impose that the real symmetric $\mat{X}'$ and $\mat{Y}'$ are traceless matrices so that $\det(\hat{\mat{g}}) = 1$.
Hence in terms of our original form of metric~\eqref{eq:metricform} we have $\mat{g} = e^{\frac{1}{2} \mat{X}} e^{i \mat{Y}}  e^{\frac{1}{2} \mat{X}} $ where,
\be
\mat{X} = \mat{X}' + \Re(\phi) \mat{1} \; , \quad \mat{Y} = \mat{Y}' + \Im(\phi) \mat{1} \; .
\ee
Now the final boundary condition we impose, in addition to fixing $\Im(\phi)$, is that the real class is fixed in the sense that we fix $\mat{X}'$.  Then at the final flow time, $\lambda = \lambda_1$, we impose,
\be
\label{eq:finalbc}
\Im(\left. \phi \right|_{\lambda_1}) = \mathrm{fixed} \; , \quad \left. \mat{X}' \right|_{\lambda_1} = \mathrm{fixed} \; .
\ee
We note that while in the real case fixing $\phi$ is equivalent to fixing the full conformal factor $e^\phi$, in this complex case they are not the same -- it is $\Im(\phi)$ we choose to fix here rather than $\Im( e^\phi )$.
Then under a variation, 
\be
\delta \mat{g} = \delta \phi \, \mat{g} + e^\phi e^{\frac{1}{2} \mat{X}'} \delta\left( e^{i \mat{Y}'}  \right) e^{\frac{1}{2} \mat{X}'}  
\ee
where $\delta \phi$ is constrained to be real. Now we write,
\be
\delta\left( e^{i \mat{Y}'}  \right) = e^{i (\mat{Y}' + \delta \mat{Y}' )}  - e^{i \mat{Y}'}  = i e^{\frac{1}{2} i \mat{Y}'}  \delta\mat{W} e^{\frac{1}{2} i \mat{Y}'}  
\ee
to linear order, and the perturbations $\delta \mat{Y}'$ and $\delta \mat{W}$ are simply related as,
\be
\left( \mat{O}  \delta \mat{W}  \mat{O}^T \right)_{ab} = \mathrm{sinc}\left( \frac{ \theta^{(a)} -  \theta^{(b)} }{2} \right) \left(  \mat{O} \delta \mat{Y}'   \mat{O}^T\right)_{ab}
\ee
where $\mat{O}$ is the real orthogonal matrix that diagonalizes the real symmetric background $\mat{Y}'$, with $\theta^{(a)}$ being the diagonal entries so,
\be
\mat{O}  \mat{Y}'  \mat{O}^T = \mathrm{diag}\left( \theta^{(0)}, \theta^{(1)}, \ldots , \theta^{(n-1)} \right)
\ee
and $\mathrm{sinc}(x) = \sin(x)/x$ where $\mathrm{sinc}(0) = 1$. Note that from this we see $\mathrm{Tr}(\delta \mat{W}) = 0$.
Furthermore, we see that since $\mat{O}$ is real, then given that $\delta  \mat{Y}' $ is real then $\delta \mat{W} $ is also real.
Hence we obtain the elegant result that,
\be
\label{eq:metboundary}
\delta \mat{g} =  \mat{A} \left(  \delta \phi \mat{1} + i \delta\mat{W} \right) \mat{A}^T
\ee
where crucially we have that both $\delta \phi$ and $\delta\mat{W}$ are real. 

Let us now compare this to the perturbation computed above~\eqref{eq:metpert}, and we note that this is given in a particular gauge choice.
The comparison implies that in this gauge, and given our final boundary condition~\eqref{eq:finalbc}, then,
\be
\label{eq:comparison}
  \psi  = \frac{\sqrt{{Q}}}{| Q |} \delta \phi \; , \quad a^{(IJ)}  \vect{N}^{(I)}  \vect{N}^{(J) T} =  i \frac{ \sqrt{{Q}}  }{| {Q} | }    \delta \mat{W} \; .
 \ee
Thus we see that for our gauge choice the perturbation at the final flow time $\lambda = \lambda_1$ implies the traceless and transverse conditions on $\delta \mat{W}$ as,
\be
\label{eq:gauge}
\mathrm{Tr}(  \delta \mat{W}  ) = 0 \; , \quad  \Re(\vect{\tilde{k}}) \cdot \delta \mat{W} = 0
\ee
at leading order in the principal part analysis.
For this to give a good boundary condition at the final flow time, we require that it kills the short wavelength growing modes that we derived in the previous section.
Let us now show that this is indeed the case for $4$ dimensions. The argument in $3$ dimensions is similar. 

We see that $\delta \mat{W}$ only has components in the $ \vect{N}^{(I)}$ directions. Projecting in these we then obtain,
 \be
 \label{eq:reln}
   \psi  = \frac{\sqrt{{Q}}}{| Q |} \delta \phi \; , \quad   a^{(IJ)} =  i \frac{ \sqrt{{Q}}  }{| {Q} | }  \vect{N}^{(I)} \cdot   \delta \mat{W} \cdot  \vect{N}^{(J)} 
 \ee
where the key point is that both  $\delta \phi$ and $ \vect{N}^{(I)} \cdot   \delta \mat{W} \cdot  \vect{N}^{(J)}$ are purely real. 
Let us consider a general 4 dimensional growing mode solution parameterized by the modes A-F above multiplied by real constants $c_{A,B,C,D,E,F}$. Without loss of generality we can consider the final flow time to be at $\lambda = 0$. Then the perturbation takes the form,
\be
\psi = \frac{1}{3} \left( \left( 2 \beta^2 - \gamma \right) c_E + i \left( 2 \beta^2 + 3 + \gamma \right) c_{F} \right)
\ee
and
\be
a^{(IJ)} = 
\left(
\begin{array}{ccc}
- 2\left( 1 + \frac{\gamma}{3}   \right) c_E + i \frac{2}{3} \gamma c_{F} & c_A & c_B \\
\cdot &  -c_D + \left( 1 + \frac{\gamma}{3}  \right) c_E - i \frac{1}{3} \gamma c_{F}  & c_C \\
\cdot & \cdot & c_D +  \left( 1 + \frac{\gamma}{3}  \right) c_E - i \frac{1}{3} \gamma c_{F} 
\end{array}
\right) \; . \nl
\ee
Now we see that provided $\Re(\sqrt{Q})\ne 0$ then~\eqref{eq:reln} implies the off diagonal modes $c_{A,B,C}$ vanish, as well as the traceless diagonal mode $c_D$ from considering,
\be
 a^{(33)} -  a^{(22)} =  2 c_D =  i \frac{ \sqrt{{Q}}  }{| {Q} | }  \left( 
 \vect{N}^{(3)} \cdot   \delta \mat{W} \cdot  \vect{N}^{(3)} -  \vect{N}^{(2)} \cdot   \delta \mat{W} \cdot  \vect{N}^{(2)}
 \right) \; .
\ee
Then writing,
\be
\sqrt{Q} = p+ i q \; , \quad  \delta W =  \vect{N}^{(1)} \cdot   \delta \mat{W} \cdot  \vect{N}^{(1)} 
\ee
for real $p, q$, we are left with the modes parameterized by $c_E$ and $c_F$ which obey,
\be
&&  \psi  = \frac{1}{3} \left( \left( 2 \beta^2 - \gamma \right) c_E + i \left( 2 \beta^2 + 3 + \gamma \right) c_{F} \right) = (p+ i q) \frac{ \delta \phi}{| {Q} |} \; , \nl
  &&   a^{(11)}  =  - 2\left( 1 + \frac{\gamma}{3}   \right) c_E + i \frac{2}{3} \gamma c_{F}   =  i (p+ i q) \frac{\delta W}{| {Q} |}  
\ee
and we recall that $\gamma$ and $\beta$ are real, as are $c_E$ and $c_{F}$, and so is $\delta \phi$ and $\delta W$.
We note that if both $\delta \phi = 0$ and $\delta W =0$ then this immediately implies $c_E = c_{F} = 0$ -- thus we assume at least one is non-vanishing.
Now separating into real and imaginary parts we may write this as the linear system,
\be
{T} \cdot \left(
\begin{array}{c}
c_E \\ c_{F} \\ \delta \phi \\ \delta W
\end{array}
\right) = 0 \; , \quad {T} = \left(
\begin{array}{cccc}
 \frac{1}{3} ( 2 \beta^2 - \gamma) & 0 & - \frac{p}{| Q|}& 0 \\
- \frac{2}{3} ( 3 + \gamma) & 0 & 0 &  \frac{q}{| Q|}\\
0 & \frac{1}{3} ( 2 \beta^2 + 3 + \gamma ) & - \frac{q}{| Q |} & 0 \\
0 & \frac{2}{3}  \gamma &0 & - \frac{p}{| Q|}
\end{array}
\right) \; .
\ee
This linear system only admits non-trivial solutions allowing $c_{E,F} \ne 0$ if the determinant of the matrix $T$ vanishes, where,
\be
\det({T}) &=& - \frac{2 }{9 | Q |^2}  \left( p^2 ( 3 + \gamma ) ( 2\beta^2 + 3 + \gamma ) + q^2 \gamma ( \gamma - 2 \beta^2 )  \right) \nl
&\le& - \frac{2 }{9 | Q |^2}   p^2 ( 3 + \gamma ) ( 2\beta^2 + 3 + \gamma )   \le  - \frac{2 }{| Q |^2}   p^2 
\ee
where the inequalities follow from the form of $\gamma$ in~\eqref{eq:gamma} which implies $\gamma \ge 2 \beta^2 \ge 0$. 
However the condition that $p = \Re(\sqrt{Q})\ne 0$ implies this cannot vanish. Hence we conclude that with the condition $\Re(\sqrt{Q})\ne 0$ then all $c_{A,B,C,D,E,F}$ must vanish provided that a gauge is chosen so that the unfixed part of the metric, $\Re(\phi)$ and $\mat{Y}'$ give rise to a fluctuation of the metric satisfying the gauge conditions~\eqref{eq:gauge}.\footnote{
The analogous result in $3$ dimensions is
$\det({T}) = - \frac{1 }{ 2 | Q |^2}  \left( p^2 ( 2 + 5 \beta^2 + 3 \beta^4 ) + q^2 \beta^4 \right)$. Then $p \ne 0$ leads to the same conclusion that $c_{E,F}$ must vanish. 
}

We reiterate that $ \Re(\sqrt{Q})\ne 0$ was precisely implied by our starting condition on the complex metric in~\eqref{eq:Qcondition} so that we could write it in the form, $\mat{g} = e^\phi  e^{\frac{1}{2} \mat{X}'} e^{i \mat{Y}'}  e^{\frac{1}{2} \mat{X}'} $. Hence we are guaranteed our flow character is of forward-backward type and, furthermore, that fixing $\Im(\phi)$ and the traceless symmetric real $\mat{X}'$ gives final boundary conditions that are compatible with the characteristic analysis of the flow equation, provided we choose an appropriate gauge.

\subsection{Examples of non-linear flows}

We conclude this section by considering some numerical examples of non-linear complex flows in 4-dimensions away from a fixed point, considering the same spacetimes as we did for the Einstein flow, namely Euclidean de Sitter and the large Schwarzschild solution in a cavity (we do not consider the small Schwarzschild solution here). In both cases we are able to reach large deformations away from the saddle point geometry that result in a significant monotonic increase in  the real part of the action $\Re(S)$ along the flow. As expected from~\eqref{eq:restriction}, we find that the KSW allowability criterion for all $p$-forms is violated well before the flow fails. 
For all these flows, we use a choice of metric ansatz that is identical to those of our earlier Einstein flows, but now our metric functions are complex-valued, and likewise for our diffeomorphism terms. 

Before proceeding, we provide a quick summary in Table~\ref{tab:endpoint-data} of how we fix our flow data. At the initial flow time $\lambda \to - \infty$ the geometry is set to be that of the fixed point. For the final flow time the simplest case is the large Schwarzschild black hole fixed point which is stable. Hence the final data is given by the imaginary conformal factor and real conformal class. Euclidean de Sitter is more complicated due to the unstable homogeneous mode. This then requires that the final imaginary conformal factor is tuned, which we do by tuning its homogeneous component so that only its inhomogeneous part, $\Im(\phi_\perp)$, is prescribed as data. Conversely one constraint is put on the real conformal factor, which we do by fixing its homogeneous component $\Re(\phi_h)$ as data. 
\begin{table}[ht]
  \centering
  \begin{tabular}{lcc}
    \hline
    Fixed Point & Stability & Final flow data  \\
    \hline
    Euclidean de Sitter & Unstable homogeneous mode & $\Im(\phi_\perp)$,\; $\Re(\phi_h)$,\; $\mat{X'}$ \\
    Large Schwarzschild & Stable  & $\Im(\phi)$,\; $\mat{X'}$ \\
    \hline
  \end{tabular}
  \caption{Data fixed at the final flow time, where $\phi_h$ is the homogeneous mode of the conformal factor~$\phi$, and $\phi_\perp$ represents the remaining modes, i.e. $\phi_\perp = \phi - \phi_h$. }
  \label{tab:endpoint-data}
\end{table}

\subsubsection{Complex flows away from the Euclidean de Sitter saddle}

We now consider flows away from the Euclidean de Sitter fixed point. Each flow is a set of complex geometries that give a path within the Lefschetz thimble of the de Sitter fixed point, and the set of all such flows will generate the Lefschetz thimble. We choose the metric to have the same form as earlier in~\eqref{eq:spheremetricform}, but now with complex $\Phi$ and $T$, and similarly take the same form for the diffeomorphism vector as in~\eqref{eq:diffvecsphere} but with complex $F$. 
Hence the flow preserves an $SO(3) \times SO(2)$ isometry. The boundary conditions where the $S^2$ shrinks, at $\theta = 0$, and where the $S^1$ shrinks, at $\theta = \pi/2$, remain the same as before.

Following from  our earlier discussion in Section~\ref{sec: complex flow character}, we  introduce the frame
\be
\hat{\theta}^{0} = \dd \theta \,, \; \; \hat{\theta}^{1,2} = \sin \theta \times \left(\text{orthonormal frame on $S^2$}\right) \,, \; \;  \hat{\theta}^{3} = \cos \theta \,\dd \chi \, ,
\ee
in which the metric components are
\be
\mat{g} = \frac{3}{\Lambda} e^{\Phi} \;\mathrm{diag}\left(e^T, e^T, e^T, e^{-3T}\right) = e^{\phi} \hat{\mat{g}},
\ee
where $\phi = \Phi + \log(3/\Lambda)$ and $\mat{\hat{g}} = \exp(T\, \mat{D})$, with $\mat{D} \equiv \mathrm{diag}(1, 1, 1, -3 )$. Given that $\mathrm{Tr} \,\mat{D} = 0$, we find $\det \mat{\hat{g}} = 1$, and so $\mat{\hat{g}}$ is a representative of the conformal class with unit determinant. We can now identify the metric in this frame with our earlier expression $\mat{g} = e^\phi  e^{\frac{1}{2} \mat{X}'} e^{i \mat{Y}'}  e^{\frac{1}{2} \mat{X}'} $, choosing $\mat{X}' = \Re (T) \mat{D}$ and $\mat{Y}' = \Im (T) \mat{D}$. Since $\mat{D}$ is diagonal, $\mat{X'}$ and $\mat{Y'}$ commute,
\be
\hat{\mat{g}} = e^{\frac{1}{2} \mat{X}'} e^{i \mat{Y}'}  e^{\frac{1}{2} \mat{X}'} = e^{\mat{X'} + i \mat{Y'}} =  e^{T\,\mat{D}}  \, ,
\ee
and so fixing the real or imaginary part of $T$ we may think  of as  fixing the real or imaginary conformal class (or the logarithm of it). Likewise,  the conformal factor $\phi$ is controlled by the boundary conditions imposed on $\Phi$. The real reference metric is then given by,
\be
\mat{\mathfrak{g}} = \frac{3}{\Lambda} e^{\Re(\Phi)} \;\mathrm{diag}\left(e^{\Re(T)}, e^{\Re(T)}, e^{\Re(T)}, e^{-3\Re(T)}\right) \; .
\ee
As for the real Einstein flow in Section~\ref{sec:EinsteinS4} the flow equations determine $F$ at each flow time. They give equations for $\Phi$ and $T$ whose highest derivative terms are,
\be
\label{eq:PSphere}
\frac{\partial}{\partial \lambda} \Phi = \frac{\Lambda}{3} e^{-\Phi} e^{ - \overline{T}} \partial^2_\theta \left( 2 \overline{\Phi} -  \overline{T} \right) + \ldots \; , \quad \frac{\partial}{\partial \lambda} T = - \frac{\Lambda}{3} e^{-\Phi} e^{ - \overline{T}} \partial^2_\theta  \overline{T}  + \ldots
\ee
where $\ldots$ are lower order terms. 

We have discussed above how perturbatively the real homogeneous mode of the conformal factor, so $\Phi$, exponentially grows in flow time, and the imaginary homogeneous mode decays, and hence the round sphere fixed point is unstable. Thus we decompose $\Phi$ at the end point of the flow as we did previously for the Einstein flow in equation~\eqref{eq:PhiDecompFinal}, so $\Phi(0,\theta) = \Phi_h + \Phi_\perp(\theta)$, although now both $\Phi_h$ and $\Phi_\perp(\theta)$ are complex valued. 
For a stable fixed point we would fix the imaginary conformal factor, so $\Im(\Phi)$, and the real class, so $\Re(T)$, at the final flow time. However the real growing behaviour of the real homogeneous conformal perturbation implies we have additional data which we can take to be the value of $\Re(\Phi_h)$ at the end of the flow. In addition, the unstable imaginary homogeneous mode implies we must tune $\Im(\Phi)$ in order to originate from the fixed point, and we do this by only specifying the data $\Im(\Phi_\perp(\theta))$, and determining $\Im(\Phi_h)$ from the flow solution.

For fluctuations $\delta \Phi$ and $\delta T$, our coordinate choice here does not lead to perturbations in the same gauge as in our analysis above, so~\eqref{eq:metpert}, where we showed that a characteristic analysis implies fixing $\Im(\phi)$ and $\mat{X}'$ is good data provided $\Re(\sqrt{Q}) > 0$. Nonetheless it is straightforward to  decompose the principal part of the equation above in~\eqref{eq:PSphere} into real and imaginary parts of $\Phi$ and $T$. Then one can deduce the growing real Fourier modes, and show that indeed fixing $\Im(\Phi)$ and $\Re(T)$ at the final flow time constrains these  precisely when the condition $\Re(\sqrt{Q}) > 0$ holds, where the wavevector $q$ is taken in the $\theta$ direction, which is the only direction that $\Phi$ and $T$ depend on. Thus whilst our form of the metric does not give rise to the gauge condition~\eqref{eq:gauge}, the coordinates we have chosen nonetheless give the same criterion for when our boundary conditions are compatible with the flow character.

We compactify flow time as for the Einstein flow from the sphere, taking a computational flow time $y$ as in~\eqref{eq:away compact flow time}\footnote{For the flows we present we have chosen the constant $a = 10$.}, discretising and solving as before, except now with complex valued functions $\Phi$, $T$ and $F$. At $y = 0$, the asymptotic past of the flow, we set the geometry to be the round sphere, so fix $\Phi = T = F = 0$.
Then as we have discussed, at the final flow time, so $y = 1$, we prescribe $\Im(\Phi_\perp)$, $\Re(T)$ and $\Re(\Phi_h)$. 
If we were to set $\Im(\Phi_\perp) = 0$ whilst perturbing $\Re(\Phi_h)$ and $\Re(T)$, then the flow solution would be real and hence we would simply 
 recover the Einstein flow we studied earlier. Hence we only consider flows with non-zero $\Im(\Phi_\perp(\lambda_1))$ to ensure they are not real.

For the flows we present we choose a deformation of the round sphere given by taking the final flow time data,
\be
\label{eq: dS final data}
\Re(T(0,\theta)) = \frac{1}{4} \delta_{\mathrm{R}}\, \cos^2\theta \, , \quad
\Re(\Phi_h) = \frac{8}{35}  \delta_{\mathrm{R'}} \, , \quad
\Im(\Phi_\perp(\theta)) = \delta_{\mathrm{I}} \left( \cos^4\theta - \tfrac{8}{35} \right) 
\ee
for amplitudes $\delta_{\mathrm{R}}, \delta_{\mathrm{R'}}$ and $\delta_{\mathrm{I}}$. We now show flow solutions for three cases summarised in Table~\ref{tab: endpoint perturbations}.
\begin{table}[h]
    \centering
    \renewcommand{\arraystretch}{1.3}
    \begin{tabular}{c l c c c}
        \hline\hline
        Case & Deformation & $\delta_{\mathrm{I}}$ & $\delta_{\mathrm{R}}$ & $\delta_{\mathrm{R'}}$ \\
        \hline
        1 & small $\Im(\Phi)$ & $0.001$ & $0$ & $0$ \\
        2 & large $\Im(\Phi)$ & $5$ & $0$& $0$  \\
        3 & large $\Im(\Phi)$ and $\Re(T)$ & $4.3$ & $4.3$  & $4.3$  \\
        \hline\hline
    \end{tabular}
    \caption{Amplitudes of the final data~\eqref{eq: dS final data} at $\lambda_1 = 0$ for the flows away from the de Sitter saddle. }
    \label{tab: endpoint perturbations}
\end{table}
We will postpone the discussion of the action for these flows until after the next section.

\begin{figure}[tbp]
\centering
    \subfigure[$\Re (T(\lambda,\theta))$.]{\dSpanel{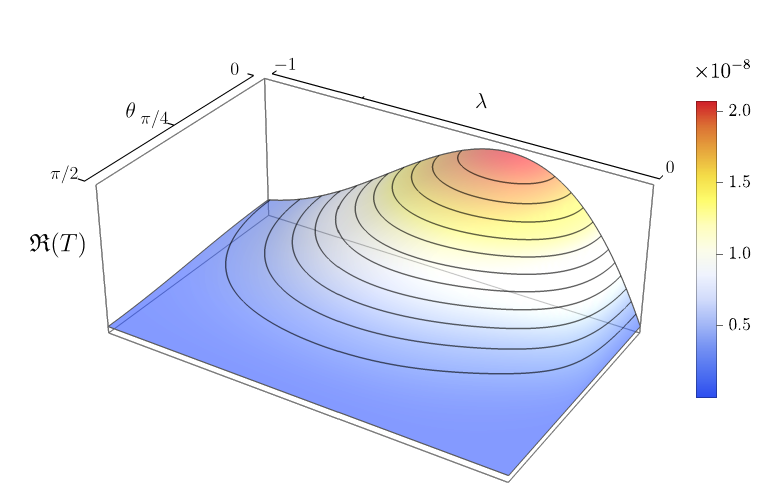}}%
    \hspace{0.02\textwidth}%
    \subfigure[$\Re(\Phi(\lambda,\theta))$.]{\dSpanel{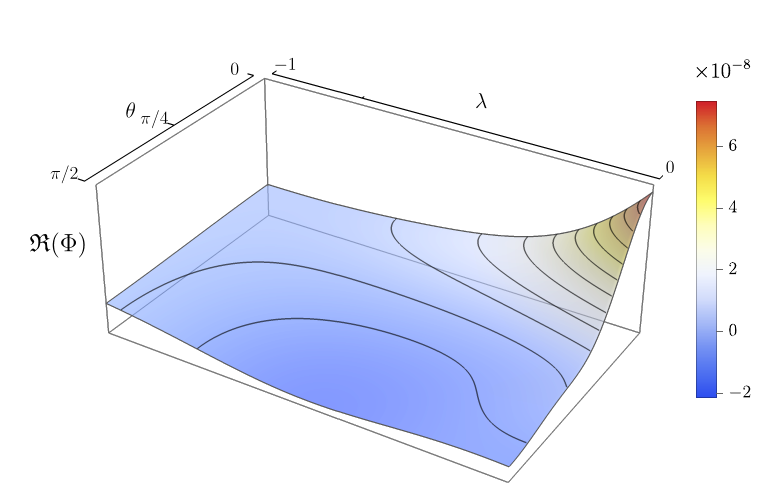}}\\[1ex]
    \subfigure[$\Im (T(\lambda,\theta))$.]{\dSpanel{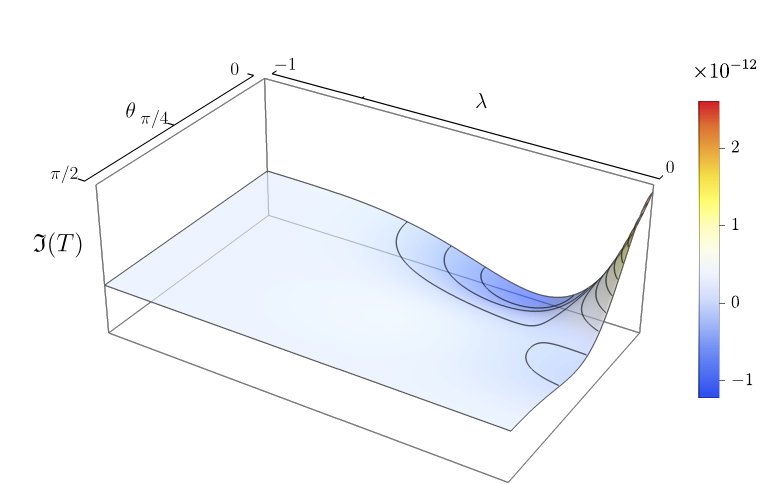}}%
    \hspace{0.02\textwidth}%
    \subfigure[$\Im( \Phi(\lambda,\theta))$.]{\dSpanel{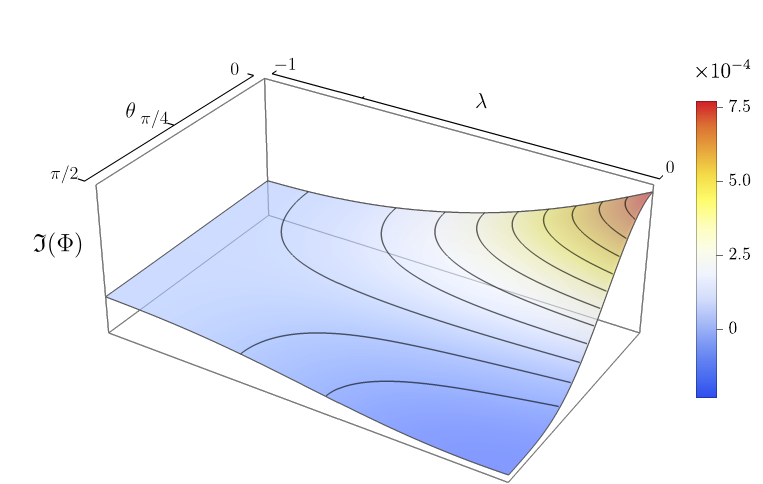}}
    \caption{
    Einstein--Lefschetz flow away from the de Sitter saddle for a small deformation of $\Im(\Phi)$ (case 1). We see that $\Re(\Phi)$ and $T$ remain much smaller than $\Im(\Phi)$ as they decouple from $\Im(\Phi)$ for linear perturbations. This flow represents a path in the thimble of the de Sitter fixed point which reproduces the perturbative GHP prescription; the conformal factor is Wick rotated to become purely imaginary. Only the range $\lambda \in [-1, 0]$ is shown (using a log scale) as the geometry is very close to that of the round sphere for earlier times. 
    }
    \label{fig:dS PL small ImPhi pert}
\end{figure}

In Figure~\ref{fig:dS PL small ImPhi pert} we show case 1, a relatively small deformation of only $\Im(\Phi_\perp)$  at the final time. We observe that all of the fields except $\Im(\Phi)$ have a suppressed contribution as we expect for a solution that is described by linear theory. 
This flow represents a path in the thimble where $T \simeq 0$ and the conformal factor $\Phi$ is approximately purely imaginary. This is precisely a part of the thimble which agrees with the perturbative GHP prescription, where we have perturbed only the conformal factor which is Wick rotated to be imaginary. We display only the range of flow times $\lambda \in [-1, 0]$  since the final data relax sufficiently quickly towards the fixed point in the past so that the flow is essentially indistinguishable from the fixed point at earlier times.

\begin{figure}[tbp]
\centering
    \subfigure[$\Re (T(\lambda,\theta))$.]{\dSpanel{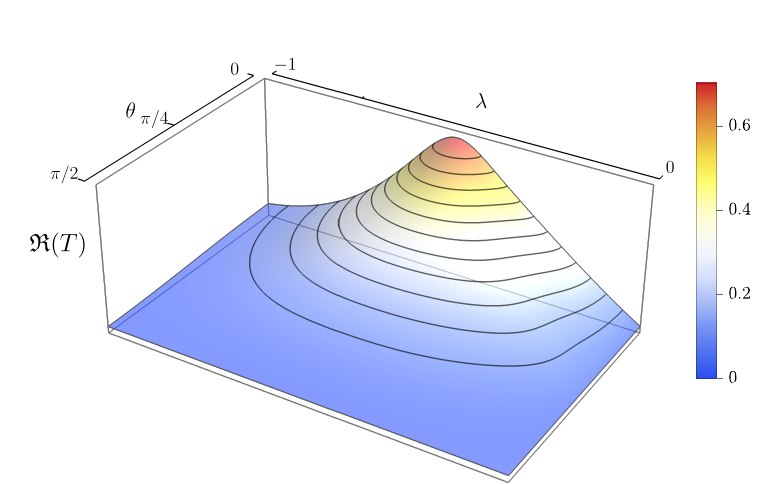}}%
    \hspace{0.02\textwidth}%
    \subfigure[$\Re(\Phi(\lambda,\theta))$.]{\dSpanel{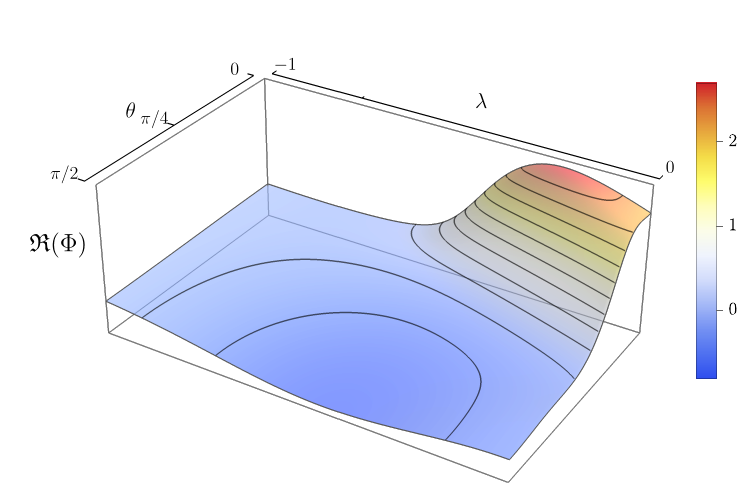}}\\[1ex]
    \subfigure[$\Im (T(\lambda,\theta))$.]{\dSpanel{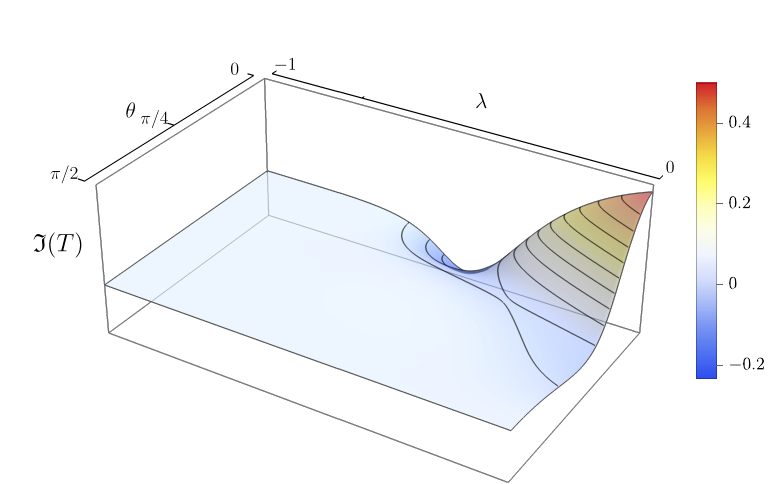}}%
    \hspace{0.02\textwidth}%
    \subfigure[$\Im( \Phi(\lambda,\theta))$.]{\dSpanel{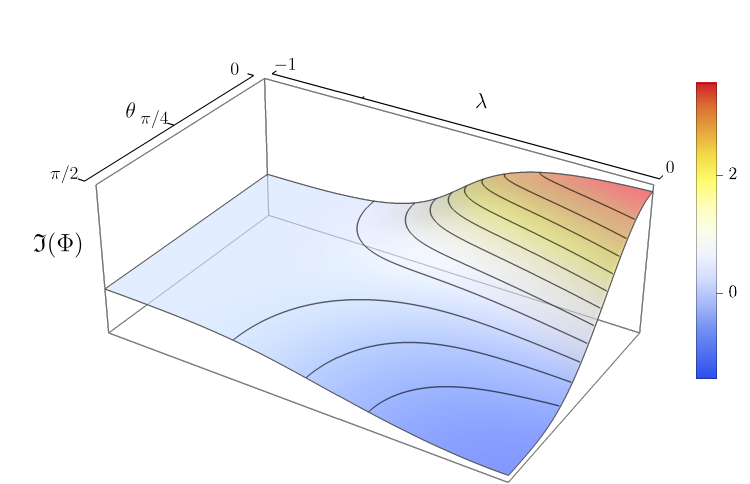}}
    \caption{Einstein--Lefschetz flow away from the de Sitter saddle for a large deformation of $\Im(\Phi)$ (case 2). The real and imaginary parts of each field are now of comparable size,
   showing their non-linear response to the deformation of $\Im(\Phi)$, and indicating that by the end of the flow it is well beyond the perturbative GHP regime. Again only the range $\lambda \in [-1, 0]$ is displayed.}
    \label{fig:dS PL large ImPhi pert}
\end{figure}

In Figure~\ref{fig:dS PL large ImPhi pert} we see case 2, a large deformation of $\Im(\Phi_\perp)$, again with $\Re(T) = 0$ and $\Re(\Phi_h) =0$ at the final time. 
Here near the fixed point in the past the perturbative GHP prescription gives the behaviour of the flow, but for larger flow times we are clearly in the non-perturbative regime and beyond the reach of the GHP prescription. 
We no longer observe the same suppression of $\Im(T)$ and $\Re(\Phi)$ as these couple non-linearly to the deformation of $\Im(\Phi)$; indeed, they are of a similar order of magnitude to  $\Im(\Phi)$ in their response to its deformation. 
This explicitly shows that beyond the GHP perturbative regime the thimble is a complicated slice through the space of complex metrics, where the real and imaginary parts of the conformal class and factor are related.
Again we display only the range $\lambda \in [-1, 0]$ as the flow is very close to the fixed point for earlier times.

\begin{figure}[tbp]
    \centering
    \includegraphics[width=0.9\linewidth]{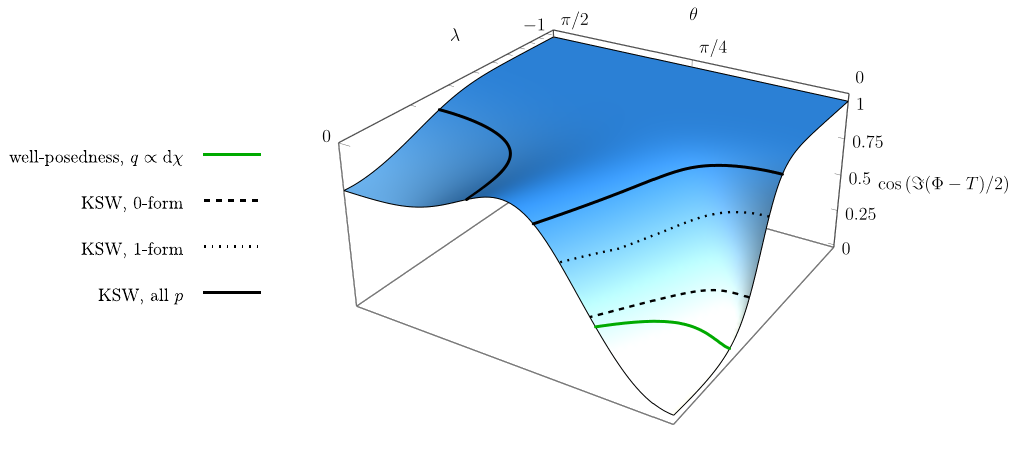}
    \caption{Well-posedness and KSW allowability along the flow from the de Sitter fixed point for  case 2. Only the range of flow time $\lambda \in [-1, 0]$ is shown. The surface plotted is $\cos\left(\Im(\Phi - T)/2\right)$, which vanishes where our well-posedness condition in equation~\eqref{eq:restriction} fails for $q \propto \dd\theta$. The KSW conditions fail well before our well-posedness condition -- we show both the individual conditions for $p=0$ and $1$ forms, as well as the full condition for all $p$-forms. We also show where the flow well-posedness would fail if we relaxed our restriction to $SO(3) \times SO(2)$, giving the example of perturbations in the $\chi$ angle direction. }
    \label{fig:KSW_WP_dS}
\end{figure}

The flow survives past where the KSW condition is violated, apparently failing only as $\Re(\sqrt{Q})\to 0$ at the end of the flow. Given that we have restricted to flows with $SO(3) \times SO(2)$ isometry, the metric functions only vary with flow time and in the $\theta$ direction. Hence when we consider our well-posedness criterion $\Re(\sqrt{Q}) \ne 0$, given in equation~\eqref{eq:restriction}, it only need apply for wavevectors $q \propto \dd \theta$, which results in the condition,
\be
\Re\left(\sqrt{e^{\Phi-T}}\right) > 0\,,
\ee
or equivalently $|\Im(\Phi - T)| < \pi$. We can therefore monitor violation of well-posedness simply by tracking whether $|\Im(\Phi - T)| \to \pi$ at any point along the flow. For case 2, we find $\max|\Im(\Phi - T)| \approx 0.98\, \pi$, which is attained at the end of the flow and is very close to a violation of this well-posedness bound. We were unable to find solutions of the form in case 2 with greater amplitude $\delta_I$ than that presented, presumably because the well-posedness of such flows would fail.
 In Figure~\ref{fig:KSW_WP_dS}, we show this well-posedness criterion for this case 2 flow. We indicate where we violate the KSW condition in its various forms, so the general condition, but also the individual $p=0$ and $p=1$ form constraints. We also show where our well-posedness criterion would be violated if we allowed more general perturbations that break the $SO(3) \times SO(2)$ isometry. We show the example of deformations in the $\chi$ angle direction, so that $q \propto \dd \chi$, where we see the flow would fail slightly sooner, but still after the KSW condition is violated.
 
\begin{figure}[tbp]
\centering
    \subfigure[$\Re (T(\lambda,\theta))$.]{\dSpanel{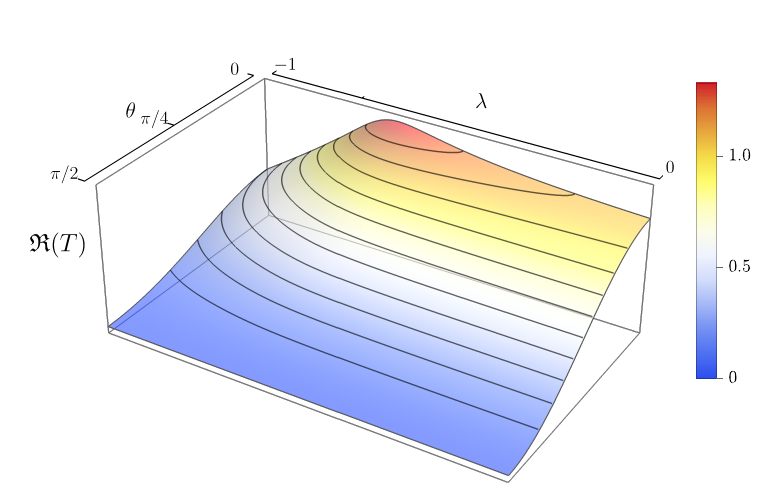}}%
    \hspace{0.02\textwidth}%
    \subfigure[$\Re(\Phi(\lambda,\theta))$.]{\dSpanel{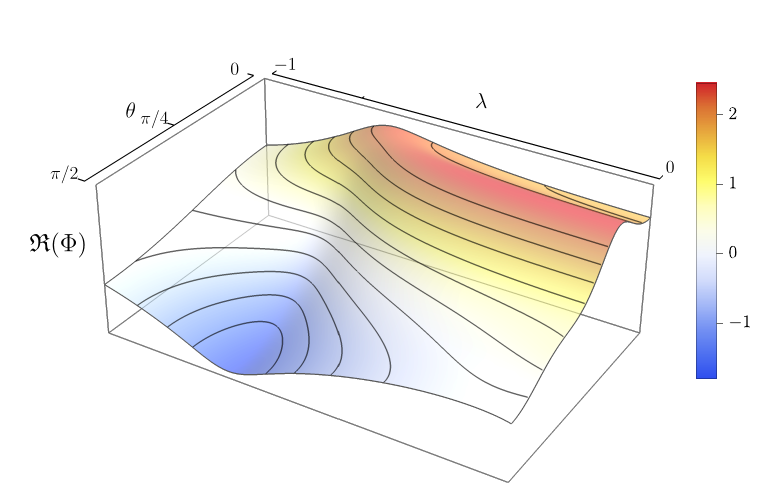}}\\[1ex]
    \subfigure[$\Im (T(\lambda,\theta))$.]{\dSpanel{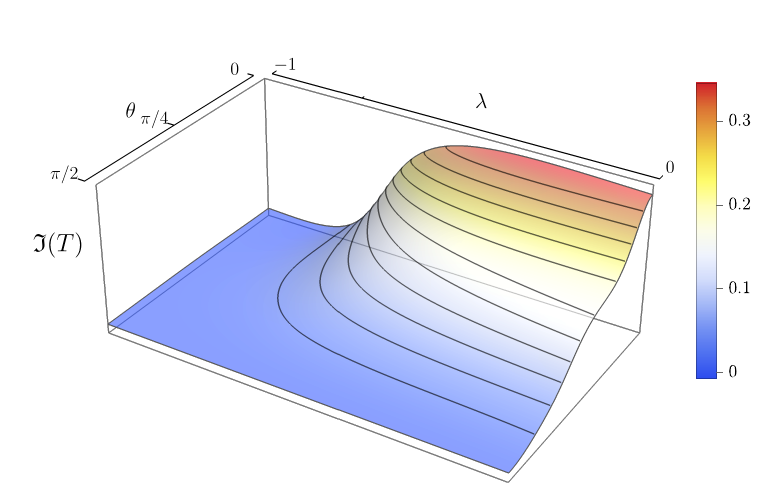}}%
    \hspace{0.02\textwidth}%
    \subfigure[$\Im( \Phi(\lambda,\theta))$.]{\dSpanel{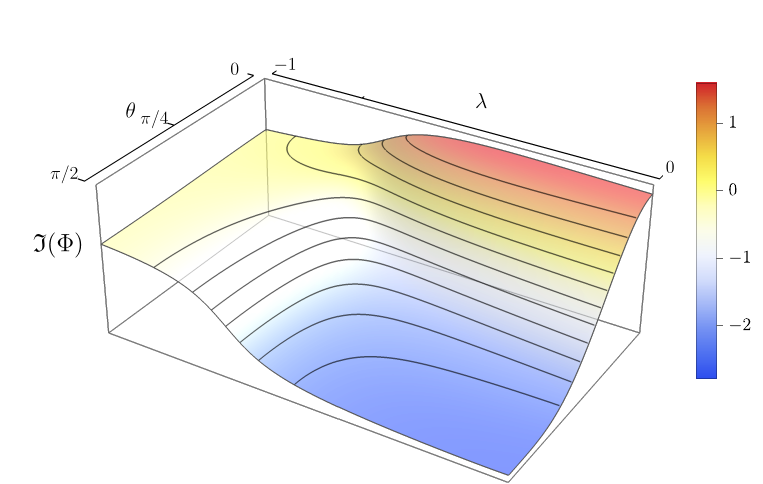}}
    \caption{Einstein--Lefschetz flow away from the de Sitter saddle for a large deformation of both $\Im(\Phi)$ and $\Re(T)$ (case 3). }
    \label{fig:dS PL ImPhi ReT pert}
\end{figure}

Figure~\ref{fig:dS PL ImPhi ReT pert} shows case 3, a non-perturbative deformation of both $\Re(T)$ and $\Im(\Phi)$. 
Once again, the deformations are large enough that we are well outside the regime where the GHP prescription holds, with the real and imaginary parts of the conformal class roughly of the same order of magnitude, and likewise with the real and imaginary parts of the conformal factor.
We show a slightly larger range of flow time here than in the previous two cases, displaying the flow for $\lambda \in [-10, 0]$.
For this strong deformation the well-posedness criterion reaches a maximum value $\max|\Im(\Phi - T)| \approx 0.88\, \pi$ at the end of the flow. Taking the amplitudes of the perturbation $\delta_{\mathrm{R, R', I}}$ to all be equal, we were unable to increase their value and find flow solutions past that presented for case 3. It is unclear if this is because the well-posedness is starting to fail, or whether it is due to the diffeomorphism vector function $F$ which develops steep gradients in flow time for these large amplitudes. In the latter case then choosing a different complex gauge might allow even stronger deformations of this form.

\subsubsection{Complex flows away from the large black hole saddle in a cavity}
\label{sec:complexBHflow}

We now consider flows away from a large black hole. We again take the same form of metric as for the Einstein flow, so equation~\eqref{eq: BH metric with T and Phi}, and with the same diffeomorphism vector, except that now the metric functions $\Phi$ and $T$, and $F$ that specifies the diffeomorphism are taken to be complex functions of flow time and the radial coordinate. Thus our flows again preserve the static spherical symmetry of the Schwarzschild fixed point. 
We introduce the frame,
\be
\hat{\theta}^{0} = r_0\, r\, \dd \tau \,, \; \; \hat{\theta}^{1} = 2 r_0 \sqrt{1 - r_0}\, g(r)\, \dd r \,, \; \; \hat{\theta}^{2,3} = r_0 \sqrt{g(r)} \times \left(\text{orthonormal frame on $S^2$}\right) ,
\nl
\ee
with $g(r)$ defined as in~\eqref{eq: BH metric}, in which the metric components are
\be
\mat{g} = e^{\Phi}\, \mathrm{diag}\left(e^{T}, e^{T}, e^{-T}, e^{-T}\right) = e^{\phi} \hat{\mat{g}} \, ,
\ee
with $\phi = \Phi$ and $\hat{\mat{g}} = \exp(T\, \mat{D})$, where now $\mat{D} \equiv \mathrm{diag}(1, 1, -1, -1)$. Since $\mathrm{Tr}\,\mat{D} = 0$, $\hat{\mat{g}}$ is again a unit-determinant representative of the conformal class, and exactly as for de Sitter, fixing the real and imaginary parts of $T$ at the appropriate endpoints fixes the conformal class, while the boundary conditions on $\Phi$ control the conformal factor. The real reference metric is,
\be
\mat{\mathfrak{g}} = e^{\Re(\Phi)}\, \mathrm{diag}\left(e^{\Re(T)}, e^{\Re(T)}, e^{-\Re(T)}, e^{-\Re(T)}\right)  \, .
\ee
The spatial boundary conditions are the same as those of the real Einstein flow, now applied to both the real and imaginary parts: regularity at the horizon is enforced by Neumann conditions on $\Phi$, $T$ and $F$, and we impose Dirichlet conditions at the cavity wall so that $\Phi$ and $T$ vanish, and the complex diffeomorphism (and hence the function $F$) vanishes there too.

The large black hole has no unstable mode, so we fix the data according to the stable row of Table~\ref{tab:endpoint-data}. At the asymptotic far past of the flow we set $\Phi$, $T$ and $F$ to vanish, so that the geometry is that of the large Schwarzschild fixed point. Then at the final flow time $\lambda_1 = 0$ we deform the real part of the conformal class, $\Re(T)$, and the imaginary part of the conformal factor, $\Im(\Phi)$.
The principal part of the flow equations take a similar form to that for the sphere in~\eqref{eq:PSphere}, and a similar characteristic analysis shows that while our metric form does not yield fluctuations at the final flow time in the gauge~\eqref{eq:gauge}, nonetheless the condition that these boundary final boundary conditions are compatible with the character is precisely $\Re(\sqrt{Q}) > 0$, where the wavevector is taken in the radial direction $q = d r$.

For the examples we present, we choose the deformations,
\be
\label{eq: cavity final data}
\Re(T(0,r)) = \delta_{\mathrm{R}} \left(1 - r^2\right)^4 , \quad
\Im(\Phi(0,r)) = \delta_{\mathrm{I}} \left(1 - r^2\right)^4 ,
\ee
where the profile $(1 - r^2)^4$ is even in $r$ and vanishes at the wall, consistent with the boundary conditions there. We consider the three cases summarised in Table~\ref{tab: cavity endpoint deformations}.
\begin{table}[h]
    \centering
    \renewcommand{\arraystretch}{1.3}
    \begin{tabular}{c l c c}
        \hline\hline
        Case & Deformation & $\delta_{\mathrm{I}}$ & $\delta_{\mathrm{R}}$ \\
        \hline
        1 & small $\Im(\Phi)$ & $0.001$ & $0$ \\
        2 & large $\Im(\Phi)$ & $3.28$ & $0$ \\
        3 & large $\Im(\Phi)$ and $\Re(T)$ & $1.75$ & $1.75$ \\
        \hline\hline
    \end{tabular}
    \caption{Amplitudes of the final data~\eqref{eq: cavity final data} at $\lambda_1 = 0$ for the flows away from the large black hole. }
    \label{tab: cavity endpoint deformations}
\end{table}

We compactify the flow time as before for complex de Sitter, using the computational flow time in equation~\eqref{eq:away compact flow time}, again choosing $a = 10$ for the flows we present here. We display the real and imaginary part of $\Phi$ and $T$ as functions of flow time and the proper distance $\rho_0$ from the cavity wall, now measured with respect to the real reference metric $\mathfrak{g}$. We  show only the  flow time period $\lambda \in [-10, 0]$, as for earlier times the flow is very close to the saddle point geometry.

\begin{figure}[tbp]
\centering
    \subfigure[$\Re (T(\lambda,r))$.]{\cavpanel{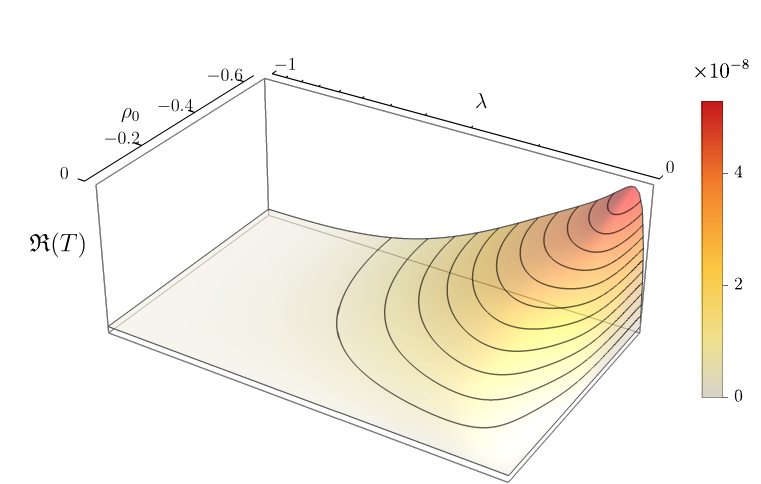}}%
    \hspace{0.02\textwidth}%
    \subfigure[$\Re(\Phi(\lambda,r))$.]{\cavpanel{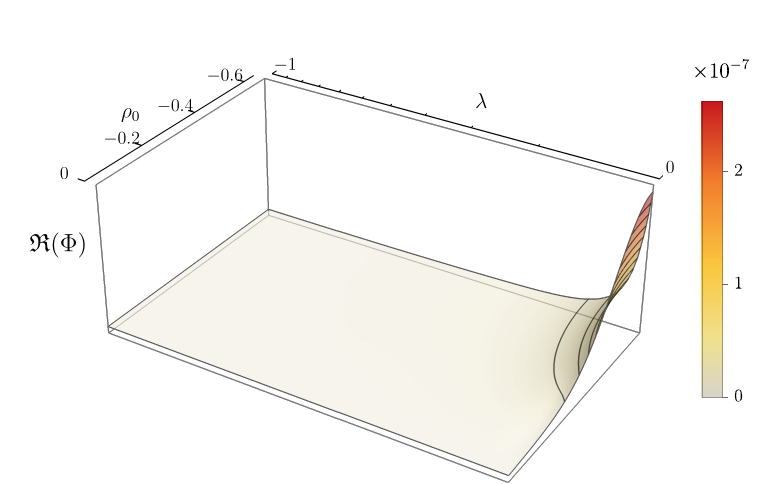}}\\[1ex]
    \subfigure[$\Im (T(\lambda,r))$.]{\cavpanel{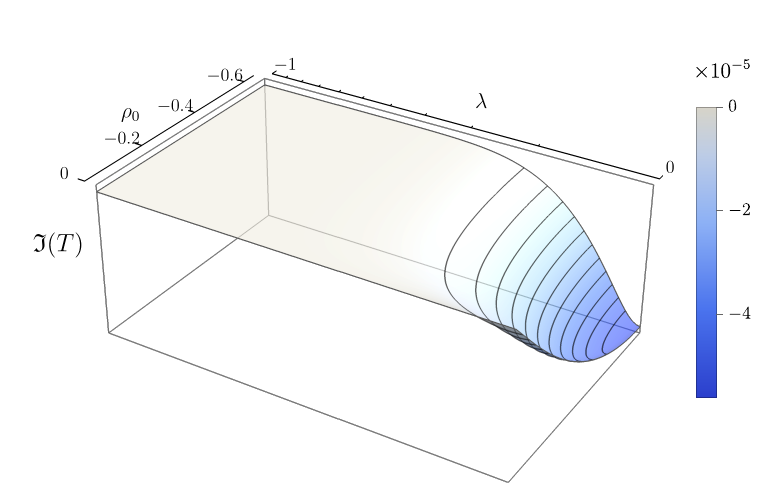}}%
    \hspace{0.02\textwidth}%
    \subfigure[$\Im( \Phi(\lambda,r))$.]{\cavpanel{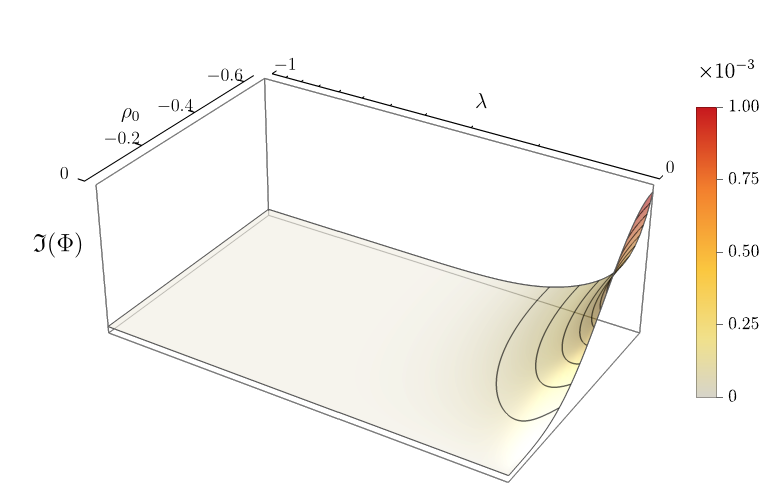}}
    \caption{Einstein--Lefschetz flow away from the large black hole for a small deformation of $\Im(\Phi)$ (case 1). While the deformation is small, and hence described by perturbation theory, since the cavity walls couple $\Phi$ and $T$ already in linear theory, this represents a path in the thimble of the large black hole that is not described by the simple GHP prescription. 
    We see this explicitly as $\Im(T)$ is excited at linear order by the boundary conditions, and hence is considerably larger than the real parts of $T$ and $\Phi$ whose response is due to non-linearity.
    }
    \label{fig:cavity PL small ImPhi pert}
\end{figure}

Figure~\ref{fig:cavity PL small ImPhi pert} shows case 1, a small deformation of $\Im(\Phi)$ with $\Re(T) = 0$ at the final time. 
Given that we are considering small deformations of the large black hole, we  find that the proper radius does not vary much along this flow. 
The remaining fields are again small, but in contrast to the de Sitter case 1, $\Im(T)$ is not as suppressed. This is due to the coupling between the conformal class and the conformal factor perturbations which is induced already at linear order by the boundary conditions at the cavity wall. While this flow stays a perturbation of the fixed point, the fact that $\Im(T)$ is  deformed explicitly shows that this flow, while being in the perturbative regime, is beyond the GHP prescription which would have purely real $T$.

\begin{figure}[tbp]
\centering
    \subfigure[$\Re (T(\lambda,r))$.]{\cavpanel{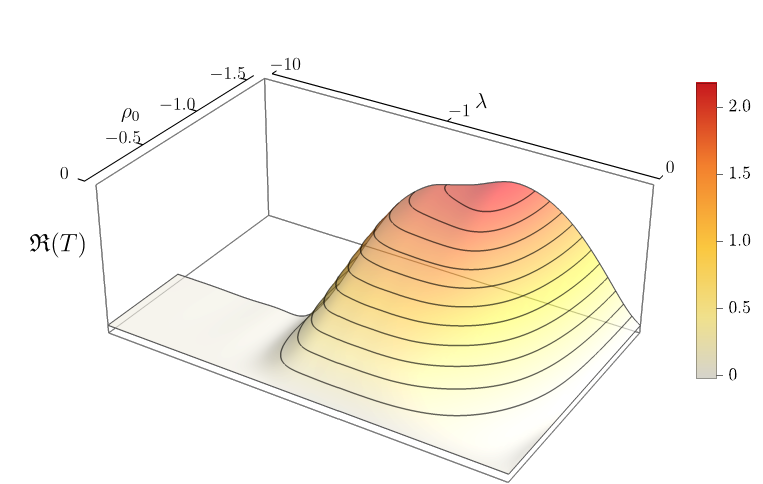}}%
    \hspace{0.02\textwidth}%
    \subfigure[$\Re(\Phi(\lambda,r))$.]{\cavpanel{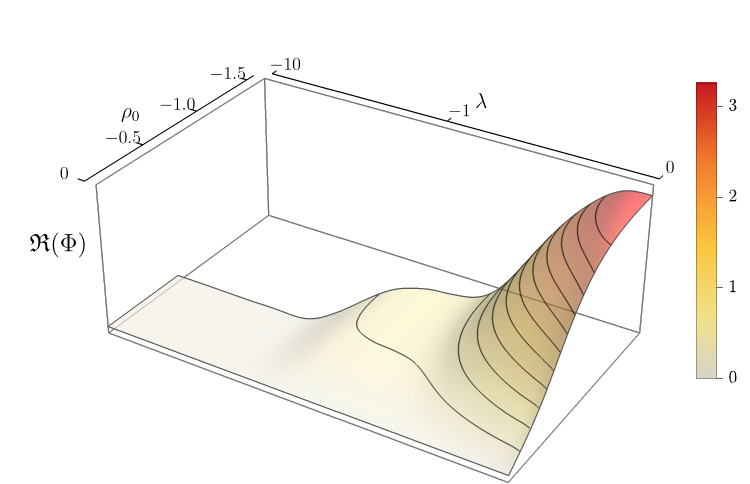}}\\[1ex]
    \subfigure[$\Im (T(\lambda,r))$.]{\cavpanel{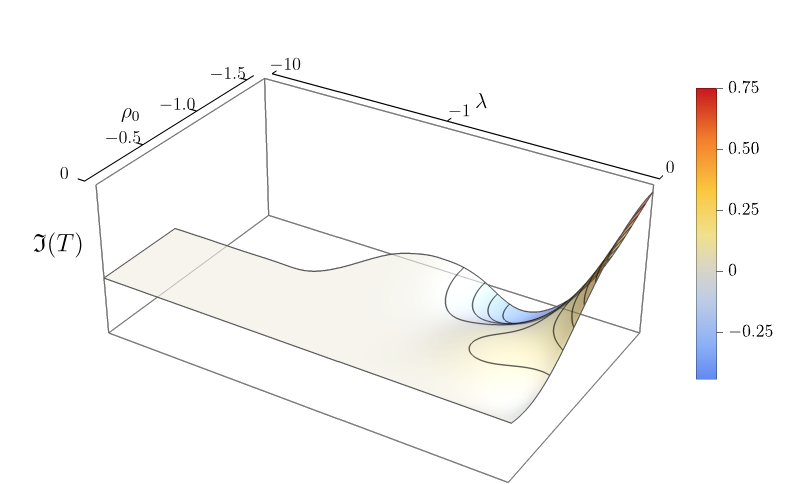}}%
    \hspace{0.02\textwidth}%
    \subfigure[$\Im( \Phi(\lambda,r))$.]{\cavpanel{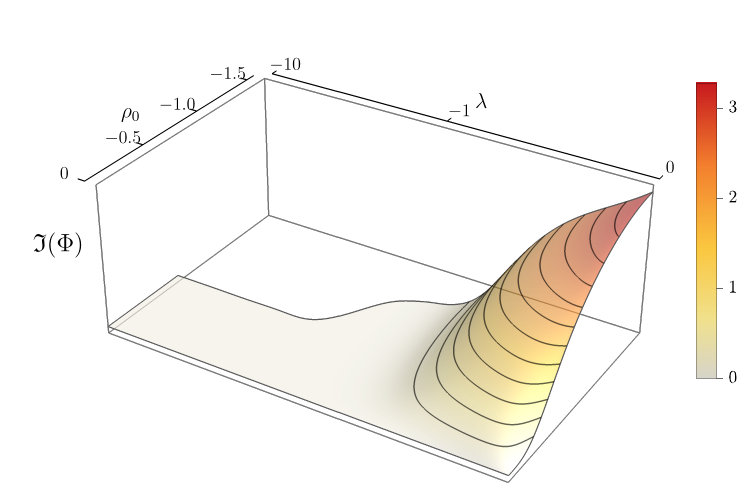}}
    \caption{Einstein--Lefschetz flow away from the large black hole for a large deformation of $\Im(\Phi)$ (case 2). We set $\Im(\Phi(0,r)) = 3.28\,(1 - r^2)^4$, with $\Re(T) = 0$ at $\lambda = 0$. All four fields are excited non-linearly and are roughly of comparable size.}
    \label{fig:cavity PL large ImPhi pert}
\end{figure}

In Figure~\ref{fig:cavity PL large ImPhi pert} we consider case 2, a large deformation of $\Im(\Phi)$, again with $\Re(T) = 0$ at the final flow time. At late flow times $\Re(\Phi)$, $\Im(T)$ are both strongly deformed in their non-linear response to  $\Im(\Phi)$. While we impose that $\Re(T)$ vanishes at the final flow time, $\lambda = 0$, we see earlier in the flow that it also has a large non-linear response. This flow has the largest amplitude $\delta_{\mathrm{I}}$ where we could find solutions for deformations of only $\Im(\Phi)$, and taking the functional form given above.

As was the case for de Sitter, we can monitor the well-posedness condition~\eqref{eq:restriction} and the KSW allowability of the complex metrics along the flow. In the frame above, the metric is diagonal with phases $\Im(\Phi + T)$ in the $(\tau, r)$ directions and $\Im(\Phi - T)$ in the $S^2$ directions. Since only radial derivatives appear in our flow equations as we have restricted to static spherical symmetry, the relevant well-posedness condition is that taking $q \propto \dd r$, which up to real positive factors, yields
\be
\Re\left(\sqrt{e^{\Phi - T}}\right) > 0 \, ,
\ee
or equivalently $|\Im(\Phi - T)| < \pi$. 
If we were to relax the symmetry restriction on the flow, we would also have to consider fluctuations in other directions. As an example we consider fluctuations in the sphere directions, where the condition~\eqref{eq:restriction} taking $q$ to be a covector in the sphere  gives the condition $|\Im(\Phi + T)| < \pi$. The KSW conditions become $|\Im (\Phi)| < \pi/4$ for $0$-forms, $|\Im(\Phi \pm T)| < \pi/2$ for $1$-forms and $|\Im (T)| < \pi/4$ for $2$-forms, while the condition for all $p$ is $2|\Im(\Phi + T)| + 2|\Im(\Phi - T)| < \pi$. Since $|x + y| + |x - y| = 2\max(|x|, |y|)$, this last condition is equivalent to the $0$-form and $2$-form conditions together. 
\begin{figure}[tbp]
    \centering
    \includegraphics[width=0.9\linewidth]{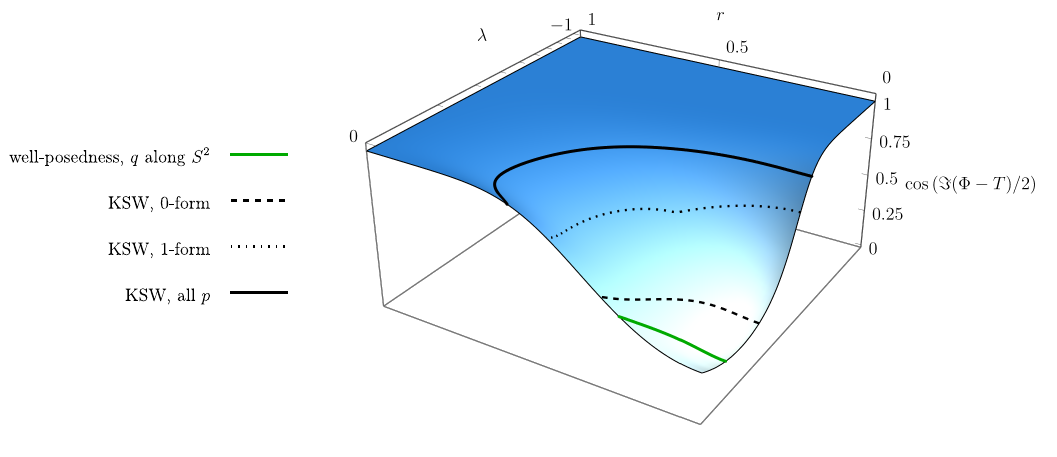}
    \caption{
    Well-posedness and KSW allowability along the flow of case 2 from the large Schwarzschild fixed point. We show only the late times in the flow $\lambda \in [-1, 0]$. The surface shown is $\cos\left(\Im(\Phi - T)/2\right)$ which vanishes where the well-posedness condition~\eqref{eq:restriction} fails for $q \propto \dd r$. As for the earlier de Sitter example, we see the KSW conditions fail well before our well-posedness condition -- again we show the individual conditions for $p=0$ and $1$ forms, as well as the full condition for all $p$-forms. We also see well-posedness would fail slightly earlier if we relaxed the restriction to static spherical symmetry, and we give the example of perturbations in the sphere directions.
     }
    \label{fig:KSW_WP_Sch}
\end{figure}
 As we show in Figure~\ref{fig:KSW_WP_Sch}, similarly to the de Sitter example earlier, the KSW conditions are violated well before our well-posedness condition, which is satisfied throughout case 2 with $\max|\Im(\Phi - T)| \approx 0.87\,\pi$.  
We see this flow is quite close to violating well-posedness, which may explain why we were unable to find greater amplitude example flows. However it is also the case that we find gradients in the diffeomorphism function $F$ become quite large, so perhaps a change of complex gauge would allow us to push to larger amplitudes, and closer to the well-posedness bound.

\begin{figure}[tbp]
\centering
    \subfigure[$\Re (T(\lambda,r))$.]{\cavpanel{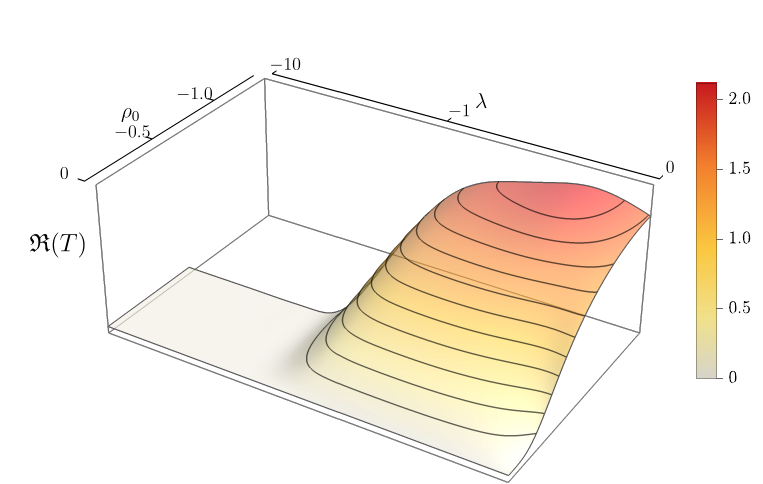}}%
    \hspace{0.02\textwidth}%
    \subfigure[$\Re(\Phi(\lambda,r))$.]{\cavpanel{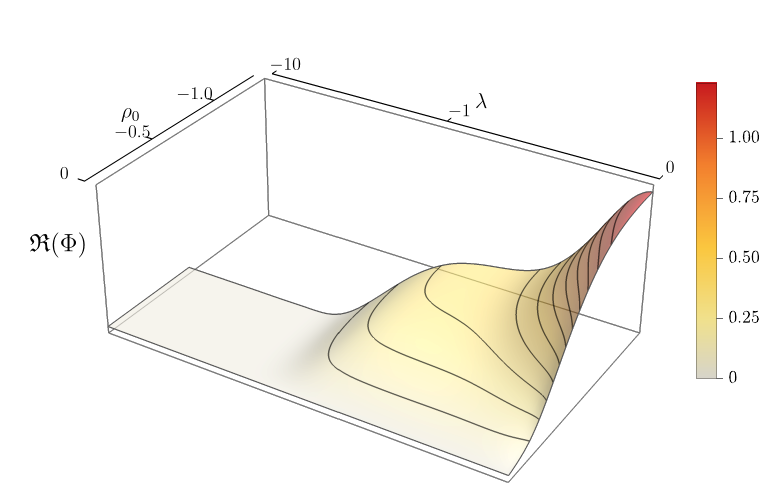}}\\[-1ex]
    \subfigure[$\Im (T(\lambda,r))$.]{\cavpanel{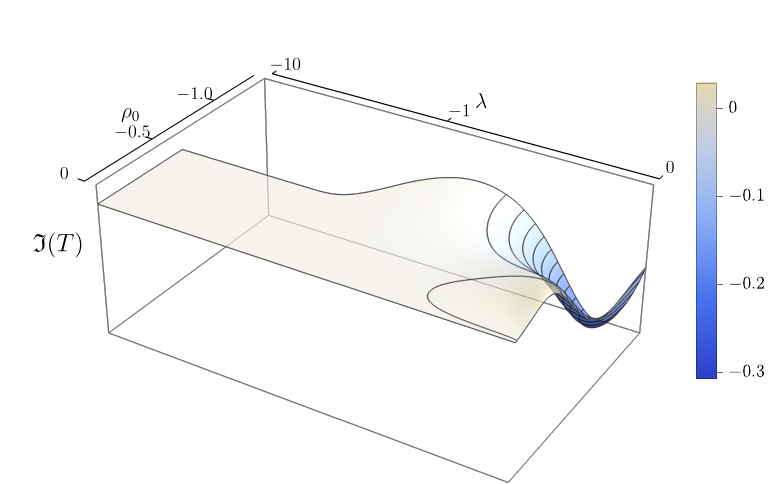}}%
    \hspace{0.02\textwidth}%
    \subfigure[$\Im( \Phi(\lambda,r))$.]{\cavpanel{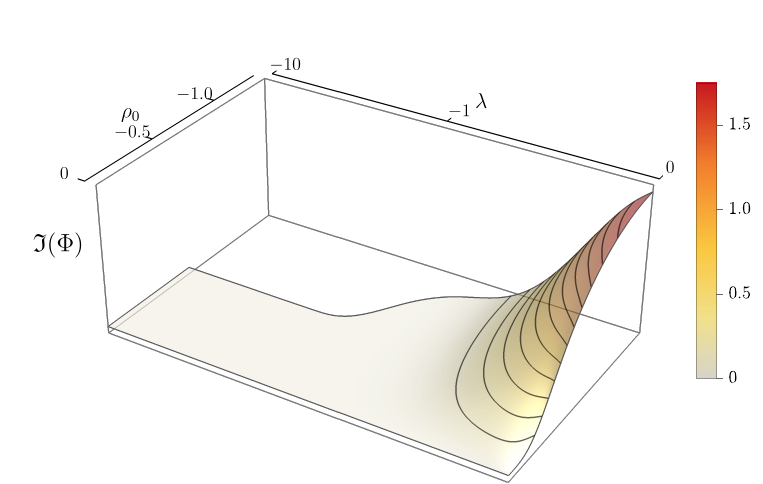}}
    \caption{Einstein--Lefschetz flow away from the large black hole for a large deformation of both $\Im(\Phi)$ and $\Re(T)$ (case 3). We set $\Re(T(0,r)) = \Im(\Phi(0,r)) = 1.75\,(1 - r^2)^4$.}
    \label{fig:cavity PL ImPhi ReT pert}
\end{figure}

Case 3, shown in Figure~\ref{fig:cavity PL ImPhi ReT pert}, deforms both $\Re(T)$ and $\Im(\Phi)$ at the final time. Again all four fields are of comparable size, and we are well beyond the GHP regime. This was the maximum amplitude of deformation that we were able to find with this functional form. We find that $\max|\Im(\Phi - T)| \approx 0.61\,\pi$ for the flow, so it remains well within the bound imposed by well-posedness. Hence the fact that we cannot find larger amplitude deformations here does not seem to be due to slightly greater amplitudes violating well-posedness, and presumably is due to our complex gauge choice.

\begin{figure}[t]
    \centering
    \subfigure[Case 2: $r_T$]{%
        \includegraphics[width=0.48\textwidth]{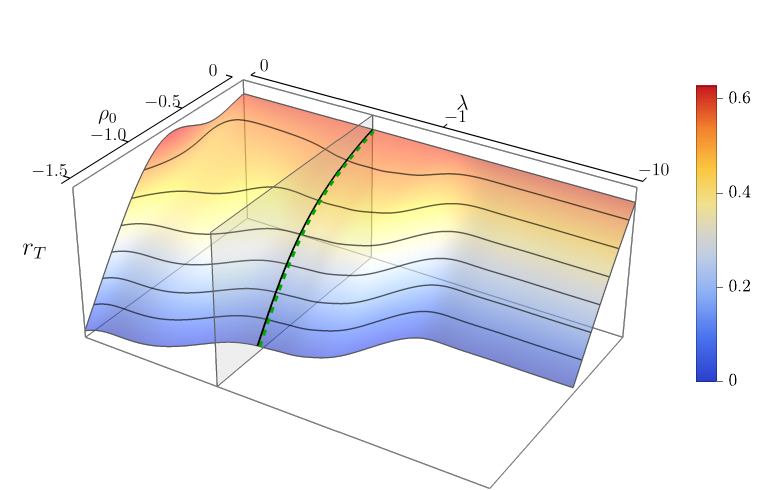}%
        \label{fig:ref_rT_case2}%
    }
    \hfill
    \subfigure[Case 2: $r_A$]{%
        \includegraphics[width=0.48\textwidth]{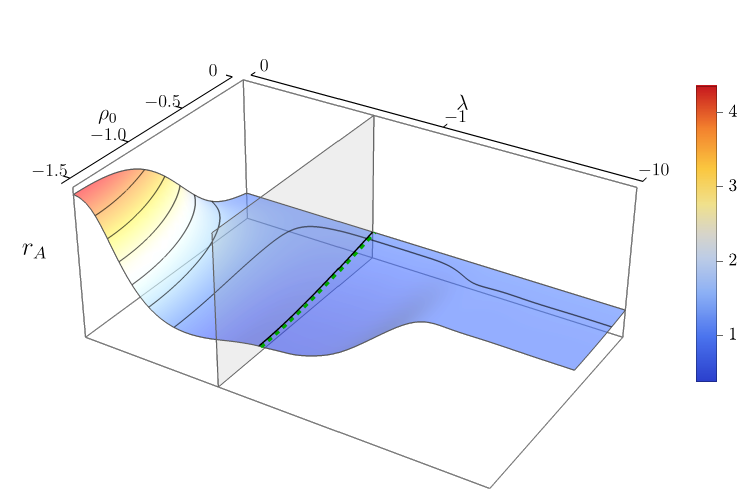}%
        \label{fig:ref_rA_case2}%
    }
    \\[2mm]
    \subfigure[Case 3: $r_T$]{%
        \includegraphics[width=0.48\textwidth]{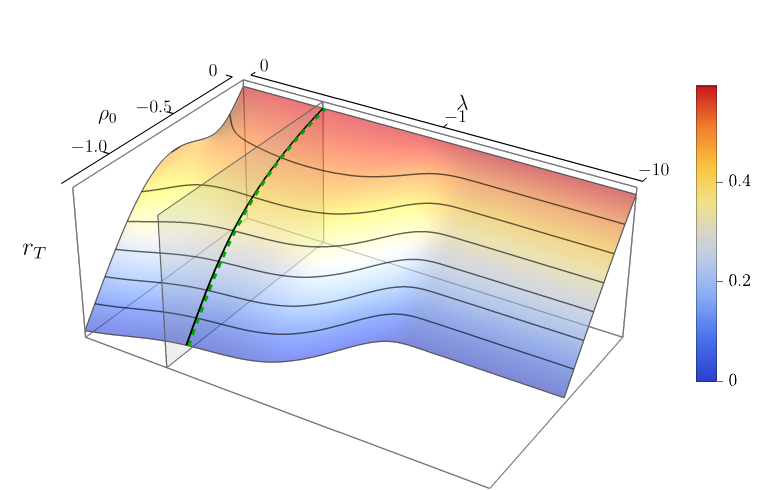}%
        \label{fig:ref_rT_case3}%
    }
    \hfill
    \subfigure[Case 3: $r_A$]{%
        \includegraphics[width=0.48\textwidth]{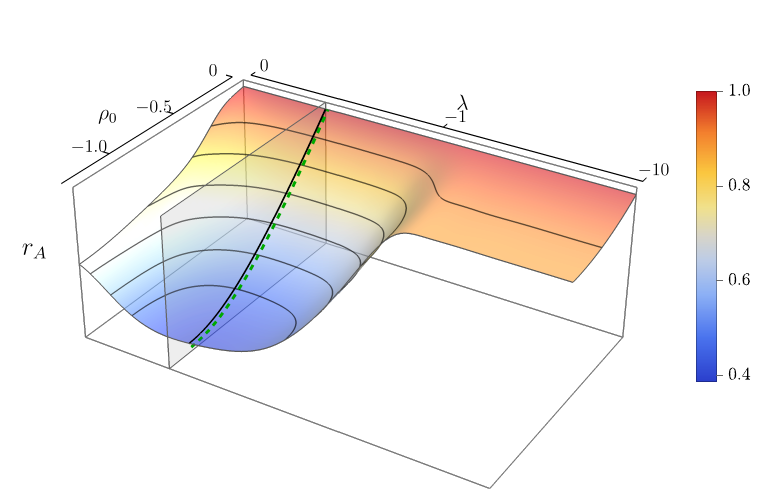}%
        \label{fig:ref_rA_case3}%
    }
    \caption{The reference metric $\mathfrak{g}$ along the complex flows away from the large black hole: the Euclidean time circle radius $r_T$ (left) and the areal radius $r_A$ (right), against the proper distance $\rho_0$ from the cavity wall in $\mathfrak{g}$ and the flow time $\lambda$. We have included a cross section where the reference metric is closest to the small BH.}
    \label{fig:g0 comparison}
\end{figure}

Before we move on to the next section where we discuss the action, it is worth considering how the real reference metric, $\mathfrak{g}$, evolves. Figure~\ref{fig:g0 comparison} shows this for cases 2 and 3. This gives a measure of the real part of the complex metric $\mat{g}$ that is evolved in the flow. Since the large black hole fixed point is real, the reference metric asymptotes to this in the far past. As the flow proceeds, the reference metric flows away from the large black hole, but very interestingly, and for both cases, then appears to flow towards the small black hole with the Hawking--Page temperature. The reference metric pauses near the small black hole geometry, before then subsequently flowing away. It is not clear to us why this behaviour occurs, and it would be interesting to understand it better.

\subsection{The action along the example flows}

Having examined examples of flows that asymptote from the de Sitter and large Schwarzschild black hole saddle points, we now consider how the Einstein--Hilbert action varies along these flows. The Einstein--Lefschetz flow is constructed so that the imaginary part of the action is constant and the real part increases monotonically. Since we are starting our flows at real fixed points, the action has vanishing imaginary part asymptotically in the past, and hence remains real during the flow. Indeed this provides an excellent global numerical check of accuracy, and in Appendix~\ref{app:convergence} we use it to show convergence for our numerical solutions.

\begin{figure}[tbp]
\centering
    \subfigure[de Sitter, cases 2 and 3.\label{fig:dS action comparison}]{%
        \includegraphics[width=0.47\textwidth]{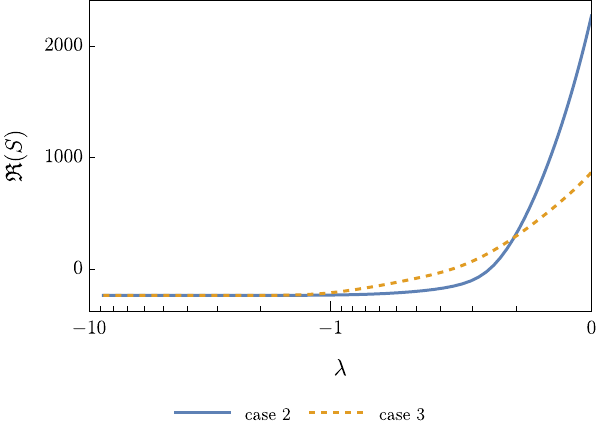}}%
    \hspace{0.02\textwidth}%
    \subfigure[Large black hole, cases 2 and 3.\label{fig:cavity action comparison}]{%
        \includegraphics[width=0.47\textwidth]{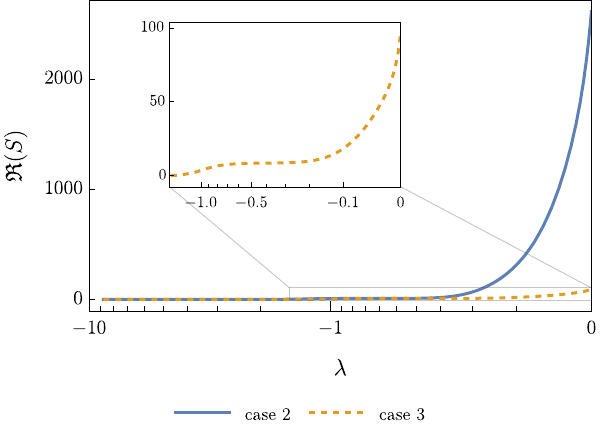}}
    \caption{Real part of the Euclidean action along the flows away from the de Sitter and large black hole saddles, for $\lambda \in [-10, 0]$. The actions start at their saddle values, $-24\pi^2$ and $0$ respectively, and increase monotonically. We observe that the deformations of just $\Im(\Phi)$ (case 2) lead to a greater increase in the action than the deformations of both $\Im(\Phi)$ and $\Re(T)$ (case 3) across both sets of flows.}
    \label{fig:action comparison}
\end{figure}

In Figure~\ref{fig:action comparison} we show the real part of the action along the example flows given by cases 2 and 3 for both de Sitter and the large Schwarzschild black hole. We show the flow time over the range $\lambda \in [-10, 0]$. For de Sitter, the action starts at its saddle point value $-24\pi^2$, while for the large black hole it starts at zero, since we have chosen to give example flows away from  the large black hole in a cavity at the Hawking--Page temperature $\beta = 32\pi/27$. This is precisely the temperature at which the large black hole has the same action as hot flat space, whose action vanishes. In this black hole case the action computed includes the Gibbons-Hawking term as we are fixing the induced metric at the cavity wall, and a subtraction so that hot flat space has zero action.
In all cases the real part increases monotonically with flow time as it should. We have seen above that cases 2 and 3 for both spacetimes represent non-perturbative deformations, and this is confirmed by the large change in the action over the course of the flow. In the $\hbar \to 0$ semiclassical limit, this implies that the geometries towards the end of the flow are very highly suppressed as they are integrated over in the path integral taken over this Lefschetz thimble.

\FloatBarrier

\section{Outlook}\label{sec:Outlook}

This work has been motivated by Witten's suggestion that the Euclidean gravitational path integral in the semiclassical regime might be evaluated as a sum of integrals over the Lefschetz thimbles of saddle points of the gravitational action~\cite{witten2022notecomplexspacetimemetrics}. At the perturbative level this is the scheme introduced by Gibbons--Hawking--Perry where the conformal factor perturbations are Wick rotated~\cite{Gibbons:1978ac}.
Here we have given a concrete prescription for computing the Lefschetz thimble for a saddle point of the Euclidean gravitational path integral, appropriate for the semiclassical limit.
As a stepping stone to achieve this we have introduced the Einstein flow for Riemannian geometries.
This is similar to the Ricci flow, although the flow is in the direction of the Einstein tensor rather than (minus) the Ricci tensor. It is not a parabolic flow, but we have shown that it can be solved as a forward-backward flow, with the conformal factor flowing forwards and the conformal class flowing backwards.
It has the property that it is a gradient ascent of the Einstein--Hilbert action, so that the action varies monotonically in flow time.

We also observed that the Lefschetz flow for a scalar field in Euclidean signature takes the form of a forward-backward flow. Indeed we expect that the Lefschetz flow for any real field theory has a forward-backward character.
Following from this we have introduced the Einstein--Lefschetz flow. This is a flow in the space of complex metrics, and is a gradient flow of the analytically continued Einstein--Hilbert action so that the real part of the action monotonically increases along the flow, and the imaginary part is constant. Hence this generates flows of complex metrics away from saddle points of the action which lie within a Lefschetz thimble of that saddle point. 
We have shown that this again has a forward-backward character, and have explained how to give data for a flow from a fixed point at a finite flow time. 
Essentially this can be thought of as specifying the real conformal class and the imaginary conformal factor at the end of the flow.
The thimble of a saddle point is then generated by the collection of all such flows given by varying this data.
Our Einstein--Lefschetz flow is constructed so that in the case that the data is chosen so the flow is real, it reduces to a solution of the real Einstein flow.

Our flow is not unique, as there is a choice we have made for the metric on the space of metrics. In addition our flow involves a real reference metric that is constructed from the complex metric, and it seems that there may be many choices one can make there. These different choices will yield different Lefschetz thimbles for a saddle point. Furthermore, away from a fixed point, the path generated by the flows in the space of complex metrics depends on the choice of complex gauge due to the real reference metric breaking the complex diffeomorphism invariance, and likewise for $\overline{\delta S/\delta g_{\mu\nu}}$. Thus given a flow solution, a complex diffeomorphism of this is no longer a solution. Hence the choice of complex gauge also changes the thimbles the flow generates.
However, small changes to the thimble, due to differing choices, should not change the value of $e^{- \frac{1}{\hbar} S}$ integrated over the thimble. This does, however, naturally lead to the question of whether the choices we have made are good ones, and whether there are choices that are optimal in some sense.

The way we have written our complex metric as $\mat{g} = e^\phi e^{\frac{1}{2} \mat{X}'} e^{i \mat{Y}' } e^{\frac{1}{2} \mat{X}'}$ in equation~\eqref{eq:ELmetric}, and have chosen the real reference metric as $\mathfrak{\mat{g}} = e^{\Re(\phi)} e^{\mat{X}'}$ yields a characteristic analysis that for $3$ and $4$-dimensions gives a very elegant condition necessary for well-posedness, namely that in equation~\eqref{eq:ELconditions}, and simple final data for flows where (for a stable saddle point) we fix $\Im(\phi)$ and the traceless matrix $\mat{X}'$ at the final flow time, and impose an appropriate gauge condition on $\Re(\phi)$ and $\mat{Y}'$. 
Curiously the condition~\eqref{eq:ELconditions} is weaker than that of KSW, in principle allowing us to construct flows outside the KSW allowable domain of complex metrics.
We have experimented with writing the complex metric in other ways, and taking different choices for the reference metric, such as the obvious choice $\mathfrak{\mat{g}} = \Re(\mat{g})$, but then imposition of the final boundary conditions, and also the condition required by well-posedness of the flow, become very complicated. A very interesting direction for future work is to understand better whether there is a natural structure that underlies our choice for the form of the metric, and our choice of reference metric, and explains the resulting simplicity that these choices yield. We have only considered $3$ and $4$-dimensions in detail here, and a future task is to understand the characteristic analysis for the case of higher dimensions. The gauge choice required for the general characteristic analysis of the boundary conditions in Sections~\ref{sec: complex flow character} and~\ref{sec:complexBC} does not appear to be simple to implement in a numerical setting. While the transverse condition in~\eqref{eq:gauge} is relatively simple in Fourier space, it naively looks unpleasant in position space. Indeed in our numerical examples we employed different choices of complex coordinates, but nonetheless these lead to the same condition~\eqref{eq:ELconditions} necessary for well-posedness. A natural question is then whether there is a more convenient complex gauge that gives this condition, or perhaps even a gauge that gives an even weaker well-posedness constraint. 

For small perturbations of a real fixed point, in simple cases, the thimble our Einstein--Lefschetz flow constructs reproduces the GHP prescription -- conformal  perturbations are Wick rotated. 
However, as emphasized by Marolf and Santos~\cite{Marolf_2022}, in the case that the manifold has boundary conditions where conformal factor and class perturbations are coupled in linear theory, such as for black holes in a cavity, the GHP prescription can not be applied. Our flows will generate the perturbative portion of the thimble in this case. We emphasize that the thimble that we generate is different to that proposed by Marolf and Santos for perturbations, and it would be interesting to better understand whether both prescriptions are simply different choices of thimble, or whether they are more fundamentally different. 

The main point we emphasize is that our construction is the first one to be fully non-perturbative. Both GHP and the proposal in~\cite{Marolf_2022} are only perturbative schemes.
Thus, at least in principle, our Einstein--Lefschetz flows allow a Lefschetz thimble to be explicitly constructed far from the fixed point in the non-linear regime.

An omission in this work is that we have not considered the dual Lefschetz flow. This is simply the Lefschetz flow in the opposite flow direction, so reversing the sign in the flow equation. It would be interesting to understand the boundary conditions for this flow, which naively are to fix the real conformal factor and the imaginary conformal class, and to construct explicit dual flows. These dual flows are very important in Picard--Lefschetz theory as they allow one to understand whether a particular saddle point contributes to an integral with a given  integration contour. Whether a given saddle point contributes is determined by considering the dual flows towards the saddle point, and in particular how those dual flows intersect the original contour of integration. 
In our work here we have only considered examples of real saddle points, and these are always expected to contribute to the Euclidean gravitational path integral. However in future work it would be very interesting to consider complex saddle points, and then it is not at all obvious whether they should contribute or not, and this might in principle be resolved by understanding the dual Einstein--Lefschetz flow.

Taking a step back, one might wonder why such a construction would be useful, given that the whole point of the semiclassical approximation is that in the end only the saddle point geometry itself is required in the leading $\hbar \to 0$ limit. While this is certainly true, at a formal level it is very important to understand what the integration contour that justifies this leading saddle point approximation is. Furthermore, if we are interested in corrections to the leading saddle point approximation, then one has to understand the structure of the thimble near the saddle. While the leading correction may only require understanding the thimble for linear perturbations, which give the leading functional determinant correction to the saddle point approximation, it is important that there is a systematic scheme for going to higher orders in perturbation theory. We may regard our non-perturbative  formulation as precisely such a framework to allow, in principle, all orders of quantum corrections to a saddle point to be deduced. Even for the leading correction from linear perturbations, in the case that the conformal modes and transverse-traceless modes are coupled so GHP does not apply, one needs a scheme to know how to perform the 1-loop integration. We emphasize that our flow constructs perturbative thimbles in such a setting, and further also in the more complicated case that one considers a saddle point which is itself a complex geometry.

There are many open questions arising from this work. 
We have not proved that the flow is well-posed at the level of mathematical rigour here, and it would clearly be important to do so.
Having given a non-perturbative prescription for construction of the integration cycle of a Euclidean saddle-point certainly does not resolve the issue of existence of the Euclidean path integral. Many questions will remain from our analysis. In particular the question of whether the increase in the real part of the action  away from a saddle point is sufficiently quick to yield a convergent integral is something we have not addressed here.
Understanding the global properties of these flows, and hence of the Lefschetz thimbles will remain a difficult task. In particular, how could the integration cycles be constructed beyond the neighbourhood where the forward-backward flows are well-posed and exist? Indeed, given that we have found the condition required for well-posedness is weaker than the KSW allowability condition, does it make sense to consider all the complex metrics these flows produce, or should the semiclassical path integral be restricted only to the KSW allowable ones?
It would be interesting to add matter fields to the path integral and presumably one can solve a combined Lefschetz flow for  gravity and matter together. With matter there may be many complex saddle points to consider. While we have only constructed flows from real saddle points, it would be worthwhile to understand how flows behave for complex ones. Lastly, in 3-dimensions gravity can be reformulated as a Chern--Simons theory~\cite{Achucarro:1986uw,Witten:1988hc}, and it would be very interesting to understand whether our flow looks natural in that case, or whether Chern--Simons suggests a more elegant or robust formulation.

While many questions remain, we emphasize that having a non-perturbative construction for the Lefschetz thimble of a gravitational saddle point allows these questions to be  explicitly addressed in future work, hopefully clarifying the status of the Euclidean gravitational path integral in the semiclassical regime.
As a final comment it would be very interesting to understand whether a similar Lefschetz flow could be constructed for Lorentzian gravity. Timelike perturbations generally flow in the opposite sense to spacelike ones, but since our Euclidean flows already have a forward-backward character, perhaps this added Lorentzian complication is straightforward to accommodate.

\section*{Acknowledgments} 
We are grateful to Jorge Santos and to Harvey Reall for useful discussions. This work was supported by the STFC Consolidated Grant ST/X000575/1. NR was supported by funding from The Sir Richard Stapley Educational Trust.
OpenAI's ChatGPT and Anthropic's Claude were used to proofread the final version of the paper, and helped with useful discussion on the gauge choice~\eqref{eq:metpert} and Appendix~\ref{app:metricform}, and typesetting of figures, but otherwise AI was not used in this work.

\appendix

\section{Einstein flow unstable mode for small black holes}
\label{app:negative}

The linearized Ricci tensor for a Schwarzschild black hole in a cavity (with Dirichlet boundary conditions) famously has a negative mode as shown in~\cite{Gregory:2001bd}, which is a generalization of the GPY mode of the asymptotically flat solution~\cite{GPY}. 
Since the GPY mode is purely transverse-traceless for the asymptotically flat Schwarzschild solution, it is also a negative eigenmode of the linearized Einstein tensor. It is this linearized Einstein tensor that is of interest here as it controls the stability of the black hole in a cavity for our Einstein and Einstein--Lefschetz flows. However in a cavity, the conformal perturbations are mixed with the transverse-traceless ones by the cavity wall boundary conditions, so that the negative mode for the linearized Ricci tensor will not be an eigenfunction of the linearized Einstein tensor. For small black holes in a cavity we again expect the linearized Einstein tensor to admit a mode that is approximately the GPY mode, since the GPY mode localizes near the horizon and hence will not `feel' the cavity.
We also know that there is a zero mode at the minimum temperature solution for both the linearized Ricci and Einstein tensors, which is simply given by the tangent to the space of Schwarzschild solutions. For a black hole with horizon radius $r_0$ in a unit radius cavity, this minimum temperature solution is at $r_0 = 2/3$.
A reasonable guess is then that the linearized Einstein tensor has a negative eigenmode that behaves similarly to that for the Ricci tensor as $r_0 \to 0$, disappears at $r_0 = 2/3$, and interpolates in a qualitatively similar manner to that for the Ricci tensor for intermediate $r_0 < 2/3$. The presence of this additional negative mode for the linearized Einstein tensor would then suggest that the small Schwarzschild fixed point will be unstable for the Einstein and Einstein--Lefschetz flows.

\begin{figure}[h]
    \centering{
        \includegraphics[width=\textwidth]{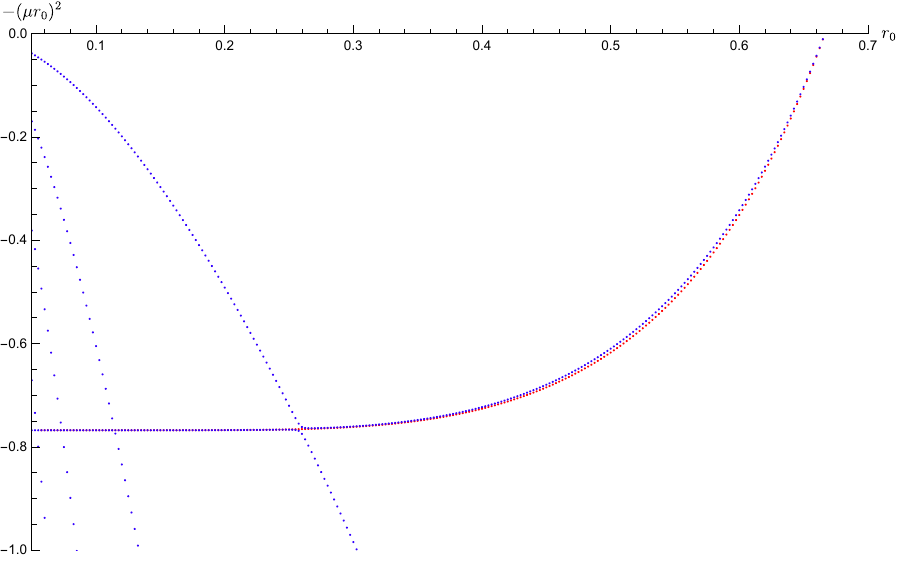}
    }
    \caption{
    The blue points show the smallest amplitude negative eigenvalues of the linearized Einstein tensor in the static spherical sector as a function of the black hole horizon size $r_0$ in a unit radius cavity.     
    The red points are the same for the linearized Ricci tensor. The eigenvalues are rescaled by $r_0^2$, and the red points show the usual single negative mode in the cavity. For small $r_0$ this tends to the GPY mode, with $\mu^2 r_0^2 = 0.768$. This mode of the linearized Ricci tensor ceases to have negative eigenvalue precisely at the point where the black hole has minimum temperature, $r_0 = 2/3$.
    We see from the blue points that the linearized Einstein tensor also has a negative mode that very closely follows that of the Ricci tensor, although as discussed in the main text, it is not exactly the same. For small $r_0$ it also tends to the GPY mode, as expected, and it also ceases to have negative eigenvalue at $r_0 = 2/3$, again as expected as it becomes the zero mode tangent to the space of Schwarzschild solutions there.
    The fractional difference between the eigenvalue of the mode for the Einstein and Ricci tensors is at most $\sim 3 \%$. 
    A key difference is that the linearized Einstein tensor has other negative modes, which on flat space are the conformal factor perturbations. Looking closely we can see these slightly disrupt the generalization of the GPY mode as they appear to cross it. In fact the modes do not cross, and instead repel each other.
    \label{fig:negativemode}
    }
\end{figure}

In Figure~\ref{fig:negativemode} we plot the smallest amplitude negative eigenvalues, $- \mu^2$, of the linearized Einstein tensor (as blue points), restricting to the static spherical sector of perturbations. We also plot the negative mode of the linearized Ricci tensor (as red points). These are computed in a similar manner to the computations of spectra in~\cite{Liu:2024ymn}.
We choose to normalize the eigenvalues by $r_0$ as $- (\mu r_0)^2$ so that as $r_0 \to 0$ we tend to the GPY behaviour where $(\mu r_0)^2 \to 0.768$.  

At first sight our naive expectation holds true, namely that for all small black holes, so $r_0 < 2/3$, there exists a static spherically symmetric negative mode of the linearized Einstein tensor that is very close in eigenvalue to the negative mode of the linearized Ricci tensor. Looking at the actual wavefunctions we see these are also similar. 

However there is a more subtle and interesting phenomenon due to the fact that the linearized Einstein tensor has negative modes associated to conformal factor perturbations. These modes appear to `cross' the path of the generalized GPY mode. As we reduce $r_0$ from the point where the negative mode first appears, $r_0 = 2/3$, then this first occurs at $r_0 = 0.26$. It then occurs again at $r_0 = 0.12$, and this appears to repeat, presumably infinitely many times as $r_0 \to 0$. On closer inspection these negative eigenvalues do not cross, but rather repel each other. Hence it appears that if one spectrally follows the negative mode from $r_0 = 2/3$ to smaller $r_0$, it is not this mode that approximates the generalized GPY mode of the linearized Ricci tensor after the interaction with the incoming negative mode associated to conformal perturbations. Instead for $r_0 < 0.26$ that role falls to another mode which originated for larger $r_0$ as a conformal perturbation. Hence the structure and behaviour of this GPY analog for the linearized Einstein tensor seems to be much more subtle than one might have naively first thought. 

For our work here this will not be relevant, as  the example flows we show  involve small black holes with $r_0 > 0.26$. For these black holes, the linearized Einstein tensor negative mode is close to that of the linearized Ricci tensor, differing in eigenvalue by at most a fractional difference of $\sim 3\%$. However, it would be very interesting to explore the Einstein flows involving small black hole fixed points with temperatures such that they are near these negative mode crossing values, so $r_0 = 0.26, 0.12, \ldots$, as there may be interesting and unexpected behaviour near these values. We leave this for future investigation.

\section{Einstein flow of Schwarzschild in a cavity}
\label{App:Sch}

\begin{figure}[h]
    \centering
    \subfigure[Time circle and areal radii]{%
        \includegraphics[width=\textwidth]{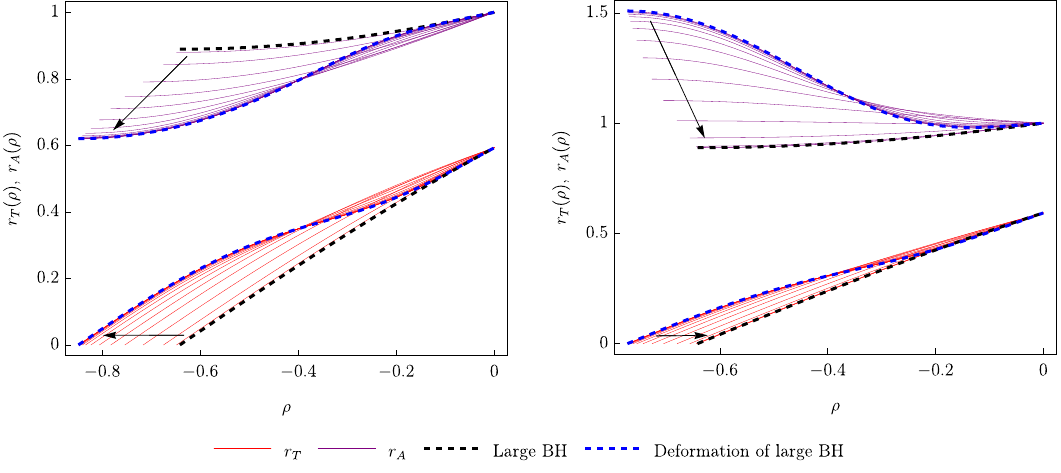}%
        \label{fig:einstein_flow_large_radii}%
    }
    \\[2mm]
    \subfigure[Action, flow away]{%
        \includegraphics[width=0.48\textwidth]{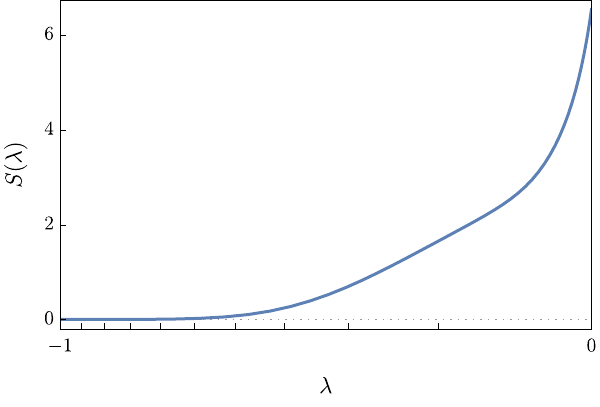}%
        \label{fig:einstein_flow_large_action_away}%
    }
    \hfill
    \subfigure[Action, flow towards]{%
        \includegraphics[width=0.48\textwidth]{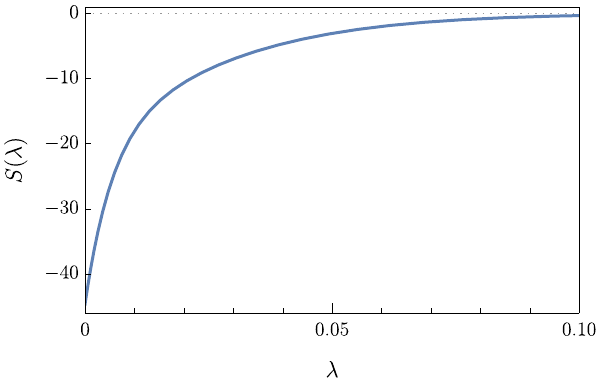}%
        \label{fig:einstein_flow_large_action_toward}%
    }
    \caption{Einstein flow away from (left) and towards (right) a large black hole at the Hawking--Page temperature in a cavity of radius $R=1$. In (a) we plot the Euclidean time circle radius $r_T$ and the areal radius $r_A$ against the proper radius $\rho$. In (b) and (c) we plot the action along the two flows; the dotted lines mark the action of the large black hole. The thin curves are spaced logarithmically in flow time, four per decade over $10^{-4} \le |\lambda| < 1$; by $|\lambda| \sim 1$ the geometry is indistinguishable from the fixed point at plotting resolution.}
    \label{fig:einstein_flow_large}
\end{figure}

For completeness, here we show example real Einstein flows towards and away from small and large Schwarzschild black hole fixed points. For both we choose the cavity to be at the Hawking--Page temperature, as in the main text, so we take $\beta = 32 \pi/ 27$. 
We  begin with the large black hole. Then taking the metric ansatz~\eqref{eq: BH metric with T and Phi} the large black hole has $r_0 = 8/9$ and vanishing $\Phi$ and $T$. This case is comparatively more straightforward than the small black hole since the large black hole is a stable fixed point, meaning that no tuning of final flow data is required. 
In Figure~\ref{fig:einstein_flow_large} we give two example flows, showing how the radius of the Euclidean time circle and the areal radius vary along the flow, as measured with respect to the proper distance $\rho$ from the cavity wall, using the same conventions as in Section~\ref{sec:bhflow} in the main text. In both cases the flow time is compactified as in the main text, so for flows towards the fixed point we compactify as  in equation~\eqref{eq:compflowtime} and for a flow away from the fixed point we compactify as ~\eqref{eq:away compact flow time}, and in both cases we choose $a=10$ for the examples given.
The first flow shown is a flow towards the large black hole, where the initial data for the flow is given by prescribing the initial conformal factor at $\lambda = 0$ as $\Phi(0,r)  = (1 - r^2)^4$. The second flow is one away from the large black hole, and we have fixed the final conformal class at $\lambda = 0$ as $T(0,r) = (1 - r^2)^4$.
Interestingly, for this latter flow  away from the large black hole, we do not see the same behaviour that was found in Figure~\ref{fig:g0 comparison} for the complex flow,  where the flow of the real reference metric seems to pass near to the small black hole saddle.

We now consider the small black hole which is more complicated as it is an unstable fixed point due to an analog of the GPY mode. As discussed in the main text, we expect that for a small black hole we have to tune the class data in order to asymptote from the fixed point. Likewise for a flow to the fixed point we should find we can fix the conformal factor and put one constraint on the class too.
Choosing the cavity to be at the Hawking--Page temperature then for our ansatz~\eqref{eq: BH metric with T and Phi} we have $r_0 = (1 + \sqrt{33})/18\approx 0.375 $ and vanishing $\Phi$ and $T$, as well as the diffeomorphism vector vanishing, so $F = 0$, at the fixed point.

\begin{figure}[h]
    \centering
    \subfigure[Time circle and areal radii]{%
        \includegraphics[width=\textwidth]{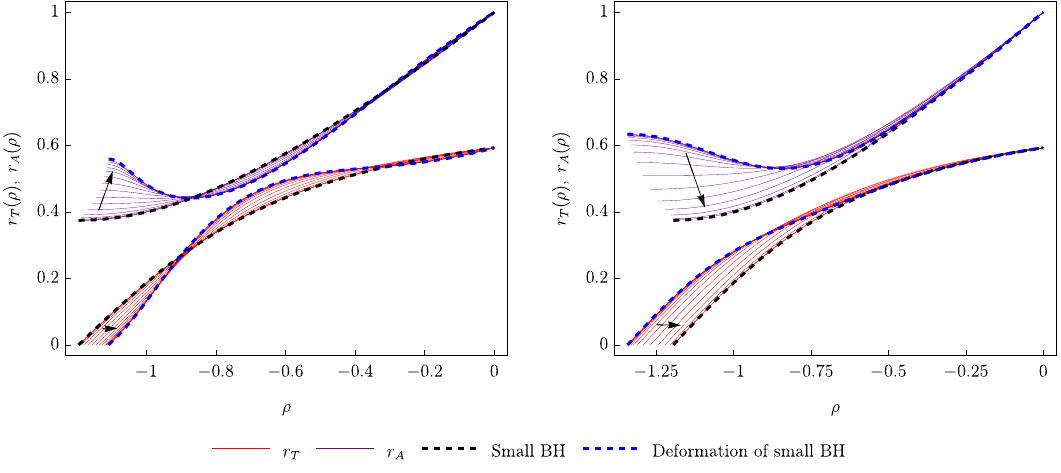}%
        \label{fig:einstein_flow_small_radii}%
    }
    \\[2mm]
    \subfigure[Action, flow away]{%
        \includegraphics[width=0.48\textwidth]{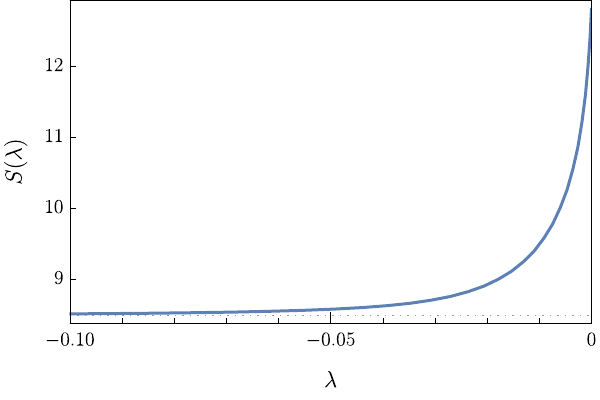}%
        \label{fig:einstein_flow_small_action_away}%
    }
    \hfill
    \subfigure[Action, flow towards]{%
        \includegraphics[width=0.48\textwidth]{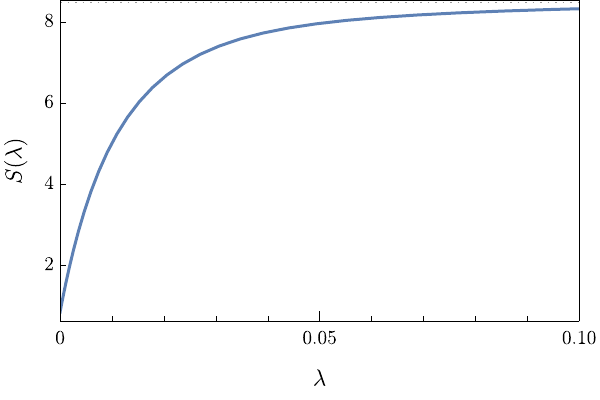}%
        \label{fig:einstein_flow_small_action_toward}%
    }
    \caption{Einstein flow away from (left) and towards (right) the small black hole at the Hawking--Page temperature in a cavity of radius $R=1$. In (a) we plot the areal radius and Euclidean time circle radius against proper distance from the cavity wall, and in (b) and (c) the action along the two flows. For the flow away from the fixed point, the final conformal class data must be tuned to allow the flow to asymptote to the fixed point in the past. The thin curves are spaced logarithmically in flow time, four per decade over $10^{-4} \le |\lambda| < 1$; by $|\lambda| \sim 1$ the geometry is indistinguishable from the fixed point at plotting resolution.}
    \label{fig:einstein_flow_small}
\end{figure}

For the flow towards the fixed point we may compactify the flow time in the usual way, and the flows are easy to find. We show an example in  Figure~\ref{fig:einstein_flow_small} of a flow where the initial conformal factor is deformed as for the large black hole case, so $\Phi(0,r) = (1 - r^2)^4$. 
The additional constraint on the initial class we choose to impose simply by fixing the value of the function $T$ at the horizon at the initial flow time. We choose $T(0,0) = -0.05$ for the example flow shown.

For the example flow away from the fixed point we must tune the final data. We choose the class as,
\be
T(0,r) &= (1 + 5 r^2)(1 - r^2)^2 + \alpha \, \delta T_{GPY}(r)
\ee
where $\delta T_{GPY}(r)$ is the numerically determined unstable mode perturbation to $T$, normalized so that $\delta T_{GPY} = 1$ at the horizon, which is the analog of the GPY mode of the linearized Ricci tensor, although now computed for the linearized Einstein flow.  The amplitude $\alpha$ is the constant that must be tuned. While the flow is not a linear perturbation of the small black hole, perturbing by the functional form of the linear unstable mode is a reasonable way to tune the final data, and is compatible with the boundary and regularity conditions.
Rather than compactifying the flow time, we instead use a computational flow time,
\be
\lambda(y) = \frac{1}{a}  -\frac{a + \epsilon}{a\left( a\, y + \epsilon\right)} 
\ee
so that $y \in [0, 1]$ is mapped to $\lambda \in [ -1/\epsilon, 0]$, so taking $\epsilon \to 0$ gives the full flow.
In practice we find solutions for small $\epsilon$, and tune the value $\alpha$ by hand to ensure the flow does not blow up in the past. Once one has a reasonable solution for modest $\epsilon$, one can use this as a seed initial guess for one with smaller $\epsilon$. In this way we find that $\alpha = - 2.165$, and we can reduce $\epsilon$ to $10^{-4}$.

\section{Derivation of form of complex metric}
\label{app:metricform}

Given our real frame and complex metric $g_{\mu\nu} = \vect{\hat{\theta}}_\mu \mat{g} \vect{\hat{\theta}}_\nu$ then the condition given in~\eqref{eq:ELconditions} can be written as,
\be
\label{eq:cond}
\Re\left( \left( ( \sqrt{\det(\mat{g})}) ( \vect{q}^T \mat{g}^{-1} \vect{q} ) \right)^{1/2} \right) > 0
\ee
for any non-zero real covector $q_\mu = \vect{q}  \vect{\hat{\theta}}_\mu$. Thus at a point this implies that,
\be
\vect{q}^T \mat{H} \vect{q} \notin ( - \infty, 0 ] \; , \quad \mat{H} = \sqrt{\det(\mat{g})}  \mat{g}^{-1} \; .
\ee
We claimed in~\eqref{eq:metricform} that the complex metric could then be written as,
\be
\mat{g} = e^{ \frac{1}{2} \mat{{X}}} e^{i \mat{{Y}}}  e^{\frac{1}{2} \mat{{X}}}
\ee
and we now show this is the case. We will then show that given a metric $\mat{g}$, while $\mat{Y}$ is not unique, the matrix $\mat{X}$ is.

We begin by decomposing the symmetric $\mat{H}$ into real and imaginary parts $\mat{H} = \mat{A} + i \mat{B}$. Then the condition above implies,
\be
\vect{q}^T \mat{B} \vect{q}  = 0 \quad \implies \quad \vect{q}^T \mat{A} \vect{q}   > 0
\ee
and Finsler's lemma then states that there exists $\theta$ such that at a point,
\footnote{We thank ChatGPT for a useful discussion on this construction of a positive matrix $\mat{R}$.}
\be
\mat{R}  = \cos(\theta) \mat{A} + \sin(\theta) \mat{B}
\ee
has strictly positive eigenvalues. Hence writing,
\be
e^{- i \theta} \mat{H} = \tilde{\mat{H}} = \tilde{\mat{A}} + i \tilde{\mat{B}}
\ee
we have that $\tilde{\mat{A}}$ is positive definite. Then we may compute,
\be
\mat{S} = \mat{\tilde{A}}^{-1/2} \mat{\tilde{B}} \mat{\tilde{A}}^{-1/2} 
\ee
where the key point is that since $\tilde{\mat{A}}$ is positive, we may take the inverse square root $\mat{\tilde{A}}^{-1/2}$.
Then we define the real symmetric $\mat{\tilde{X}}$ from,
\be
e^{-\mat{\tilde{X}}} =  \mat{\tilde{A}}^{1/2} \left( 1 + \mat{S}^2 \right)^{1/2}  \mat{\tilde{A}}^{1/2}
\ee
and we note that since $\mat{S}$ is symmetric and real, then $(1 + \mat{S}^2)$ is positive definite, so again the square roots on the righthand side can be taken. Since the righthand side is positive definite, then we may take the log to obtain the real symmetric $\mat{\tilde{X}}$.

Having found $\mat{\tilde{X}}$ then we may write,
\be
\mat{U} = e^{ \frac{1}{2} \mat{\tilde{X}}} \tilde{\mat{H}} e^{ \frac{1}{2} \mat{\tilde{X}}}  \; .
\ee
This is symmetric, and one can also check that it is unitary given our form for $e^{-\mat{\tilde{X}}}$ above. Thus we can write $\mat{U} = e^{-i \mat{\tilde{Y}}}$ for real symmetric $\mat{\tilde{Y}}$. Then we have,
\be
\sqrt{\det(\mat{g})}  \mat{g}^{-1} = {\mat{H}}  = e^{ i \theta} e^{-\frac{1}{2} \mat{\tilde{X}}} e^{-i \mat{\tilde{Y}}}  e^{-\frac{1}{2} \mat{\tilde{X}}}
\ee
and inverting we find,
\be
 \mat{g}  = \sqrt{\det(\mat{g})} e^{ - i \theta} e^{ \frac{1}{2} \mat{\tilde{X}}} e^{i \mat{\tilde{Y}}}  e^{\frac{1}{2} \mat{\tilde{X}}}
\ee
and then defining,
\be
\label{eq:XY}
\mat{{X}} =  \Re\left( \log\left( \sqrt{\det(\mat{g})} \right) \right) \mat{1} + \mat{\tilde{X}} \; , \quad \mat{{Y}} = \left(  \Im\left( \log\left( \sqrt{\det(\mat{g})} \right) \right) -  \theta \right) \mat{1} + \mat{\tilde{Y}}
\ee
gives the form in equation~\eqref{eq:metricform}. 

Let us now show that given $\mat{g}$ then writing it as $\mat{g} = e^{ \frac{1}{2} \mat{{X}}} e^{i \mat{{Y}}}  e^{\frac{1}{2} \mat{{X}}}$ there is a unique choice for $\mat{X}$. Firstly we note that,
\be
\mat{g} e^{- \mat{X}} \mat{g}^\dagger = e^{\mat{X}} \; .
\ee
Suppose that $\mat{g}$ can also be written as $\mat{g} = e^{ \frac{1}{2} \mat{{X}}'} e^{i \mat{{Y}}'}  e^{\frac{1}{2} \mat{{X}}'}$ for different real symmetric $\mat{X}'$ and $\mat{Y}'$. Then the above relation becomes,
\be
e^{ - \frac{1}{2} \mat{{X}}'} e^{\mat{X}} e^{-\frac{1}{2} \mat{{X}}'} = e^{i \mat{{Y}}'}  e^{\frac{1}{2} \mat{{X}}'} e^{- \mat{X}}  e^{ \frac{1}{2} \mat{{X}}'} e^{-i \mat{{Y}}'}   \; .
\ee
We now write $\mat{Z} = e^{ - \frac{1}{2} \mat{{X}}'} e^{\mat{X}} e^{-\frac{1}{2} \mat{{X}}'}$. This is real symmetric, but importantly is positive definite. Hence it can unambiguously be written as $\mat{Z} = e^{\mat{C}}$ for real symmetric $\mat{C}$. Thus the above can be rearranged to,
\be
e^{\mat{C}} = e^{i \mat{{Y}}'} e^{-\mat{C}} e^{-i \mat{{Y}}'} \; .
\ee
Taking the log of this expression then gives,
\be
\mat{C} =  \log\left( e^{i \mat{{Y}}'} e^{-\mat{C}} e^{-i \mat{{Y}}'} \right) = e^{i \mat{{Y}}'} \log\left(  e^{-\mat{C}}  \right) e^{-i \mat{{Y}}'} = - e^{i \mat{{Y}}'} \mat{C} e^{- i \mat{{Y}}'} \; .
\ee
We may diagonalize $\mat{Y}'$ by an orthogonal $\mat{O}$ so $\mat{O} \mat{Y}' \mat{O}^T = \mathrm{Diag}\left( \theta_1, \theta_2 , \ldots \right)$ for real phases $\theta_i$, and then writing $\mat{C}' = \mat{O} \mat{C} \mat{O}^T$ we have,
\be
\mat{C}' = - 
\left(
\begin{array}{cccc}
e^{i \theta_1} & 0 & \ldots & 0\\
0 & e^{i \theta_2} & \ldots & 0 \\
\vdots & & & \vdots \\
0 & 0 & \ldots & e^{i \theta_n}
\end{array}
\right)
\mat{C}' 
\left(
\begin{array}{cccc}
e^{-i \theta_1} & 0 & \ldots & 0\\
0 & e^{-i \theta_2} & \ldots & 0 \\
\vdots & & & \vdots \\
0 & 0 & \ldots & e^{-i \theta_n}
\end{array}
\right) \; .
\ee
Considering a component,
\be
\mat{C}'_{ab} = - e^{i \theta_a} \mat{C}'_{ab} e^{-i \theta_b} 
\ee
we see that either $\mat{C}'_{ab}$ vanishes or else $e^{i ( \theta_a - \theta_b )} = - 1$ does. However our condition~\eqref{eq:cond} implies the latter cannot occur. If it did we could use the orthogonal diagonalization to choose the first two phases on the diagonal such that, $e^{-i \theta_1} + e^{-i \theta_2} =0$. But then if we consider a covector,
\be
\vect{q} = e^{  \frac{1}{2} \mat{X}' } \mat{O}^T 
\left(
\begin{array}{c}
1 \\ 1 \\ 0 \\ \vdots \\ 0
\end{array}
\right)
\ee
we would have $\vect{q}^T \mat{g}^{-1} \vect{q} = e^{-i \theta_1 } + e^{-i \theta_2 } = 0$, violating the condition~\eqref{eq:cond}. Thus we conclude that $\mat{C}' = 0$ and hence $\mat{C} = 0$ and so,
\be
e^{ - \frac{1}{2} \mat{{X}}'} e^{\mat{X}} e^{-\frac{1}{2} \mat{{X}}'} = \mat{1}
\ee
which implies $\mat{X} = \mat{X}'$. Therefore we conclude that $\mat{X}$ is indeed unique in this decomposition of the complex metric. The matrix $\mat{Y}$ is not unique, as we can already see from equation~\eqref{eq:XY} as the imaginary part of the log is not uniquely defined.

\section{Numerical Convergence}
\label{app:convergence}

In this work we have solved various flows which depend non-trivially on a flow time and one coordinate. The domains for the resulting two dimensional PDE problem are rectangular and we have discretised using Chebyshev pseudospectral differencing. The discrete system is then solved using the Newton method with double precision arithmetic, so that we start with an initial guess and iterate until we have reached the desired accuracy. We choose the tolerance to be sufficiently small that it is limited by machine precision. 
We have implemented the linear solve step in the Newton method using various direct solvers;  JAX, LAPACK and also UMFPACK. In the latter case we note that the linearization of the equations is sparse since there are no mixed two derivative terms -- the only two derivative terms are in the coordinate direction, not in flow time.

We observe an exponential convergence with increasing resolution for the flows considered as we expect for pseudospectral differencing where the solution is  smooth.
The results shown have been produced using $60$ grid points in the flow time direction, and $50$ points in the non-trivial coordinate direction. Accuracy is no longer improved going to higher resolutions due to the limitation of double precision arithmetic, and to obtain more accurate solutions we would be required to use higher precision arithmetic.

The solutions converge pointwise, but for the complex flows there is a particularly nice global measure of accuracy. The imaginary part of the Einstein--Hilbert action should be constant for the Einstein--Lefschetz flow. However numerical inaccuracy spoils this. For the real fixed points we consider the imaginary part of the action vanishes, so we may compare the imaginary action computed at the end of the flow to zero. It is important to note that while the imaginary action should vanish, the imaginary Lagrangian density certainly does not, so this provides an excellent consistency check.

\begin{figure}
    \centering
    \includegraphics[width=0.48\linewidth]{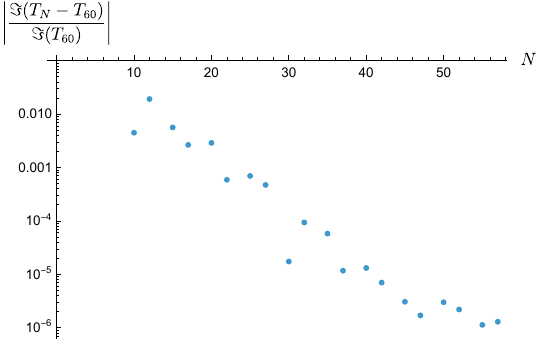} \hspace{0.25cm}
    \includegraphics[width=0.48\linewidth]{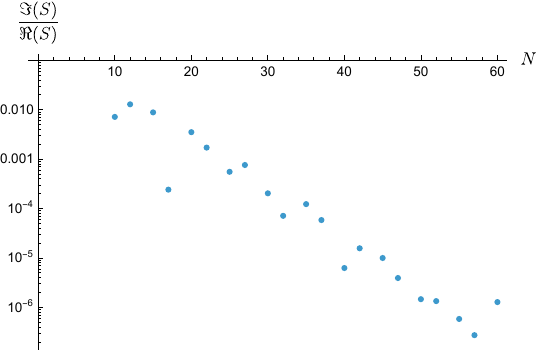}
    \caption{The lefthand plot shows the fractional difference of $T(0,0)$, so $T$ at the horizon and at the final flow time, computed for various resolutions $N \times N$, relative to that for $60 \times 60$. The righthand plot shows the imaginary part of the action, $\Im(S)$, at the final flow time normalized by the real part of the action for various resolutions $N \times N$. This should vanish in the continuum. We see that both $T(0,0)$ and $\Im(S)$ are consistent with exponential convergence to their continuum values with increasing $N$.}
    \label{fig:convergence}
\end{figure}

As an example numerical flow we consider the complex Einstein--Lefschetz flow from the large black hole at the Hawking--Page temperature, as considered in Section~\ref{sec:complexBHflow}, and here choose the final data $\Re(T) = \Im(\Phi) = (1 - r^2)^4$. This is a reasonably large amplitude perturbation, so certainly non-linear, but not as large as some of the deformations considered in the main text. We compute the flow on a variety of Chebyshev grids with $N \times N$ points. In the lefthand frame of Figure~\ref{fig:convergence} we show the pointwise convergence of the flow solution. We consider the largest value of $\Im(T)$, which occurs at the horizon, $r=0$, and the final flow time $\lambda = 0$. We emphasize this is not fixed by the final data, but instead is the response to it. We show the fractional difference between this value computed at a variety of resolutions, $N < 60$, against the value for the resolution at $N = 60$. Noting the log scale of the vertical axis, we see that the approach to the highest resolution value is exponential. In the righthand frame of the Figure we show the imaginary part of the Einstein--Hilbert action (plus Gibbons Hawking term) at the final flow time, normalized by its real part. This should vanish in the continuum limit, and we again see a nice exponential decrease with $N$ as we would hope for with pseudospectral differencing if the flow solution is smooth.

\bibliographystyle{JHEP}
\bibliography{references}

\end{document}